\documentclass{emulateapj}

\newcommand{\noprint}[1]{}

\newcommand{\myemail}{ebuenzli@email.arizona.edu}

\slugcomment{Draft version \today}

\shorttitle{Brown Dwarf Photospheres are Patchy}
\shortauthors{Buenzli et al.}

\begin{document}

\title{Brown Dwarf Photospheres are Patchy: A Hubble Space Telescope Near-infrared Spectroscopic Survey Finds Frequent Low-level Variability}

\author{Esther Buenzli$^{1,2}$, D\'aniel Apai$^{1,3}$, Jacqueline Radigan$^4$, I. Neill Reid$^4$ and Davin Flateau$^3$}

\affil{$^1$Department of Astronomy and Steward Observatory, University of Arizona, Tucson, AZ 85721, USA, \myemail}
\affil{$^2$Max-Planck Institute for Astronomy, 69117 Heidelberg, Germany}
\affil{$^3$Department of Planetary Sciences and Lunar and Planetary Laboratory, University of Arizona, Tucson AZ 85721, USA}
\affil{$^4$Space Telescope Science Institute, Baltimore, MD 21218, USA}


\begin{abstract}
Condensate clouds strongly impact the spectra of brown dwarfs and exoplanets. Recent discoveries of variable L/T transition dwarfs argued for patchy clouds in at least some ultracool atmospheres. This study aims to measure the frequency and level of spectral variability in brown dwarfs and to search for correlations with spectral type. We used HST/WFC3 to obtain spectroscopic time series for 22 brown dwarfs of spectral types ranging from L5 to T6 at 1.1-1.7~$\mu$m for $\approx$40~min per object. Using Bayesian analysis, we find 6 brown dwarfs with confident $(p>95\%)$ variability in the relative flux in at least one wavelength region at sub-percent precision, and 5 brown dwarfs with tentative $(p>68\%)$ variability. We derive a minimum variability fraction $f_{min}=27^{+11}_{-7}\%$ over all covered spectral types. The fraction of variables is equal within errors for mid L, late L and mid T spectral types; for early T dwarfs we do not find any confident variable but the sample is too small to derive meaningful limits. For some objects, the variability occurs primarily in the flux peak in the J or H band, others are variable throughout the spectrum or only in specific absorption regions. Four sources may have broad-band peak-to-peak amplitudes exceeding 1\%. Our measurements are not sensitive to very long periods, inclinations near pole-on and rotationally symmetric heterogeneity. The detection statistics are consistent with most brown dwarf photospheres being patchy. While multiple-percent near-infrared variability may be rare and confined to the L/T transition, low-level heterogeneities are a frequent characteristic of brown dwarf atmospheres.
\end{abstract}


\section{Introduction}

Brown dwarfs, objects with masses below $\approx$75 M$_{J}$, cool with increasing age because their cores cannot reach the temperatures required to sustain hydrogen fusion. As these objects cool, they undergo significant spectral evolution that is largely driven by the formation and dispersal of condensate clouds and changing molecular opacities. These changes define the spectral sequence through the M, L, T, and Y spectral classes \citep[e.g.][and references therein]{kirkpatrick05}. Detailed photometric and spectroscopic observations of hundreds of objects over a wide wavelength range 
\citep[e.g.][]{geballe02,knapp04,leggett07,cushing08,stephens09} combined with models of dust cloud evolution and gas-phase chemistry in brown dwarf atmospheres \citep[e.g.][]{allard01,ackerman01,lodders02,cooper03,tsuji04, burrows06, helling08} encompass our current knowledge of the atmospheric properties and evolution of L and T dwarfs. 

Silicate dust grains are thought to be the most significant condensates that form opaque cloud layers in the photospheres of L dwarfs. Towards late L spectral types, the cloud optical thickness increases in the visible photosphere. As the effective temperature of the brown dwarfs falls below T$_{eff}\approx1,300$ K, dramatic changes in the spectra suggest that these clouds disappear below the visible photosphere. This marks the L/T transition, where near-infrared colors change from red (J-K$\approx$1-2 mag) to blue (J-K$\approx$0 mag) with T$_{eff}$ decreasing by only $100-200$ K. The J-band flux brightens and peaks at spectral type $\approx$T5 \citep{dupuy12}. Absorption features, in particular H$_2$O and CH$_4$, then dominate the near-infrared spectra. T dwarfs beyond T4 (T$_{eff}<1,000$ K) are modeled fairly well with clear atmospheres, but some evidence suggests a reappearance of condensates, potentially sulfide and alkali clouds (Morley et al. 2012). 

The L/T transition poses the largest challenge to models \citep[e.g.][]{saumon08}, and the physical mechanism behind cloud dispersal in T dwarfs is not yet well understood. One possibility is cloud thinning: increasing cloud particle sizes lead to rapid rain out that thins and eventually removes the clouds \citep[e.g.][]{tsuji04, knapp04, burrows06}. Another is the appearance of holes in clouds, where flux emerges from deeper, hotter regions \citep{ackerman01, burgasser02, marley10}. Because age, gravity, metallicity, and cloud properties affect the emergent spectra of field brown dwarfs, it is difficult to distinguish between different models of cloud evolution. Young giant planets have red colors that indicate significant clouds \citep{barman11, skemer12, marley12}, potentially because their surface gravity is lower than for the clear field brown dwarfs of similar temperatures. 

Recent discoveries of early T dwarfs which exhibit photometric variability at the multiple percent level \citep{artigau09, radigan12, gillon13, girardin13} have opened a new window into the cloud structure of brown dwarfs. The variability is thought to arise from patchy cloud structure that results in a modulated light curve as the object rotates. Furthermore, evolving
light curves over time scales of hours \citep{apai13}, days \citep{artigau09, radigan12, gillon13} and years \citep{metchev13} suggest a variety of weather phenomena. First attempts at two- and three-dimensional dynamical modeling of brown dwarf atmospheres were made by \citet{freytag10} and \citet{showman12} and suggest that cloud cover evolution is possible on short timescales. 

Spectroscopic variability can simultaneously provide longitudinal and vertical information on the atmospheric structure. Searches for spectroscopic variability from the ground \citep{nakajima00, bailerjones08, goldman08} proved challenging and were inconclusive, but recent space-based studies yielded detailed and surprising results on the nature of the heterogeneous atmospheric structure of brown dwarfs. Spectroscopic time series with HST/WFC3 for two highly-variable early T dwarfs \citep{apai13} and a T6.5  \citep{buenzli12} revealed that the characteristics of the variability are significantly different for the two L/T transition objects than for the object beyond the L/T transition. Predominantly gray spectral variation indicates two cloud components of different thickness in the early T dwarfs. For the T6.5, light curves for different narrow wavelength regions have different phases and the phase difference correlates with the probed pressure at a given wavelength, indicating complex horizontal and vertical structure. 
 
From this very small sample of variables it is difficult to draw conclusions about the mechanism of cloud dispersal through the L/T transition. Surveys to constrain the occurrence rate of near-IR broad-band variability for brown dwarfs \citep{enoch03, koen04, clarke08, khandrika13, girardin13} find variability frequencies between $\approx$10-40\% depending on what studies were included and what were the amplitude limits set to qualify as a detection. However, the significance of these studies is limited since the samples were not chosen in a uniform and unbiased way; they included several objects as detections that are actually presented only as tentative variables in need of confirmation, and non-detections were not evaluated in terms of potential for long-period or low-amplitude variability.

While high-amplitude ($\gtrsim3\%$) periodic variability thus far appears to be limited to L/T transition objects, lower amplitude variability has been found both for earlier type L and later type T dwarfs \citep{heinze13, clarke08, buenzli12}. Also, I band surveys found transient and non-periodic variability for several early L dwarfs \citep{bailerjones01, gelino02, koen03, koen05, koen13}. However, most of these detections are not robust enough to allow a statistical analysis of variability as a function of spectral type. It remains unclear how widespread patchiness in brown dwarf atmospheres actually is. Lower-level variability on the order of $\sim1\%$ or in only narrow spectral regions may be occurring frequently but would be missed by precision-limited broad-band photometric surveys. It is the goal of our study to fill this gap in order to better estimate the true frequency of photospheric patchiness.

In this paper, we present an unbiased HST snapshot spectroscopic survey for near-IR variability in brown dwarfs from mid L to mid T spectral types. Each target is surveyed for only 30-40 min but with point-to-point precision of 0.1-0.2\%, spanning the J and H near-IR flux peaks and several absorption features. We identify several new variable brown dwarfs in various spectral bands and discuss the frequency of near-IR variability. In Section 2 we describe the survey, the observations and data reduction. In Section 3 we present the data and show new detections of variability and confidence intervals or upper limits derived from a bayesian analysis. In Section 4 we discuss the occurrence rate of variability as a function of spectral type and wavelength. Our conclusions are presented in Section 5. 

\section{HST Snapshot survey}

An HST snapshot\footnote{User Information Report
UIR-2012-003} program consists of a large number of targets that are evenly
distributed on the sky that require only short visits of parts of one orbit. From these targets,
a subset of targets that are optimal to fill gaps in the HST schedule are selected to increase the observing efficiency of the telescope. For our snapshot program (GO12550, PI Apai) we selected 60 brown dwarfs\footnote{Full initial target list available at \url{http://www.stsci.edu/hst/phase2-public/12550.pro}} between spectral types L5 and T6.5, of which 22 targets, selected practically randomly, were observed. For each target, we obtained spectral time series of $\approx30-45$ minutes with Wide Field Camera 3 (WFC3).

\subsection{Target selection}

The initial target sample was selected from the DwarfArchives\footnote{http://spider.ipac.caltech.edu/staff/davy/ARCHIVE/index.shtml}, a compendium of over 1,000 brown dwarfs, to uniformly span the spectral subtypes between L5 and T6, sampling effective temperatures between $\approx1,700-800$~K and very different stages of cloud evolution. No special selection for or against known young low-gravity objects were made. The selection was unbiased in the sense that prior knowledge about variability was not a selection criterion except for the exclusion of the three 
known variable T dwarfs previously observed with HST. We excluded objects that were known resolved binaries (two known binaries were mistakenly included, one of them in the observed sample) and objects with known 2MASS sources within 20\arcsec \, in order to minimize the risk for overlapping spectra. We prioritized the targets by their J band 
brightness, but ensured that the selected sources were evenly distributed in color-magnitude space and spectral type. Because of the random selection of the subset of targets that were actually observed from this sample, the final spectral type distribution of the observed sample is not entirely uniform. 

We binned the subset of 22 targets that were observed in the SNAP survey into 4 spectral bins that correspond to distinct evolutionary stages: mid L
(L5-L7), late L (L7.5-L9.5), early T (T0-T3) and mid T (T3.5-T6). The early T dwarfs, which host the
strongest known near-infrared variables \citep{artigau09, radigan12, apai13, gillon13}, 
are the smallest sample with only three objects. The mid L and mid T sample has 7 objects each, the late Ls 5 objects (cf. Fig. \ref{fig:specfreq}).
The target properties are summarized in Table \ref{tab:sources}.

\begin{deluxetable*}{lccccccc}
\tablecaption{Properties of target brown dwarfs\label{tab:sources}} 
\tablehead{
\colhead{Target name} & \colhead{SpT} & \colhead{SpT}  & \colhead{SpT}  &  \colhead{2MASS J}  & \colhead{2MASS H} & \colhead{Dist.} & \colhead{Dist.} \\ 
\colhead{           } & \colhead{(IR)} & \colhead{(Opt)} & \colhead{ref.}  &   \colhead{[mag]} &  \colhead{[mag]}  &  \colhead{[pc]} &  \colhead{ref.}
}
\startdata
2MASS J00001354+2554180 & T4.5	&		& B06 & 15.06$\pm$0.04 & 14.73$\pm$0.07  & 14.1$\pm$0.4 & D12	\\ 
2MASS J02431371-2453298 & T6 	&		& B06 & 15.38$\pm$0.05 & 15.14$\pm$0.11  & 10.7$\pm$0.4 & V04	\\
2MASS J03105986+1648155 & L9+L9 & L8	& B06, K00, S10 & 16.03$\pm$0.08 & 14.93$\pm$0.07  & 27.1$\pm$2.5 & S13	\\
2MASS J04210718-6306022 &  		& L5$\beta$	& C09 & 15.57$\pm$0.05  & 14.28$\pm$0.04  & 22.8$\pm$3.4$^a$ 	&  \\ 
2MASS J05395200-0059019 & L5 	& L5 	& K04, F00 & 14.03$\pm$0.03  & 13.10$\pm$0.03  & 12.2$\pm$0.4 & A11	\\ 
2MASS J05591914-1404488 & T4.5 	& T5 	& B06, B03 & 13.80$\pm$0.02  & 13.68$\pm$0.04  & 10.4$\pm$0.1 & D12	\\ 
2MASS J06244595-4521548 &  		& L5 	& R08, C07 & 14.48$\pm$0.03  & 13.34$\pm$0.03  & 11.9$\pm$0.6 & F12	\\ 
2MASS J08014056+4628498 &  		& L6.5 	& K00 & 16.27$\pm$0.13  & 15.45$\pm$0.14  & 31.1$\pm$4.6$^a$ 	&  \\
2MASS J08173001-6155158 & T6	&  		& A10 & 13.61$\pm$0.02  & 13.53$\pm$0.03  & 4.9$\pm$0.3 & A10	\\
2MASS J08251968+2115521 &  		& L7.5 	& K00 & 15.10$\pm$0.03  & 13.79$\pm$0.03  & 10.7$\pm$0.1 & D02	\\
2MASS J09083803+5032088 & L9$\pm$1 & L7	& K04, C07 & 14.55$\pm$0.02  & 13.48$\pm$0.03  & 9.7$\pm$1.4$^a$ &    \\ 
2MASS J09090085+6525275 & T1.5 &  		& C06 & 16.03$\pm$0.09  & 15.21$\pm$0.10  & 19.0$\pm$2.8$^a$ & 	\\
2MASS J10393137+3256263 & T1 	& 		& C06 & 16.16$\pm$0.03  & 15.47$\pm$0.03  & 21.8$\pm$2.4$^a$ &  	\\
2MASS J12195156+3128497 & L8 	&		& C06 & 15.91$\pm$0.08  & 14.91$\pm$0.07  & 20.3$\pm$3.0$^a$ & ß	\\
2MASS J13243553+6358281 & T2.5pec &		& K10 & 15.57$\pm$0.07  & 14.58$\pm$0.06  & 13.6$\pm$1.5$^a$ &  	\\
2MASS J15150083+4847416 & L6 & L6 		& W03, C07 & 14.11$\pm$0.03  & 13.10$\pm$0.03  & 11.3$\pm$1.7$^a$ & 	\\ 
2MASS J16241436+0029158 & T6 	&  		& B06 & 15.49$\pm$0.05  & 15.52$\pm$0.10  & 11.0$\pm$0.2 & T03	\\
2MASS J16322911+1904407 & L8 	& L8 	& B06, K99 & 15.87$\pm$0.07  & 14.61$\pm$0.04  & 15.2$\pm$0.5 & D02 	\\ 
2MASS J17114573+2232044 &  	& L6.5 		& K00 & 17.09$\pm$0.18  & 15.80$\pm$0.11  & 30.2$\pm$4.3 & V04	\\
2MASS J17502484-0016151 & L5.5 	&  		& K07 & 13.29$\pm$0.02  & 12.41$\pm$0.02  & 9.2$\pm$0.2 & A11 \\ 
2MASS J17503293+1759042 & T3.5 	&  		& B06 & 16.34$\pm$0.10  & 15.95$\pm$0.13  & 27.6$\pm$3.4 & V04	\\
2MASS J23391025+1352284 & T5 	&  		& B06 & 16.24$\pm$0.10  & 15.82$\pm$0.15  & 18.8$\pm$3.8$^a$ & 	\\
\enddata
\tablenotetext{a}{Spectrophotometric distances for sources where no parallax data is available, calculated from the relation of spectral type vs H band absolute magnitude given in \citet{dupuy12}.}
\tablecomments{References: A10: \citet{artigau10}, A11: \citet{andrei11}, B03: \citet{burgasser03}, B06: \citet{burgasser06}, C06: \citet{chiu06}, C07: \citet{cruz07}, C09: \citet{cruz09}, D02: \citet{dahn02}, D12: \citet{dupuy12},
F00: \citet{fan00}, F12: \citet{faherty12}, K99: \citet{kirkpatrick99}, K00: \citet{kirkpatrick00}, K04: \citet{knapp04}, K07: \citet{kendall07}, K10: \citet{kirkpatrick10}, R08: \citet{reid08}, S10: \citet{stumpf10}, S13: \citet{smart13}, T03: \citet{tinney03}, V04: \citet{vrba04}, W03: \citet{wilson03}}
\end{deluxetable*}

\subsection{Observations}

Observations were taken between October 2011 and October 2012 with HST/WFC3 in the infrared channel.
We used the G141 grism which provides slitless spectra for wavelengths 1.1-1.7 $\mu$m. The detector
is a Teledyne HgCdTe with $1024 \times 1024$ pixels. The pixel size is 0.13\arcsec. All observations were executed in the $256 \times 256$ subarray
mode with a field of view of approximately $30 \times 30$\arcsec. In this mode, for our exposure times all images acquired in a
visit could be stored without the need for a WFC3 buffer dump, ensuring maximum observing efficiency.
The first order of the spectrum is fully captured on the subarray, while the zeroth and second
orders are not recorded. The spectrum has a dispersion of 4.65 nm pixel$^{-1}$ and spans $\approx$
140 pixels. For wavelength calibration, we obtained a direct image at the beginning of the visit
through the F132N or F127N narrowband filter that provides an accurate measurement of the location
of the source on the detector.

We used the SPARS25 readout mode for all targets. In this mode, a sequence of non-destructive reads
is taken in one exposure. At the beginning of each exposure, a 0 s read and a 0.27 s read are taken.
Then, a number of reads of 22.34 s are obtained. The time for each read is fixed, but the number of
reads can be chosen between 1 and 15 depending on the brightness of the target. We set the number of
reads to 1, 2, 5 or 10  such that the maximum number of counts acquired by the brightest pixel was between
$5,000-15,000$ counts. This ensured that there was no significant image persistence, which can become relevant
if a pixel is exposed above about half-well ($\approx$40,000 counts). The number of exposures
varied between 9 and 67 depending on the number of reads (effective exposure time) and duration of
the visit. The spectroscopic time series lasted between 32 to 45 minutes and the cadence (exposure
time plus overhead) was between 41 and 242 s. The observation details for each source are summarized
in table \ref{tab:obsdetail}.

\begin{deluxetable*}{lccccccc}
\tablecaption{Log of observations \label{tab:obsdetail}} 
\tablehead{
\colhead{Target name} & \colhead{Obs. date} & \colhead{Visit dur.$^a$}  & \colhead{Exp. time}  &  \colhead{N$_{exp}$}  & \colhead{N$_{read}$$^b$} & \colhead{Cadence} & \colhead{Notes} \\
\colhead{           } &                     & \colhead{[min]} 		& \colhead{[s]}   	 &       &    &   \colhead{[min]} & 
}
\startdata
2MASS J00001354+2554180	&	2012-09-23	&	41.7	&	45.0	&	40	&	2	& 1.05	&	\\
2MASS J02431371-2453298	&	2011-12-31	&	40.9	&	112.0	&	19	&	5	& 2.17	&	\\
2MASS J03105986+1648155	&	2012-08-25	&	40.0	&	223.7	&	10	&	10	& 4.03	& c \\
2MASS J04210718-6306022	&	2012-03-20	&	40.9	&	112.0	&	19	&	5	& 2.17	&	\\
2MASS J05395200-0059019	&	2012-03-01	&	38.6	&	45.0	&	37	&	2	& 1.05	&	\\
2MASS J05591914-1404488	&	2011-10-16	&	42.1	&	22.6	&	62	&	1	& 0.68	&   \\
2MASS J06244595-4521548	&	2012-05-08	&	39.6	&	45.0	&	38	&	2	& 1.05	&	\\
2MASS J08014056+4628498	&	2011-11-10	&	44.1	&	223.7	&	11	&	10	& 4.03	&	\\
2MASS J08173001-6155158	&	2011-10-09	&	45.5	&	22.6	&	67	&	1	& 0.68	& d \\
2MASS J08251968+2115521	&	2012-05-09	&	45.2	&	112.0	&	21	&	5	& 2.17	&	\\
2MASS J09083803+5032088	&	2011-12-09	&	41.7	&	45.0	&	40	&	2	& 1.05	&	\\
2MASS J09090085+6525275	&	2012-08-21	&	40.0	&	223.7	&	10	&	10	& 4.03	&	\\
2MASS J10393137+3256263	&	2012-05-08	&	44.1	&	223.7	&	11	&	10	& 4.03	&	\\
2MASS J12195156+3128497	&	2012-06-18	&	36.0	&	223.7	&	9	&	10	& 4.03	&	\\
2MASS J13243553+6358281 &	2012-02-25	&	45.2	&	112.0	&	21	&	5	& 2.17	&	\\
2MASS J15150083+4847416	&	2012-02-23	&	37.5	&	45.0	&	36	&	2	& 1.05	&	\\
2MASS J16241436+0029158	&	2012-07-13	&	36.5	&	112.0	&	17	&	5	& 2.17	& e \\
2MASS J16322911+1904407	&	2012-08-11	&	44.1	&	223.7	&	11	&	10	& 4.03	&	\\
2MASS J17114573+2232044	&	2012-08-01	&	36.0	&	223.7	&	9	&	10	& 4.03	& f \\
2MASS J17502484-0016151	&	2012-06-15	&	43.4	&	22.6	&	64	&	1	& 0.68	&	\\
2MASS J17503293+1759042	&	2012-10-05	&	32.0	&	223.7	&	8	&	10	& 4.03	&	\\
2MASS J23391025+1352284	&	2012-08-21	&	44.1	&	223.7	&	11	&	10	& 4.03	&	\\
\enddata
\tablenotetext{a}{Visit duration not including acquisition and direct image}
\tablenotetext{b}{Number of non-destructive reads per exposure not including the zero read and first very short read}
\tablenotetext{c}{Object is a resolved binary (Stumpf et al. 2010); wavelengths between 1.18-1.26 $\mu$m and $>1.65$ $\mu$m include a large number of bad pixels (see text)}
\tablenotetext{d}{Spectrum is cut off at 1.61 $\mu$m (see text)}
\tablenotetext{e}{Overlap with 1st order spectrum of a background star for $\lambda<1.18$ $\mu$m and with 2nd order for $\lambda \gtrsim 1.5$ $\mu$m. (see text)}
\tablenotetext{f}{Overlap with a faint background star for $\lambda \gtrsim 1.65$ $\mu$m. (see text)}
\end{deluxetable*}

We kept the location of the spectra on the detector fixed and did not dither during the
observations. This avoided issues with pixel-to-pixel sensitivity variations that cannot be
corrected to sub-percent precision by flatfielding. The positional stability was better than 0.06
pixels for all targets as verified by cross-correlation of the images.

For 4 targets, parts of the spectrum could not be used: For 2M0310, the proper motion given in \citet{faherty09} was found to be incorrect\footnote{A new proper motion measurement by \citet{smart13} is consistent with our data. The new $\mu_{RA}$ value differs from the one given in \citet{faherty09} by $\sim1$\arcsec/yr, the $\mu_{DEC}$ value by $\sim$0\farcs1/yr, while given errors are only 0.02\arcsec/yr. Because of this unusually large discrepancy, and because there were no issues with 6 other proper motion values from \citet{faherty09} that we used, we assume this problem is inherent to the 2M0310+16 value only, perhaps due to a mistake when creating the table.} Due to the resulting location of the spectrum near the upper edge of the subarray, wavelengths from 1.18 to 1.26 $\mu$m and $>1.65$ $\mu$m, were affected by a large number of bad pixels and had to be cut from the spectrum. For 2M0817, a mistake with the proper motion value in the observing preparation file resulted in the spectrum being located across the right edge of the subarray, and wavelengths $>1.61$~$\mu$m were not read out. For target 2M1624, the first and second order spectra of a background star that was missed in the target selection overlap partially with the target spectrum for $\lambda<1.18$ $\mu$m and $\lambda>1.65$ $\mu$m. For target 2M1711, the spectrum of a very faint background star which is not in the 2MASS Point Source Catalogue partially overlaps for $\lambda\gtrsim1.65$ $\mu$m.

\subsection{Data reduction}

We reduced the data that is output by the standard WFC3 pipeline with custom IDL routines and the
PyRAF software package aXe\footnote{http://axe-info.stsci.edu}, a tool developed for extracting and
calibrating slitless spectroscopic data. The WFC3 pipeline calfw3 delivers two dimensional spectral
images that are zero-read and dark subtracted, corrected for non-linearity and gain and include
flags for bad pixels.

We started from the .flt files that already include the combined images from all subreads of an exposure. 
Unlike in Apai et al. (2013), we do not extract the individual subreads because the signal-to-noise
ratio of the spectra in the individual subreads is too low. We therefore do not apply the correction
for the small flux loss within one exposure as a function of subreads. This correction would slightly increase the average flux
of each exposure by a common factor and would not introduce any relative change in the time series. 

Cosmic ray hits were identified as $>5\sigma$ outliers for a given pixel compared to the same pixel in the nearest 8 frames in the time series and
replaced by the median. We corrected bad pixels that were flagged by the pipeline by interpolating over nearest neighbors in the same row.
Like \citet{berta12} we found that only flag numbers 4 (dead pixel), 32 (unstable pixel), 256 (saturated) and
512 (bad in flatfield) impact the flux in a pixel in a significant way, we therefore only corrected bad pixels
with those flags. 

Because the aXe pipeline cannot handle subarray images, we embedded the
frames into full-frame images and flagged the extra pixels with a data quality flag to exclude them
from further processing. With the axeprep routine we subtracted the background that is determined
by scaling a master-sky frame. We then used the axecore routine that flatfields the frames, performs wavelength calibration,
extracts the two-dimensional spectra and flux-calibrates with the G141 sensitivity curve. We chose
the extraction width to be 6 full-width-half-maxima (FWHM). For smaller values, we found that for some objects an artificial variability slope could be introduced that increased for narrower extraction widths. We did not find significant differences when using 6 to 8 FWHM, therefore we used 6, which minimizes the noise. The FWHM is measured on the direct image at the beginning of the observations. 

We calculated the error in each pixel as a combination of the photon noise of the source, the error from the sky subtraction and 
the readout noise. In the final step we corrected a ramp effect that we 
previously identified and discussed in \citet{apai13}. There, we corrected the effect by using
an analytical function fitted to data from a non-variable star. Because our current program includes
several additional non-variable sources, we refine the ramp correction as elaborated in the next
section. 

\subsection{Correction of the ramp effect}
\label{sect:ramp}

\begin{figure}
\epsscale{1.0}
\plotone{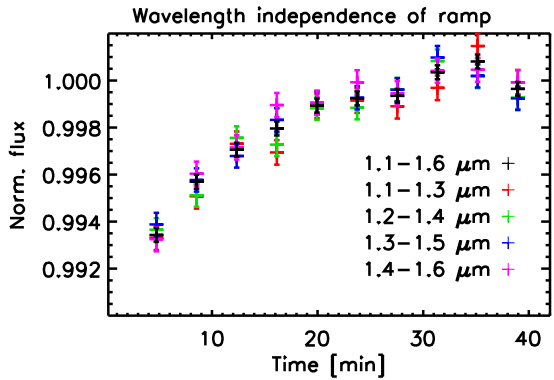}
\caption{The ramp derived from the non-variable star from GO12314 for 4 wavelength bins: 1.1-1.3 $\mu$m (red), 1.2-1.4 $\mu$m (green), 1.3-1.5 $\mu$m (blue), 1.4-1.6 $\mu$m (purple). 
Also shown is the ramp integrated over the whole spectrum (black). Different reads from the same exposure were averaged to increase the SNR. 
\label{fig:rampwav}}
\end{figure}

\begin{figure}
\epsscale{1.1}
\plotone{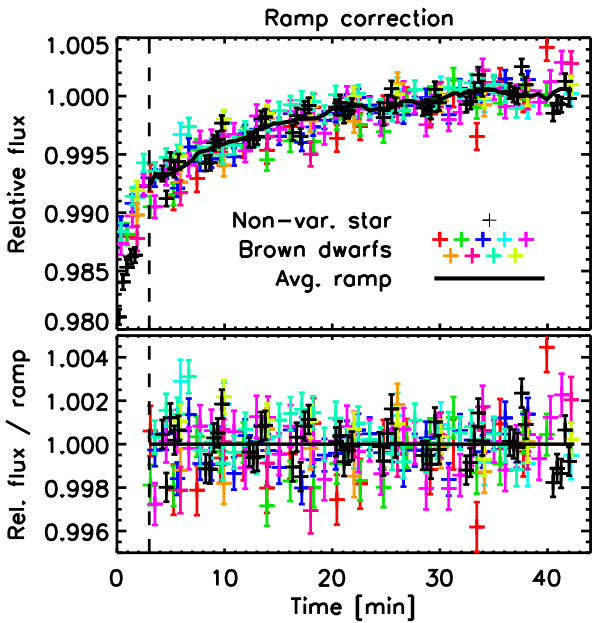}
\caption{Top: Relative flux integrated from 1.05 to 1.7 $\mu$m for 9 brown dwarfs and a comparison star. The flux ramp
is very similar for all objects, indicating that the ramp is a detector effect and these objects are intrinsically 
non-variable. The line is a moving average. We exclude the first 180 s due to large scatter. Bottom: Residual flux after dividing by the average ramp.
The standard deviation of the residuals is 0.00115. 
\label{fig:ramp}}
\end{figure}

\begin{figure}
\epsscale{0.8}
\plotone{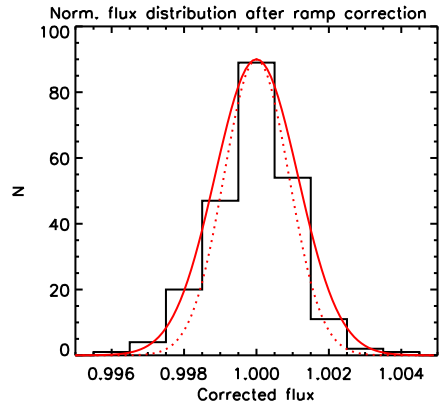}
\caption{Histogram for the distribution of the normalized flux for non-variable objects after the ramp correction (see Fig. \ref{fig:ramp}, bottom). Overplotted is a gaussian
with a standard deviation of 0.00115 (standard deviation of residuals, solid line) and a gaussian with a standard deviation of 0.00090 (average random error, dotted line).  
\label{fig:reshist}}
\end{figure}

Time series observations with the WFC3/IR channel have shown the presence of a ramp effect where the
measured flux increases strongly at the beginning of an orbit and then flattens out. The effect is
present for several readout modes, but characteristics differ. The ramp was first analyzed for the RAPID
readout mode by \citet{berta12}, and characterized in more detail for different subarray sizes and sampling
modes in RAPID and SPARS10 readout mode by \citet{swain13} and \citet{mandell13}. They find a correlation
of the ramp features with the length of the WFC buffer dump. Here we discuss the ramp for time series data taken in the SPARS25 mode 
and the 256 x 256 subarray, which allows observations of faint targets without intra-orbit buffer dumps. 

The flux increase corresponds to $\approx$2\% during the first orbit of a visit, and only $\approx 0.7$\%
for subsequent orbits in the same visit \citep{apai13}. Because all data in this program were taken in single orbits, we
focus here only on the ramp characteristics of the first orbit. 

In a first step, we corrected the data from our targets with the ramp derived from the non-variable star
data from program GO12314 \citep{apai13}. There, we had already found that the ramp depends neither
on wavelength nor count rate. Figure \ref{fig:rampwav} shows the ramp of the star for different wavelength regions.
It agrees very well for all wavelengths. \citet{berta12} also found their ramp to be achromatic. We therefore derived the ramp for each target by 
integrating over the full spectral range from 1.12 to 1.66 $\mu$m to maximize the SNR. We excluded the 4 sources with
missing wavelength segments or overlap with a background star in order to avoid other systematics.
It was immediately evident that 6 of the remaining 18 objects showed clearly different behavior 
than the non-variable star, and different for each object. This 
indicates inherent time-variability for these objects on top of the detector systematic. We therefore 
excluded these objects from further analysis for the ramp. From the other 12 objects and the
non-variable star, we calculated a moving average with time steps of 30 s and bin width of 180 s. Subtracting
this ramp from the data, we found small trends for 3 additional sources, again with different trends for
each source. We therefore removed those as well for the final ramp correction. For the remaining 9 objects (2M0243, 2M0421, 2M0539, 2M0801, 
2M0908, 2M0909, 2M1039, 2M1515, 2M1632), the agreement of the 
ramp with the non-variable star was very good for all times after $\approx$180 s
after the beginning of the observations (Fig. \ref{fig:ramp}). There is a large scatter in the first 3 min where the ramp is steepest. We therefore disregard 
the first 180 s of each time series from all further analysis in order not to introduce artificial trends
at the beginning of the time series. Beyond 43 min, only a few objects have data. From there, we extend the ramp
as a horizontal line. Comparing with the previous 10 min, this is valid to $\approx0.1$\% level. 

The final ramp is shown in Fig. \ref{fig:ramp}. We find evidence
for small fluctuations that deviate from the analytical fit made in \citet{apai13}. 
If we create the ramp by averaging different subsets of these objects, the differences are $<0.1\%$. 
After dividing the data by the derived average ramp, we find a standard deviation of $0.115\%$ for the
residuals. Because of the small-scale fluctuations, there may be correlated errors on the order of $0.1\%$ 
over the span of 5-10 minutes. However, the overall distribution of residuals is reasonably close to gaussian
(Fig. \ref{fig:reshist}) with the standard deviation similar to the average random error of individual points ($0.090\pm0.015\%$).  
We add an error of $0.1\%$ in quadrature to account for the uncertainties in the ramp correction. 
For points shortwards of 5 min and longwards of 39 min, where the ramp correction is less certain, we add an error of $0.2\%$ in 
quadrature. 

We tested whether the ramp or the residuals after the correction for the different objects correlate with the positional shifts of $0.01-0.06$ pixels perpendicular to the spectral trace within one orbit. We do not find any correlation and conclude that the ramp and the trends after the ramp correction are not caused by positional instabilities as it is the case for warm Spitzer photometry \citep[e.g.][]{heinze13, lewis13}.  We note however, that for some objects, small shifts ($\lesssim0.1$ pixels) parallel to the spectral trace can result in small wavelength shifts. This can produce an artificial variability trend if the spectra are integrated over a narrow wavelength interval with a strong gradient in counts, in particular at $> 1.66$ $\mu$m where the grism sensitivity drops strongly. We cannot reliably fit for the shifts because their influence on the other parts of the spectrum is too small. Similarly, \citet{mandell13} used only the spectral edges to fit for these shifts in their data of transiting planets. We therefore cannot reliably disentangle the effect of the shifts from potential true spectral variability that occurs only in that region. We exclude these outermost wavelength regions from our analysis and check for all detected variables that wavelength shifts cannot be responsible for the observed trends.

Finally, we used the data of a background star from a newer HST program (13176, PI Apai) that used the same observing mode to verify our reduction and calibration. We reduced and analyzed it in the same way as all the data in our program. For all wavelength bins we find the star to be non-variable to very good precision, as expected. The results are shown in Figure \ref{fig:v98} in the online appendix.

\section{Results}

\begin{figure*}
\epsscale{0.8}
\plotone{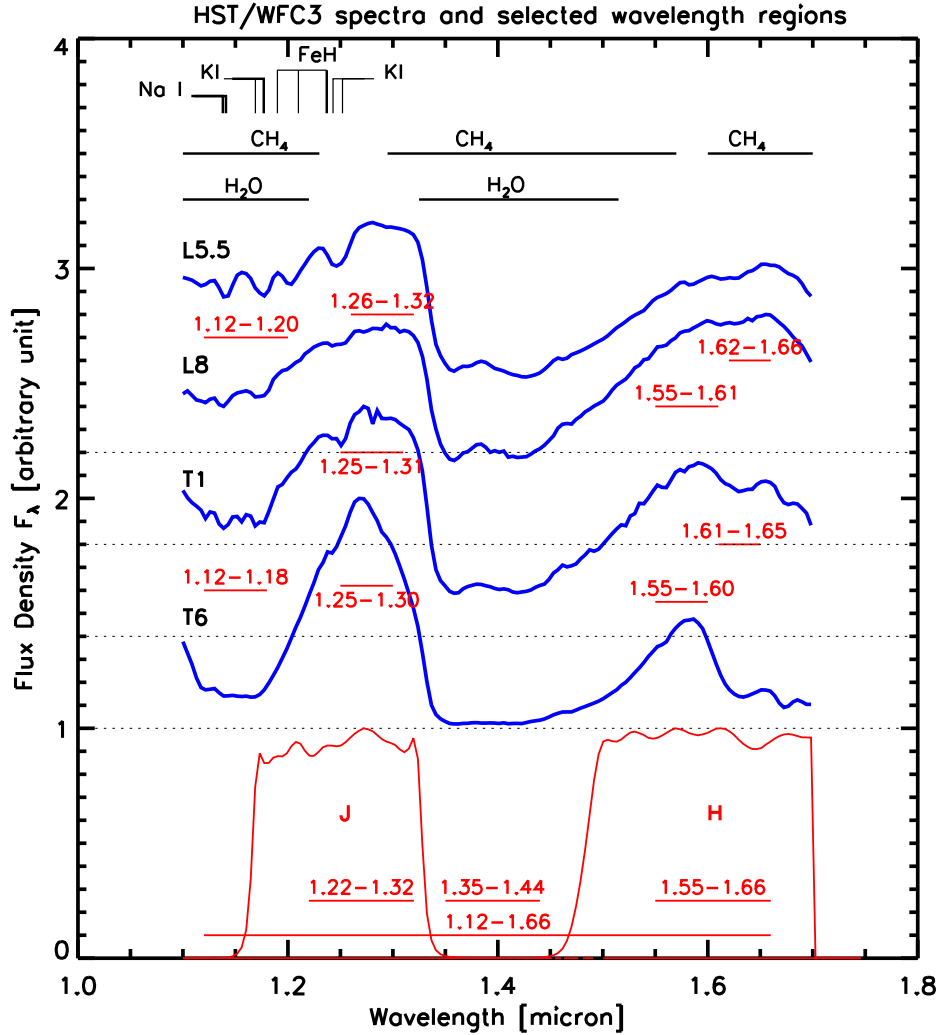}
\caption{HST/WFC3 spectra of four selected sources that are representative for their respective spectral type bins. Spectra are normalized by the maximum and shifted for clarity. 
The dotted line indicates the zero line for the four spectra. Red lines are the filter curves used to create light curves: MKO J and H (truncated at 1.7 $\mu$m) and flat filters in narrow wavelength regions. 
Notable absorption features are indicated.
\label{fig:spectra}}
\end{figure*}

To assess the time variability of each source as a function of wavelength, we integrate the spectral time series over several wavelength regions
in order to derive light curves. We use the J and H MKO filter profiles (H filter cut off at 1.7 $\mu$m) to get the broad-band variability,
and several flat filter profiles for narrow wavelength regions covering either a flux peak or an absorption band. 
Notable absorption features include the water band at $\approx1.35-1.44$ $\mu$m that becomes deeper with later spectral types, alkali features mixed
with water and methane in the $1.12-1.18$ $\mu$m region, potassium and FeH at $1.19-1.26$ $\mu$m and methane at $\sim1.62$ $\mu$m for the T dwarfs.

Figure \ref{fig:spectra} shows representative spectra for the different spectral types together with the chosen wavelength regions. 
We then fit a linear function to each light curve. For most sources, a linear function is an appropriate fit to the variability. This is not surprising because the duration of the observations is
significantly shorter than the rotation period of these objects, which are likely on the order of a few hours. For two objects we find that they appear to go through a minimum, and a linear
fit is very poor in terms of least squares residuals. However, we find a linear function is still a good fit to the curve when starting at the minimum and we derive the slope from that part of the light curve only.

\begin{figure}
\epsscale{1.1}
\plotone{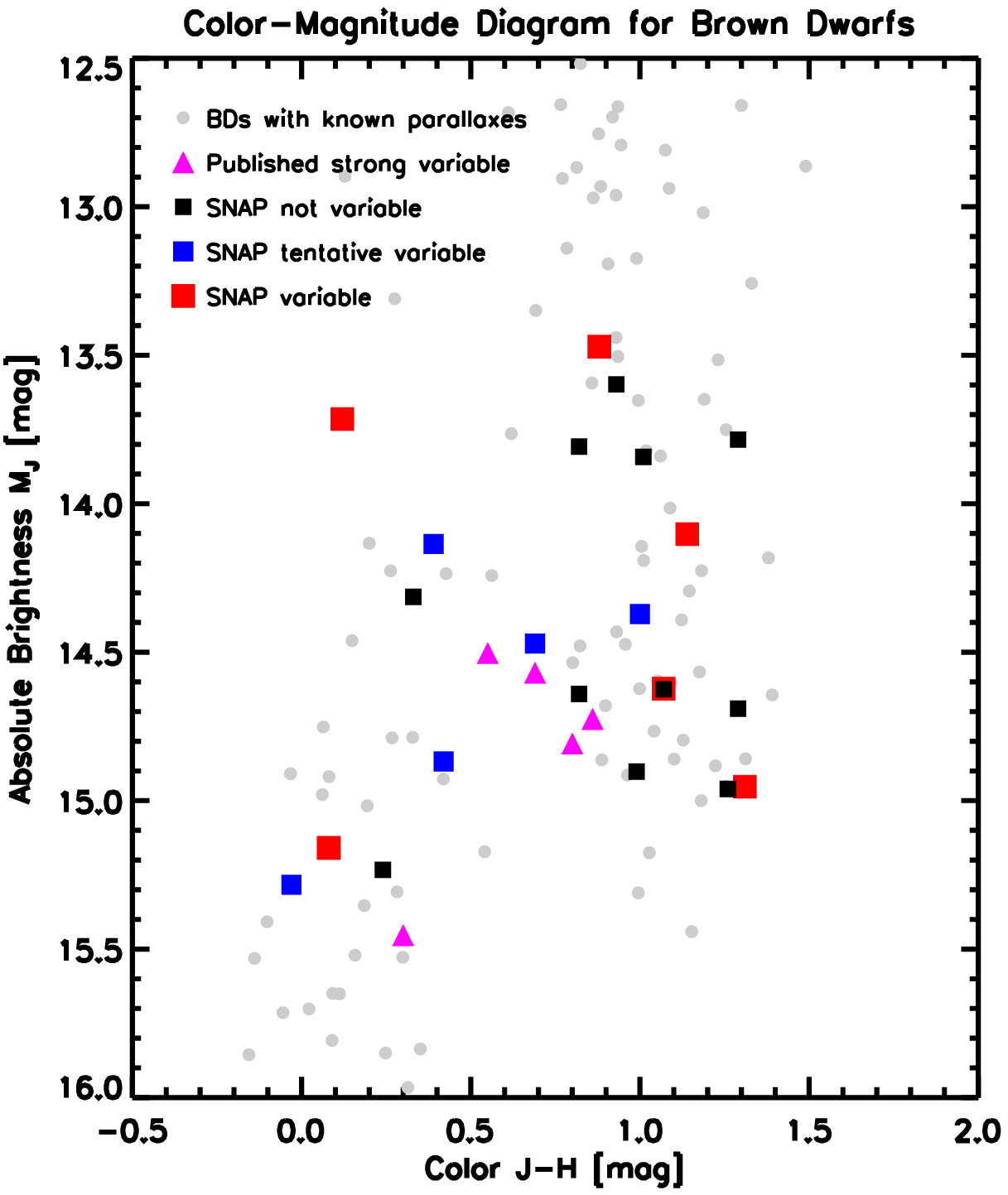}
\caption{Color-magnitude diagram J-H vs M$_J$ (2MASS) that shows the location of our sources with respect to L and T field dwarfs with known parallaxes (grey dots). Also shown are the 5 published brown
dwarfs with significant variabiliy ($>3\%$ in at least one wavelength band, purple triangles). Sources from our survey are divided into 3 groups, selected by wavelength region with the strongest
variability signal: confident variables ($>95\%$ probability, red squares), tentative variables ($>68\%$ probability, blue squares) and non-variables (rest, black squares). 
\label{fig:cmd}}
\end{figure}

Figures \ref{fig:v01} to \ref{fig:v67} in the appendix show the light curves for all sources and all wavelengths with the best-fit slopes. 
Figure \ref{fig:cmd} shows the location of the sources in a color-magnitude diagram together with field L and T brown dwarfs. They are color
coded by the confidence in the variability in the best wavelength band. We find 6 confident ($>95\%$) and 5 tentative ($>68\%$) variables, as elaborated in the following sections. 
Most of them are variable only in particular regions of the spectrum. For the remaining 11 sources we find no evidence for variability above the uncertainty level on the time scale of 40 min. 

We also inspect the direct images to look for potential binarity. 2M0310+16 is a known resolved binary with 0\farcs2 separation \citep{stumpf10}. It is marginally resolved in the direct image. However, because of the HST roll angle, the separation of the spectra on the chip is only about 0\farcs14 or 1~pixel and a separate extraction of the two spectra is not possible. For all sources, we fit a two-dimensional Gaussian to search for potential elongation of the PSF. Typically, the FWHMs are 0.15-0\farcs2 ($\sim$1-1.5 pixels, i.e undersampled PSF) along both axes and only for 2M0310+16 there is clear elongation. We rule out binarity at a separation of  $\gtrsim$0\farcs2 for all other sources. Better limits could perhaps be set by careful PSF subtraction and tests with fake companions, but this goes beyond the scope of our paper.

\subsection{Bayesian analysis of variability slope}

Because the error bars in each point are of similar order as the measured flux changes, the random noise can influence the slope that
is ultimately measured. To determine the probability density function for the true slope $a_t$ given the measured slope $a_m$ and error bars, we 
use Bayes' theorem, that states 
\begin{equation}
p(a_t | a_m) =  \frac{ p(a_m | a_t) p (a_t) }{ \int_0^\infty p(a_m | a_t) p(a_t) \mathrm{d} a_t },
\end{equation}
where $p(a_t | a_m)$ is the posterior probability distribution, $p(a_m | a_t)$ the likelihood function, and $p(a_t)$ the prior probability density function for variability slopes. 

\begin{figure}
\epsscale{1.2}
\plotone{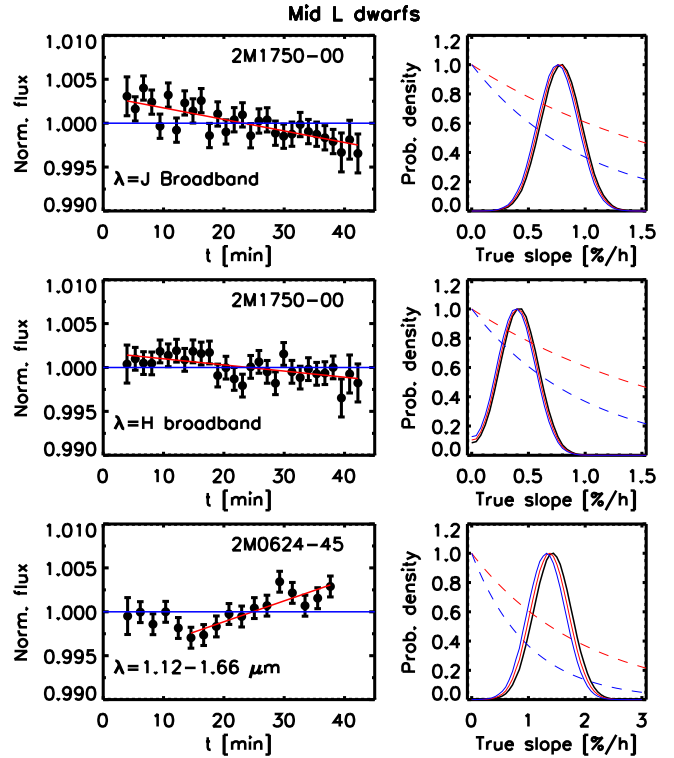}
\caption{Selected light curves for the 2 mid L dwarfs that show variability. The left panels are the observations with a best-fit linear slope (red). The right panel gives the probability (normalize to 1 for maximum probability) for the value of the true slope calculated from Bayes' theorem. The black line is the likelihood function for the slope, the dashed red and blue line two prior distributions, and the red and blue solid lines the corresponding posterior distribution functions. For each target, the most relevant wavelength ranges were selected, all others are shown in Figures \ref{fig:v01} to \ref{fig:v67} in the online appendix. For 2M1750-00, a significant trend is found in the J band (top) but not in the H band (middle). For 2M0624-45 (bottom), the curve appears to go through a minimum and we only fit the rising slope. Here, the whole spectrum is integrated because the variability is similar across all wavelengths. 
\label{fig:midL}}
\end{figure}

\begin{figure}
\epsscale{1.2}
\plotone{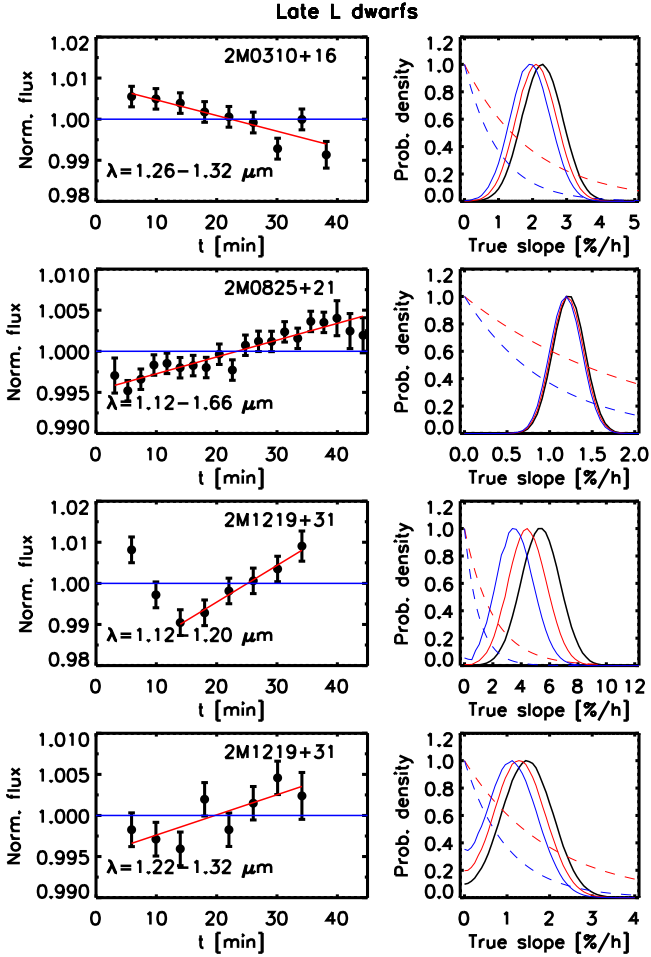}
\caption{Selected light curves for the 3 late L dwarfs that show variability (confident or tentative). Lines are like in Fig. \ref{fig:midL}.  For 2M1219+31, which appears to go through a minimum for the 1.12-1.2 $\mu$m wavelength range, we only fit the rising slope. However, this does not happen for the 1.12-1.32 $\mu$m range.
\label{fig:lateL}}
\end{figure}

\begin{figure}
\epsscale{1.2}
\plotone{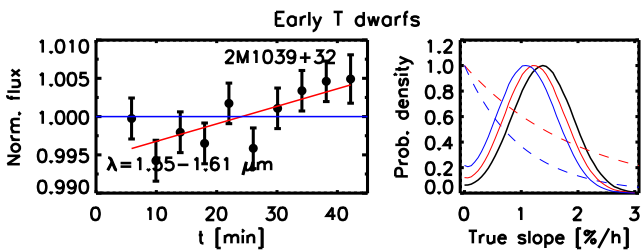}
\caption{Light curve for the only early T dwarf that shows tentative variability. Lines are like in Fig. \ref{fig:midL}.
\label{fig:earlyT}}
\end{figure}

\begin{figure}
\epsscale{1.2}
\plotone{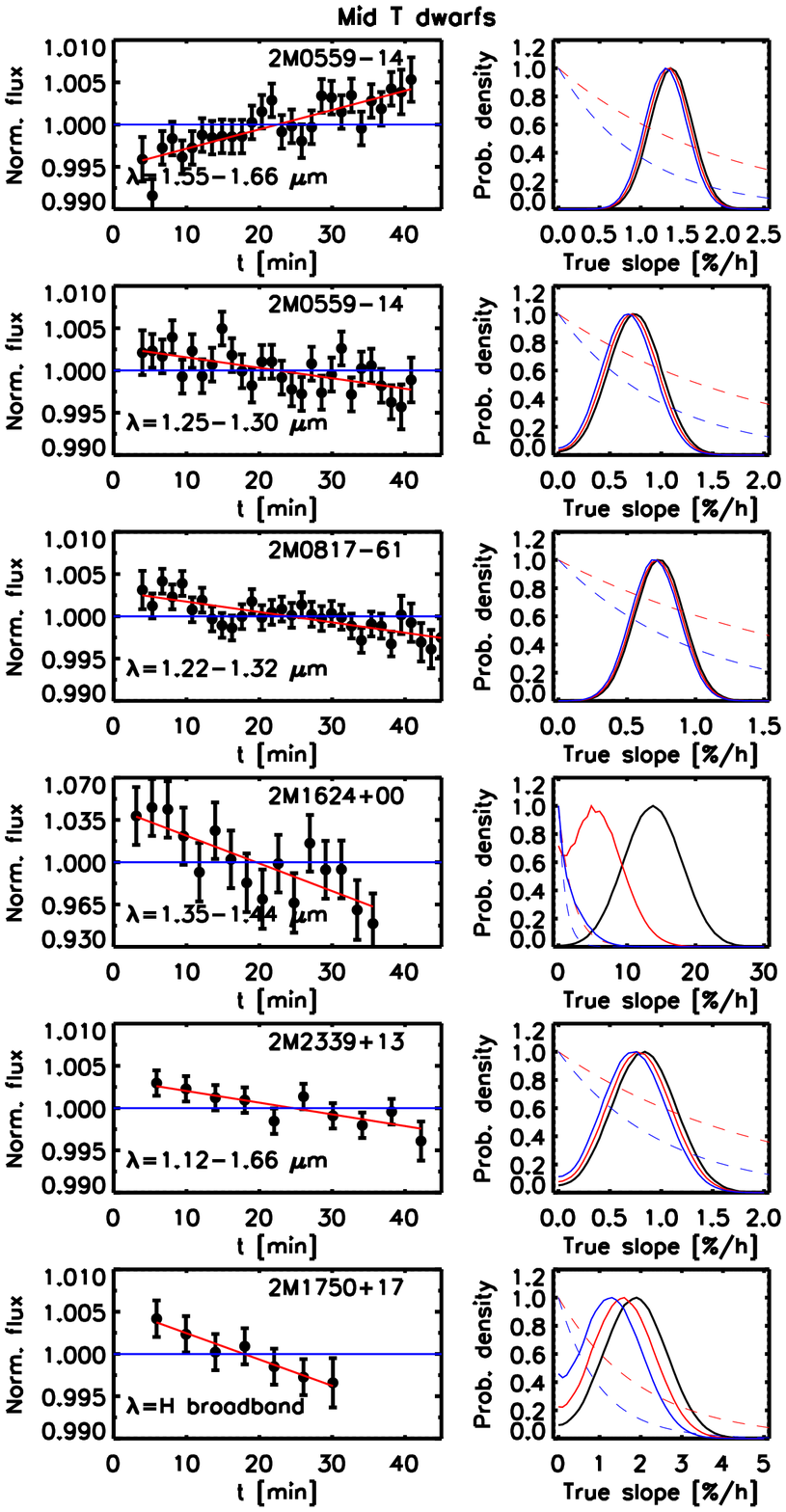}
\caption{Selected light curves for the 5 mid T dwarfs that show variability (confident or tentative). Lines are like in Fig. \ref{fig:midL}.
\label{fig:midT}}
\end{figure}

We perform a Monte Carlo simulation to determine the likelihood function for the slope for each source and each wavelength range. For a range of slopes, we add random gaussian noise with the appropriate standard deviation and measure the new slopes. Repeating this 50,000 times, we measure how often the measured slope occurs, where we accept the slope to be equal if it is within a 1$\sigma$ interval of the measured slope. For an smaller acceptance interval, the likelihood function would be slightly narrower. The resulting likelihood function, normalized to the maximum likelihood, is shown in Figures \ref{fig:midL} to \ref{fig:midT} as the solid black line.

Choosing an appropriate prior distribution is not obvious because this is what we ultimately want to measure in this study. Using a flat prior (which would equate the posterior probability
distribution with the likelihood function) is inappropriate in this case because $a_t$ does not have an upper bound. We know from earlier studies that strong variability is rare. We therefore adopt an exponentially declining function as prior $p(a_t) \propto e^{-ba_t}$ where we use two different values of $b$ to explore a more optimistic and a more pessimistic case and study the resulting differences. The priors are plotted as dashed blue or red lines in Figures \ref{fig:midL} to \ref{fig:midT}, and the resulting posterior probability distribution as solid blue or red lines (both normalized to their maximum value).  Because none and very small variability slopes are favored for this prior, which may not necessarily be true, the derived amplitudes may be slightly underestimated. However, the likelihood function corresponds to the posterior probability for a flat prior with a large upper bound, which can be used for comparison. In any case, the choice of the prior has only minimal influence on the solid detections of variability, where the likelihood function is very small for slopes near zero. However, the measured large slopes in wavelength regions with large error bars, most notably for the water band of 2M1624, are overwhelmed by the prior. Our short duration observations are therefore not able to draw firm conclusions about variability in deep absorption bands, where the signal to noise ratio is intrinsically lower. 

From the resulting posterior probability distributions, we compute the 68\% and 95\% credible intervals (highest density) for the slopes. These bayesian confidence intervals denote the range in which the parameter lies with 68\% or 95\% probability. They do no correspond to frequentist confidence intervals because our prior is not uninformative. These intervals, together with the maximum likelihood and maximum probability for the slope, are tabulated in Table \ref{tab:all} for the different wavelength bands for all sources. All slopes are given in units of \% relative flux change per hour in order to be able to compare between sources with different visit durations. We classify a detection of variability as significant if the 
lower boundary of the 95\% confidence interval level for the strong prior (L95, P2 in table \ref{tab:allvar}) is $>0.3\%/h$, and as tentative if the lower boundary of the 68\% confidence interval for the strong prior (L68, P2 in table \ref{tab:allvar} is $>0.3\%/h$. An exception to this rule is elaborated in the next section for a case where the strong prior overwhelms even a strong likelihood function.  
The slopes and confidence intervals for sources with confident or tentative variability are listed in table \ref{tab:allvar}. In the following discussion of the variability, we generally refer to the slope $a$ as column Max P2 in table \ref{tab:allvar} unless otherwise noted. 

As an alternative, we also calculate $\chi^2$ values. We derive the $\chi^2$ probability $p(null)$ of the null hypothesis that the data can be explained by a flat line. We also provide the reduced $\chi^2$ value for the best-fit linear model. On average, the reduced $\chi^2$ is 0.82, indicating that our error bars may be slightly too large. This is mainly due to the additional 0.1\% error that we add to all points for the uncertainty of the ramp correction, and 0.2\% for points at the beginning and end of the time series. For all confident variables from our Bayesian analysis, $p(null) < 0.01$ in at least one wavelength band, most often significantly lower. For the bright objects and the broader filters, where the ramp correction error dominates the total error, the p(null) value is overestimated. On the other hand, for a few non-variable cases with larger scatter than explained by the error bars, p(null) may be underestimated. We do not consider the results from the $\chi^2$ analysis for our further discussion, but provide them in tables \ref{tab:allvar} and \ref{tab:all} for comparison.

Figures \ref{fig:midL} to \ref{fig:midT} show the observations as well as the likelihood, prior distribution and posterior distribution for the true slope for each target for selected wavelength regions with a significant signal for confident and tentative variability. In the appendix, figures \ref{fig:v01} to \ref{fig:v67} include all selected wavelength bands for all sources and table \ref{tab:all} all confidence intervals including upper limits for non-variables. 

\begin{deluxetable*}{lcccccccccccccccc}
\tablecaption{All sources with confident or tentative linear variability\label{tab:allvar}} 
\tablehead{
\colhead{Target name} &  \colhead{Filter} &\colhead{ML} & \colhead{L95}  & \colhead{L68}  &  \colhead{Max}  & \colhead{U68} & \colhead{U95} & \colhead{{\bf L95}}  & \colhead{{\bf L68}}  &  \colhead{{\bf Max}}  & \colhead{U68} & \colhead{U95} &  \colhead{sg} & \colhead{p} & \colhead{Red.} \\
\colhead{           } &  [$\mu$m] &  &    \colhead{P1}		&    \colhead{P1}	 &    \colhead{P1}   &  \colhead{P1}  &  \colhead{P1}  &  \colhead{{\bf P2}}  &  \colhead{{\bf P2}}  &  \colhead{{\bf P2}}  &  \colhead{P2}  &  \colhead{P2} & & \colhead{null} & \colhead{$\chi^2$} 
}
\startdata
2MASS J06244595-4521548 & 1.35-1.44 & 1.90 & 0.49 & 1.07 & 1.70 & 2.29 & 2.88 & 0.32 & 0.88 & 1.50 & 2.10 & 2.68 & + & 0.031  &  0.36 \\
 & 1.55-1.61 & 1.20 & 0.03 & 0.55 & 1.02 & 1.58 & 1.96 & 0.00 & 0.41 & 0.96 & 1.43 & 1.79 & + & 0.24  &  0.71 \\
 & 1.62-1.66 & 1.76 & 0.46 & 1.01 & 1.60 & 2.18 & 2.74 & 0.30 & 0.84 & 1.44 & 1.99 & 2.55 & +  & 0.14  &  0.41 \\
 & 1.55-1.66 & 1.38 & 0.45 & 0.85 & 1.26 & 1.70 & 2.10 & 0.37 & 0.76 & 1.20 & 1.61 & 2.01 & + & 0.046  &  0.73 \\
 & 1.12-1.66 & 1.44 & 0.73 & 1.04 & 1.38 & 1.70 & 2.02 & 0.68 & 0.99 & 1.32 & 1.64 & 1.96 & +  & 0.002  &  0.61 \\
 & J   & 1.26 & 0.18 & 0.63 & 1.08 & 1.57 & 2.00 & 0.05 & 0.52 & 1.02 & 1.46 & 1.82 & + & 0.11  &  0.9 \\
 & H   & 1.32 & 0.52 & 0.87 & 1.26 & 1.63 & 1.99 & 0.45 & 0.80 & 1.14 & 1.56 & 1.92 & +  & 0.022  &  0.85 \\
2MASS J17502484-0016151 & 1.12-1.20 & 0.96 & 0.43 & 0.67 & 0.92 & 1.19 & 1.43 & 0.39 & 0.64 & 0.87 & 1.15 & 1.40  & + & 0.003 & 1.12 \\
 & 1.22-1.32 & 0.68 & 0.31 & 0.49 & 0.68 & 0.86 & 1.04 & 0.29 & 0.47 & 0.64 & 0.84 & 1.03  & + & 0.27 & 0.47 \\
 & 1.26-1.32 & 0.80 & 0.36 & 0.56 & 0.76 & 0.98 & 1.18 & 0.34 & 0.54 & 0.76 & 0.96 & 1.16 & + & 0.13 & 0.61  \\
 & 1.12-1.66 & 0.56 & 0.25 & 0.39 & 0.52 & 0.69 & 0.83 & 0.24 & 0.38 & 0.52 & 0.67 & 0.82 & + & 0.57 & 0.25 \\
 & J   & 0.80 & 0.43 & 0.60 & 0.76 & 0.95 & 1.12 & 0.41 & 0.58 & 0.76 & 0.93 & 1.10  & + & 0.018 & 0.65 \\
\hline
2MASS J03105986+1648155 & 1.26-1.32 & 2.30 & 0.97 & 1.52 & 2.10 & 2.70 & 3.28 & 0.79 & 1.35 & 1.90 & 2.52 & 3.10 & - & 0.0004 & 0.96 \\
2MASS J08251968+2115521 & 1.12-1.20 & 1.60 & 0.85 & 1.19 & 1.53 & 1.95 & 2.31 & 0.78 & 1.13 & 1.53 & 1.87 & 2.23 & + & 0.002 & 0.85  \\
 & 1.22-1.32 & 1.02 & 0.47 & 0.73 & 1.02 & 1.28 & 1.54 & 0.43 & 0.69 & 0.97 & 1.24 & 1.50 & + & 0.003 & 1.04 \\
 & 1.26-1.32 & 1.15 & 0.39 & 0.74 & 1.09 & 1.47 & 1.83 & 0.32 & 0.67 & 1.04 & 1.41 & 1.76 & + & 0.001 & 1.24 \\
 & 1.35-1.44 & 1.24 & 0.52 & 0.85 & 1.18 & 1.55 & 1.88 & 0.47 & 0.79 & 1.12 & 1.49 & 1.82 & + & 0.068 & 0.56 \\
 & 1.55-1.61 & 1.04 & 0.48 & 0.75 & 1.04 & 1.31 & 1.58 & 0.45 & 0.71 & 0.99 & 1.27 & 1.54 & + & 0.012 & 0.82 \\
 & 1.62-1.66 & 1.04 & 0.38 & 0.68 & 0.99 & 1.33 & 1.64 & 0.33 & 0.63 & 0.94 & 1.28 & 1.59 & + & 0.010 & 1.06 \\
 & 1.55-1.66 & 1.05 & 0.59 & 0.81 & 1.05 & 1.28 & 1.50 & 0.57 & 0.78 & 1.00 & 1.25 & 1.48 & + & 0.001 & 0.72 \\
 & 1.12-1.66 & 1.20 & 0.84 & 1.02 & 1.20 & 1.40 & 1.58 & 0.83 & 1.00 & 1.20 & 1.38 & 1.56 & + & $<10^{-4}$ & 0.53 \\
 & J   & 1.11 & 0.62 & 0.85 & 1.11 & 1.35 & 1.59 & 0.59 & 0.82 & 1.05 & 1.32 & 1.56 & + & 0.0004 & 0.96 \\
 & H   & 0.93 & 0.52 & 0.71 & 0.93 & 1.14 & 1.34 & 0.50 & 0.69 & 0.93 & 1.11 & 1.32 & +& 0.003 & 0.65 \\
2MASS J12195156+3128497 & 1.12-1.20 & 5.46 & 1.83 & 3.08 & 4.42 & 5.77 & 7.10 & 1.03 & 2.22 & 3.38 & 4.84 & 6.16 & + & 0.0001 & 0.09 \\
 & 1.22-1.32 & 1.44 & 0.10 & 0.69 & 1.28 & 1.89 & 2.38 & 0.00 & 0.52 & 1.12 & 1.70 & 2.11 & + & 0.024 & 1.03 \\
 & J   & 1.44 & 0.09 & 0.68 & 1.28 & 1.88 & 2.35 & 0.00 & 0.51 & 1.12 & 1.68 & 2.09 & + & 0.011 & 1.15\\
 \hline
2MASS J10393137+3256263 & 1.55-1.61 & 1.38 & 0.22 & 0.71 & 1.20 & 1.75 & 2.24 & 0.08 & 0.57 & 1.08 & 1.61 & 2.03 & + & 0.029 & 0.98 \\
\hline
2MASS J05591914-1404488 & 1.25-1.30 & 0.72 & 0.22 & 0.46 & 0.72 & 0.97 & 1.21 & 0.18 & 0.42 & 0.68 & 0.93 & 1.17 & - & 0.17 & 0.82 \\
 & 1.55-1.60 & 0.72 & 0.05 & 0.36 & 0.68 & 0.98 & 1.23 & 0.00 & 0.31 & 0.64 & 0.93 & 1.14 & + & 0.10 & 1.06 \\
 & 1.61-1.65 & 1.97 & 0.99 & 1.42 & 1.88 & 2.34 & 2.78 & 0.89 & 1.31 & 1.70 & 2.23 & 2.68 & + & 0.0095 & 0.79 \\
 & 1.55-1.66 & 1.36 & 0.85 & 1.08 & 1.36 & 1.59 & 1.84 & 0.81 & 1.05 & 1.30 & 1.56 & 1.80 & + & 0.0003 & 0.70 \\
 & H   & 1.24 & 0.77 & 0.99 & 1.24 & 1.45 & 1.67 & 0.74 & 0.96 & 1.18 & 1.42 & 1.64 & + & 0.0004 & 0.67 \\
2MASS J08173001-6155158 & 1.22-1.32 & 0.76 & 0.39 & 0.55 & 0.72 & 0.90 & 1.07 & 0.37 & 0.54 & 0.72 & 0.89 & 1.06 & - &  0.070 & 0.58  \\
 & 1.25-1.30 & 0.80 & 0.36 & 0.56 & 0.76 & 0.99 & 1.19 & 0.33 & 0.54 & 0.76 & 0.96 & 1.17 & - & 0.005 & 1.08\\
 & 1.55-1.60 & 0.72 & 0.16 & 0.42 & 0.68 & 0.97 & 1.23 & 0.12 & 0.38 & 0.68 & 0.93 & 1.19 & - & 0.017 & 1.20 \\
 & J   & 0.60 & 0.27 & 0.43 & 0.60 & 0.77 & 0.94 & 0.25 & 0.41 & 0.60 & 0.76 & 0.92 & - & 0.37 & 0.47 \\
2MASS J16241436+0029158 & 1.35-1.44 & 13.80 & 0.00 & 1.44 & 4.80 & 8.67 & 12.12 & 0.00 & 0.00 & 0.00 & 2.61 & 6.06 & - & 0.058 & 0.62 \\
2MASS J17503293+1759042 & 1.55-1.66 & 2.00 & 0.00 & 0.78 & 1.60 & 2.47 & 3.06 & 0.00 & 0.43 & 1.30 & 2.04 & 2.70 & - & 0.11 & 0.33 \\
 & H   & 1.90 & 0.07 & 0.81 & 1.60 & 2.32 & 2.90 & 0.00 & 0.54 & 1.30 & 2.01 & 2.55 & - & 0.13 & 0.10 \\
2MASS J23391025+1352284 & 1.22-1.32 & 0.84 & 0.00 & 0.39 & 0.76 & 1.18 & 1.44 & 0.00 & 0.31 & 0.68 & 1.09 & 1.36 & - & 0.045 & 1.08 \\
 & 1.12-1.66 & 0.84 & 0.18 & 0.47 & 0.76 & 1.09 & 1.38 & 0.14 & 0.42 & 0.76 & 1.04 & 1.33 & - & 0.082 & 0.55\\
 & J   & 0.84 & 0.09 & 0.44 & 0.78 & 1.16 & 1.47 & 0.02 & 0.38 & 0.72 & 1.09 & 1.35 & - & 0.058 & 0.91 
\enddata
\tablecomments{Columns are: Maximum likelihood, lower limit of 95\% and 68\% confidence interval, maximum of probablity distribution, upper limit of  68\% and 95\% confidence interval, all intervals for prior 1 and prior 2. All slopes are given in \%/h. The sg column gives the sign of the slope, where + means rising and - decreasing.  p(null) is the p-value for the null hypothesis that there is no variability using a $\chi^2$-test. Red. $\chi^2$ is the reduced $\chi^2$ value of the best fitting linear model. For 2M0624 (all wavelengths) and 2M1219 (only 1.12-1.20) the best-fit model was derived from the partial light curve due to non-linearity. The horizontal lines divide the spectral type bins. From top to bottom: L5-L7, L7.5-L9.5,T0-T3 and T3.5-T6. Table \ref{tab:all} in the appendix is the full table with the data for all sources and wavelengths with limits for non-detections.}
\end{deluxetable*}

\subsection{Mid L dwarfs}

Out of 7 objects with spectral types between L5 and L7, we find 2 which show significant variability. The most obvious is 2M1750-00 (L5.5, Fig. \ref{fig:midL} and \ref{fig:v61}), one of the brightest objects in our sample. It shows a clear downward trend with $a\sim0.75$\%/h in J band that is significant at $>95\%$ level. The same result is found when looking at narrower wavelength regions between 1.12-1.32 $\mu$m, indicating that the amplitude is quite uniform across this wavelength range. In the H band, a smaller trend ($<0.5$\%/h) remains, but it is not statistically significant or even tentative. 

The second object showing clear variability is 2M0624-45 (L5, Fig. \ref{fig:midL} and \ref{fig:v19}). For this object, a linear model is a very poor fit. Both in J band H band (as well as the narrower flux peaks), the light curve appears to reach a minimum at $\approx15$ min. The variability is of similar shape and strength at all wavelengths within error bars. We therefore discuss only the white light curve integrated over the full spectrum from 1.12-1.66 $\mu$m in order to minimize the error bars. We divide the curve into two linear slopes and apply our bayesian analysis only to the second, longer slope. We find a $>95\%$ significant slope with $a>1$\%/h. 

The other 5 objects do not show any indication of variability above the uncertainties (see online Figures). The $68\%$ upper limits are below $0.5-1\%$/h for most wavelengths for 2M0539-00, 2M0801+46, 2M0421-63 and 2M1515+48 and 2M1711+22 (Table \ref{tab:all}). The small trends in the longest wavelength bins for 2M0421-63 were found to originate from small shifts in the wavelength at the spectrum edge (see Sect. \ref{sect:ramp}) and are therefore not real. 

\subsection{Late L dwarfs}

We observed 5 objects with spectral types between L7.5 and L9.5 and find 2 objects with significant, and one object with tentative variability. An interesting significant detection is made for the binary 2M0310+16 (L9+L9, Fig. \ref{fig:lateL} and \ref{fig:v12}), where we find a very strong slope $a\sim 2\%$/h in the $1.26 - 1.32$~$\mu$m J band flux peak with $>95\%$ confidence. A separate extraction of the two spectra was not possible. Since it is likely that the variability stems from only one component, the true amplitude for this object is likely even larger. This is reminiscent of the variable component in the very nearby binary WISEJ104915.57-531906.1AB \citep{gillon13}. For 2M0310, we could not derive the broad-band J and H variability because of a large number of bad pixels shortward of 1.26 $\mu$m and longward of 1.65 $\mu$m. 

Another significant variable is 2M0825+21 (L7.5, Fig. \ref{fig:lateL} and \ref{fig:v26}). For this object we find trends with $a\sim1$\%/h that are similar over the whole wavelength range. Finally, 2M1219+31 (L8, Fig. \ref{fig:lateL} and Fig. \ref{fig:v44}), is a curious case. At 1.12-1.20 $\mu$m, we find a quick $\sim$2\% drop and then a slower $\sim$2\% rise. Formally, the bayesian analysis on only the rising slope shows that it is significant at $>95\%$ confidence, with a slope of $3-6\%$/h depending on the prior. However, for this faint source the analysis is dependent on only very few points. In the 1.22-1.32 $\mu$m region, we also find an upward slope at $>68$\% confidence, but no downward slope. At longer wavelengths, there is no evidence of variability. We therefore classify this source only as a tentative, and not as a confident variable. 

The remaining two objects are not found to be variable: for both 2M0908+50 and 2M1632+19 the 68\% upper limits are mostly $<0.6\%$/h. The tentative trend in the H band for 2M1632+19 is not real for the same reason as for 2M0421-63. 

\subsection{Early T dwarfs}

Our observed sample contains only 3 early T dwarfs, and none show variability at $>95\%$ significance. We find tentative trends ($>68\%$ significance) for one of the three objects, but only in one particular narrow wavelength region (Fig. \ref{fig:earlyT}). For 2M1039+32 (T1, Fig. \ref{fig:v35}) we find a tentative trend at 1.55-1.61 $\mu$m, which corresponds to the H band peak. For 2M1324+63 (T2.5pec, Fig. \ref{fig:v44}), the upper limit on the variability at all wavelengths is $<0.6\%$ (68\% confidence). For the faint 2M0909+65 (T1.5, Fig. \ref{fig:v32}), there is a trend in the 1.61-1.65 $\mu$m CH$_4$ dip in the H band, but it does not fall into our classification of tentative variability.  

\subsection{Mid T dwarfs}

With 2 confident and 3 tentative variables out of 7 objects, the mid T dwarfs show the greatest variety in variability (Fig. \ref{fig:midT}). The most prominent variable is 2M0559-14 (T4.5, Fig. \ref{fig:v18}). It shows significant variability in the H band with $a>1$\%/h. The strongest change is seen in the CH$_4$ dip in the 1.61-1.65 $\mu$m region, but the H band peak shows tentative variability as well. There is also a small tentative variability trend in the J band peak, but in opposite direction. If real, this would indicate that the light curves are out of phase. 

The second significant variable is 2M0817-61 (T6, Fig. \ref{fig:v24}), where we find confident variability in the peak region of the J band at 1.22-1.30 $\mu$m with a slope of $\sim0.7$\%/h. There is also tentative variability in the H band peak in the same direction, while wavelengths beyond 1.60 $\mu$m are missing for this source. 

An interesting source is 2M1624+00 (T6, Fig. \ref{fig:v24}), which shows a strong maximum likelihood slope of $a\sim14$\% in the deep water absorption band at 1.35-1.44 $\mu$m. Because the individual error bars are about 2\%, the likelihood function is broad. For our chosen priors such large slopes are very unlikely, therefore the posterior probability is strongly influenced by the prior for this case, making the variability formally tentative for the weaker prior and non-existent for the stronger case. However, because there is no prior data on the occurrence rate of variability in deep water absorption bands, the prior, which is based on broad-band observations, may be too stringent here. We know that a slope of several percent is not impossible: for 2M2228-43, \citet{buenzli12} found a maximum slope of $a=7.6\%$/h at the same wavelengths. With a flat prior the variability would be confident. Because errors are large, the flux is very low and the prior important, we adopt this variable as tentative. We do not find variability at other wavelengths, but several regions are missing due to overlap with a background star. 

We find two other tentative variables. One is 2M2339+13 (T5, Fig. \ref{fig:v67}), which shows a tentative downward trend for the whole integrated spectrum, most of which is due to variability in the J band peak. On the other hand, 2M1750+17 (T3.5, Fig. \ref{fig:v62}) shows a tentative trend in the H band with $a>1\%$. Because of the sources' faintness and the very short duration of the visit, the prior influences the resulting probability distribution. 

The remaining two sources, 2M0000+25 (T4.5, Fig. \ref{fig:v01}) and 2M0243-24 (T6, Fig. \ref{fig:v10}) are not variable above the uncertainty level. For 2M0423-24, the small trends in the H band are artificial and due to wavelength shifts.

\section{Discussion}

\subsection{Comparison to previous surveys}

Four of our targets have previously been monitored for variability in the same wavelength range and here we compare our findings to earlier results. 

A target of particular interest is 2M0559-14. This bright mid-T dwarf is overluminous across the near- and mid-IR  \citep[e.g.][]{dahn02, dupuy12} and has therefore been suspected to be an unresolved flux-equal binary. \citet{liu08} speculate that the object may be marginally resolved in their HST/WFPC2 image. Our direct image with an FWHM of 0\farcs15 shows no indication of binarity, but the PSF is undersampled and we cannot rule out a very tight binary. Alternatively, it may represent the peak of the cloud clearing, however this does not explain the mid-IR overluminosity. 2M0559-14 was monitored for variability by \citet{clarke08} in J band. They find it to be non-variable at a level of 0.5\%. While we see a small trend in J band, the observed flux change is <0.5\% and tentative only when integrated over the narrow flux peak. The confident variability is seen only in H band and most strongly in the absorption band at 1.61-1.65 $\mu$m, which has not been monitored before. It seems unlikely that the variability stems solely from remnant clouds, because a cloud impacting an absorption band at low pressure level would also affect the deeper layers probed by the J band. Alternatively, circulation patterns may result in temperature perturbations in some layers as suggested for 2M2228-43 by \citet{buenzli12}. 

For 2M1711+22, \citet{khandrika13} report very strong ($\sim20\%$) but only marginally significant variability in a short 0.7 h sequence. Our observations cannot confirm this detection. We find no evidence of variability with an upper limit of $\approx2\%$. They also monitored 2M0825+21 in J band, but for only 0.5 h and did not find variability, whereas we find a change of about 0.8\% in 40 min. This is likely below the precision level of their very short measurement. 

\citet{nakajima00} attempted to measure spectral variability for 2M1624+00 over an 80 min time-span. They find tentative evidence for variability in water absorption features between 1.5-1.6 $\mu$m. We find potential strong water variability between 1.35-1.44 $\mu$m for this source, a wavelength region that is not accessible at sufficient precision from the ground. For the same source, \citep{koen04} did not find signs of variability in J band, consistent with our findings, although we are missing part of the J band due to overlap with a background star. 

\subsection{Variability occurrence rate}

We derive the occurrence rate of brown dwarf variability for mid-L to mid-T spectral types from our sample by determining the binomial confidence interval considering $k$ detections out of a sample of $n=22$ brown dwarfs. We found a total of 6 confident ($>95\%$) and 5 tentative ($>68\%$) detections of variability. Because our observations only span fractions of a rotation period and are still precision-limited, the derived variability fraction must be considered to be a lower limit. We derive the 1$\sigma$ confidence interval following the appendix in \citet{burgasser03b}. With 6 confident detections for a sample of 22, the minimum variability fraction is $f_{min} = 27^{+11}_{-7}\%$. Including the tentative detections, if they are all real, the minimum fraction raises to $f_{min} = 50\pm10\%$. The true number of the minimum fraction of variables from our sample is therefore likely to lie between about one third to half of brown dwarfs. 

Furthermore, it must be noted that the fraction of brown dwarfs with intrinsic heterogeneities is expected to be larger than the measured variablity fraction: we can only detect rotationally asymmetric components and the variability signal is reduced for inclinations towards pole-on.  Additionally, the survey may miss variables with very long periods or amplitudes below the photometric precision. We therefore expect that heterogeneity is a very important property of the condensate clouds and one that should be accounted for in ultracool atmosphere models.

\subsection{Limits on amplitudes and periods}

\begin{figure}
\epsscale{0.9}
\plotone{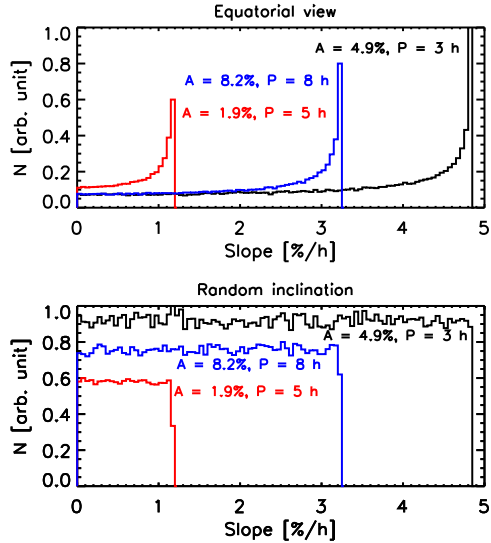}
\caption{Top: Distribution of slopes measured in 37 min simulated observations for a sinusoidal variability with given amplitude and period but random phase, seen from an equatorial viewpoint. Amplitude $A$ is the peak-to-peak amplitude, period $P$ one rotation period. (black), high variable, long period (blue) and low amplitude, medium period (red) model. 
Shown are a medium amplitude, short period. Bottom: the same but accounting for random orientation of the
rotation axis in the sky. Here, $A$ is peak-to-peak amplitude if the object were seen from an equatorial viewpoint, and $A\cos{i}$ the actual observable amplitude. For all combinations, the slope distribution is uniform between 0 and a maximum slope. 
\label{fig:slopehist}}
\end{figure}

In order to derive limits on the possible periods and amplitudes for our sample, we determine how the measured slope in an observation of 37 min (the average length of our observations) duration relates to the
actual amplitude and period of the object. We simulate sinusoidal light curves for a grid of amplitudes and periods at random phases and measure the distribution of different slopes. We find a strong peak in the frequency of slopes close to the maximum possible slope when assuming an equatorial view (Fig. \ref{fig:slopehist}, top panel). The maximum slope for any period / amplitude combination for a sine curve
is given analytically by the derivative at zero phase: $a_{max} = \pi A /  P$ where A is the peak-to-peak amplitude and P the period. For any measured slope $a$, we can therefore estimate the amplitude to be at minimum $A_{min}$ = $a P / \pi$ when assuming a period $P$. For example, for 2M0559-14, with $a=1.2\%/h$ in H band, we expect a peak-to-peak amplitude $A > 1\%$ for $P > 2.7$ h. At 1.61-1.65 $\mu$m it is even $A > 1.5\%$. Similarly, it is likely that $A\gtrsim1\%$ for 2M0825+21 and 2M0624-45 in J and H bands if $P>3$ h and for 2M1750-00 in J band if $P>4$ h. However, these calculations assume variability with a sine curve shape, which does not necessarily have to be the case.

If a brown dwarf in not seen from an equatorial view, the observed amplitude will be reduced. We therefore also account for the inclination of the brown dwarf that reduces the amplitude of the light curve by $\cos{i}$, assuming random orientations in the sky. In this case, we find that for any combination of period and true amplitude $A^\star$, the probability of measuring a particular slope is distributed uniformly between 0 and the maximum possible slope (Fig. \ref{fig:slopehist}, bottom). Although unlikely, an inclination near pole-on will yield a non-detection of variability even if the brown dwarf atmosphere is highly patchy. 

\begin{figure}
\epsscale{1.1}
\plotone{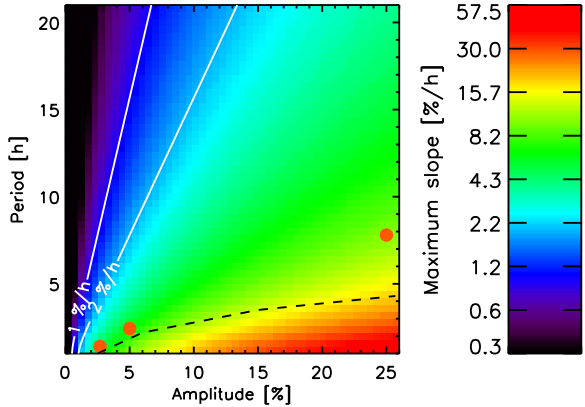}
\caption{Maximum slope (color coded) that can be measured for 37 min simulated observations of a brown dwarf with sinusoidal variability as a function of amplitude (peak-to-peak) and period at optimum phase and inclination. Lines indicate typical
measured slopes in our survey. For any measurement, all combinations to the right of the contour line (larger amplitudes or shorter period) are allowed. Non-detections exclude the area
below the dashed line, because near maximum or minimum phase significant curvature would be seen even in our short observations. Red dots indicate the known highly variables 2M2139+02 (J band), SIMP0136+09 (J band) and 2M2228-43 (H band), assuming edge-on orientation. 
\label{fig:slopes_ampper}}
\end{figure}

Figure \ref{fig:slopes_ampper} shows the maximum possible slope as a function of both period and amplitude. Overplotted are contours that correspond to typical measured slopes. We find slopes of at most $\sim2\%/h$, with the exception of 2M1624+00 in the deep water band, where the slope could be $>10\%/h$ if we do not adopt a prior that strongly favors smaller slopes. Period/amplitude combinations to the upper left of a contour line are not possible for that particular slope under the assumption that curves are sinusoidal. Combinations to the lower right, i.e. shorter period and larger amplitude, while not impossible, become increasingly unlikely further away from the contour. Extreme combinations with low periods and high amplitudes can generally be excluded for robust non-detections because they would exhibit significant curvature, even at maximum and minimum phases. This range is indicated by the dashed line. 

\begin{figure}
\epsscale{0.9}
\plotone{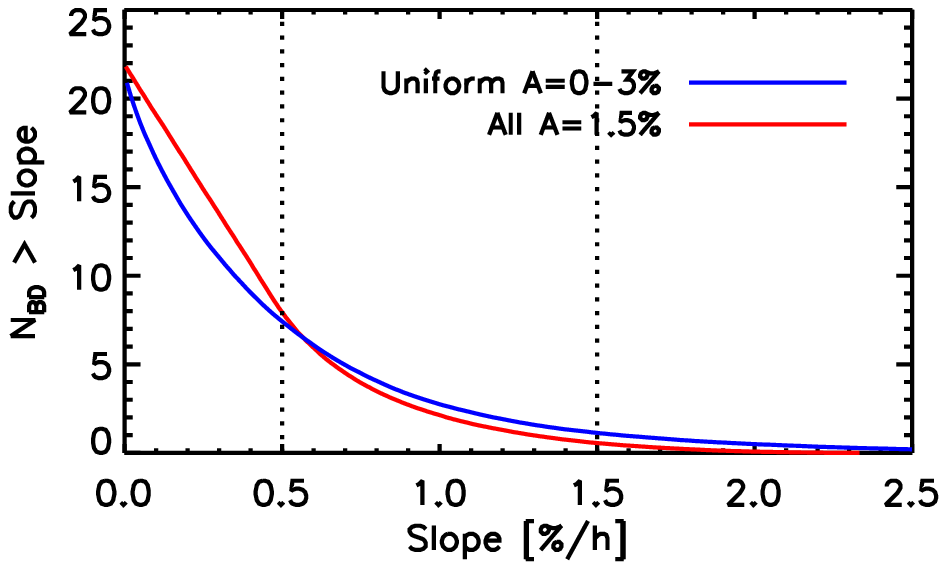}
\caption{Number of objects expected with a slope larger than a given number (on x axis) for two cases: all brown dwarfs are variable with an amplitude of 1.5\% when seen equatorially (red), or the brown dwarfs have a uniform amplitude distribution between
0 and 3\% (blue). Dotted lines indicate the approximate lower and upper limits of our measured slopes. 
\label{fig:slopecumprob}}
\end{figure}

We calculate the number of detections one would expect in our survey for two cases: all brown dwarfs have heterogeneities that result in variability with $A=1.5\%$ when seen equatorially, or brown dwarf amplitudes are uniformly distributed between $0-3$\% when seen equatorially. We assume a uniform period distribution between 2-10 h and calculate the distribution of slopes for a large sample of 10,000 objects with random orientation in the sky. We then calculate for 22 targets the number of objects expected with a slope larger than a given number (Fig. \ref{fig:slopecumprob}). For both cases, the distribution is similar and  $\approx7-8$ targets would have slopes $>0.5\%$ , while about 1 object would have a slope $>1.5\%$ (this is consistent with our detections). Our results therefore are consistent with {\em all} brown dwarfs hosting low-level variability, but are also consistent with a uniform distribution of amplitudes from zero to a few percent. 

\subsection{Variability as a function of spectral type and wavelength}

\begin{figure}
\epsscale{0.9}
\plotone{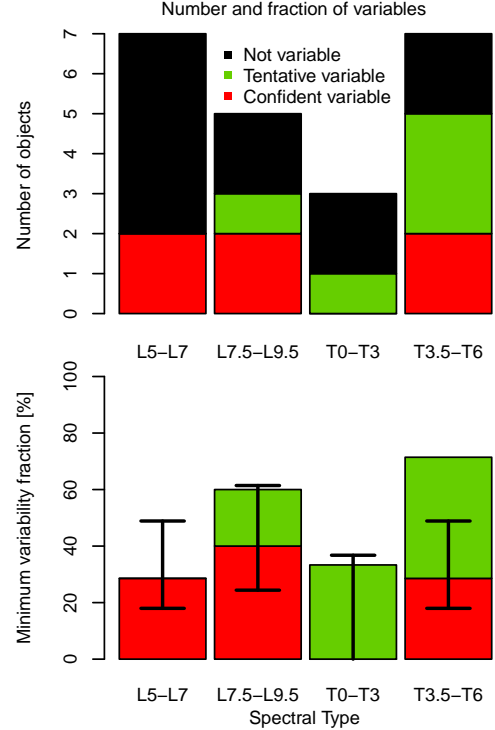}
\caption{Top: Number of confident (red), tentative (green) and non-variable (black) sources in our survey divided into four spectral type bins. Bottom: Minimum variability fraction $f_{\mathrm{min}}$ per spectral bin derived from confident variables (red) with $1\sigma$-error bars, and with the addition of tentative variables (green).
\label{fig:specfreq}}
\end{figure}

Previous surveys have discovered a handful of confident high-amplitude near-infrared variable brown dwarfs, nearly all of these with spectral types between T0 and T2.5 at the L/T transition. However, when including less confident detections of variability, the combination of previous ground-based surveys does not indicate an excess of variables in the L/T transition with respect to before and after \citep{khandrika13}. 
One notable confident variable is the T6.5 dwarf 2M2228-43 \citep{clarke08, buenzli12} which lies beyond what is generally regarded as the L/T transition region. Ground-based surveys are typically limited by their precision to detect
broad-band variables with amplitudes $A \gtrsim 1-2$\%. With broad-band filters, which cover a broad pressure range in the atmosphere, a strong signal originating from a thin atmospheric layer may be diluted. This is true in particular for 2M2228-43, where narrow wavelength regions through the J and H bands have significantly different phases and amplitudes. On the other hand, the two early T dwarfs that have been studied with HST/WFC3 spectroscopy \citep{apai13}, SIMP0136+09 and 2M2139+02, both show variability with a surprisingly weak wavelength dependence. There, the amplitudes and phases in narrow wavelength regions do not differ much (except in the deep water absorption bands where the variability is lower), therefore broad-band observations are not at a disadvantage for these objects  (not surprisingly, as these were both identified from precision-limited broad-band surveys). 

In this survey, we reach sub-percent precision and are able to study narrow wavelength regions for the first time over a statistically significant sample of brown dwarfs. It is therefore instructive to determine whether the trend of finding the majority of confidently variable brown dwarfs in the L/T transition holds, or whether we can confirm a relatively uniform distribution of variability throughout the L and T spectral types. As shown in Fig. \ref{fig:cmd}, our survey supports the latter. Indeed, we do not find a convincing variable brown dwarf in our (arguably small) sample of early T dwarfs, although
one of the three objects shows a tentative trend in the 1.5-1.7 $\mu$m wavelength region. On the other hand, we find convincing variables in all other spectral bins: 2 confident mid Ls (out of 7), 2 confident and one tentative late Ls (out of 5), and 2 confident and 3 tentative mid Ts (out of 7). Counting only the confident variables we derive the (minimum) frequency of variabilty in each spectral bin: $f_{\mathrm{min,mid-L}} = 29^{+20}_{-10}\%$, $f_{\mathrm{min,late-L}} = 40^{+21}_{-16}\%$, $f_{\mathrm{min,early-T}} = 0^{+37} \%$, and $f_{\mathrm{min,mid-T}} = 29^{+20}_{-10}\%$, with error bars giving the 1$\sigma$ confidence interval. Fig. \ref{fig:specfreq} shows the number and fraction of variables per spectral bin. We do not find significant differences between the different spectral bins, but with 3-7 objects per spectral bin the uncertainties are still large. Furthermore, our result is consistent with the statement that high-amplitude variability (several percent) is rare both inside and outside the L/T transition.

A second notable result is the fact that we find diverse variability as a function of wavelength. For the L dwarfs 2M0825+21 and 2M0624-45 we find similar variability levels across the whole spectrum, this is similar to the known variable early T dwarfs. For the other two confidently variable L dwarfs, 2M0310+16 and 2M1750-00, the variability originates in the J band while it is not evident in the H band. 
For the T dwarfs, we find variability sometimes in the J band peak, the H band peak, or methane or water absorption bands. It is not clear where these differences arise from.  It may simply be that the amplitude of variability is lower in one or the other wavelength region, perhaps because the perturbations stem from a specific pressure level in the atmosphere. Or the variability is shifted in phase, and it is therefore not detected in our short observations for some wavelengths. In order to understand the color dependence of the variability, which can pose strong constraints on atmospheric models, longer observations that cover most of a rotation period are required. 

Because of the low amplitudes and sometimes limited wavelength regions where variability occurs, broad-band photometric surveys are likely to miss it for many brown dwarfs and may therefore underestimate the occurrence rate of variability. High-precision spectroscopy from space-based instrument such as HST/WFC3 makes it possible to use spectral mapping to study clouds and weather phenomena in three dimensions for many brown dwarfs across the L- and T spectral types. In the near future, with next generation adaptive optics systems and with the James Webb Space Telescope (JWST), the same technique can be applied to extrasolar planets \citep[e.g.][]{kostov13}.

What is the origin of this low-level variability that seems to be a common occurrence for many brown dwarfs? Within the L/T transition, models have shown that patchy cloud cover can explain the optical and near-infrared spectra of brown dwarfs \citep[e.g.][]{marley10}. Asymmetric distribution of patches then results in variability at the several percent level. However, we also clearly find variability for mid to late L-type dwarfs where models generally predict a thick cloud cover, and for mid- to late-T dwarfs where silicate clouds are expected to be below the visible photosphere. 
For the L dwarfs, it seems reasonable to think that the variability could stem from heterogeneous cloud cover of thicker and thinner cloud patches, which \citet{apai13} also suggested for the early T dwarfs 2M2139+02 and SIMP0136+09. For these two objects, the variabilty amplitudes are either constant as a function of wavelength or slightly larger in the J band than the H band, in agreement to what we find for our variable L dwarfs. While for early L dwarfs magnetic spots may also contribute to variability \citep[e.g.][]{lane07}, there is no indication that this may also be true for later type brown dwarfs. For example, \citet{gelino02} argue that these photospheres are largely neutral, and they do not find any correlation between I band variability and H$\alpha$ emission. For the later T dwarfs, cloud types other than silicates may appear that could also be patchy \citep{morley12}. For the known variable T6 dwarf 2M2228-43 models with sulfide and chromium clouds, as well as with clouds of a species with similar optical properties to iron, provided a good match to the average spectrum \citep{buenzli12}, while cloud free models did not. However, the light curve phase shifts as a function of wavelength show that not only clouds may be responsible for variability: other possibilities include temperature fluctuations from circulation patterns or gas opacity perturbations. The diversity in variability that we find as a function of wavelength may point towards several mechanism where different ones dominate for different objects. Considering that even the solar system gas giants appear variable at some wavelengths \citep[e.g.][]{gelino00, karkoschka11}, it may be reasonable to expect low level heterogeneities in all atmospheres where condensates form. 

\section{Conclusions}

We conducted an unbiased near-infrared spectroscopic survey of 22 brown dwarfs spanning mid-L through mid-T spectral types in order to search for short term ($\approx 40$ min), sub-percent flux variability. Our main results are:
\begin{itemize}
\item We find 6 confident $(p>95\%)$ and 5 tentative $(p>68\%)$ variable brown dwarfs, resulting in a minimum variability fraction $f_{min} = 27^{+11}_{-7}\%$. All are newly discovered variable brown dwarfs. 
\item The fraction of brown dwarfs with patchy photospheres is likely to be significantly higher than about a third for three reasons: 
1) long-period variables will only lead to a very small signal in our short observations, 2) objects with inclinations near pole-on will have significantly lower measured amplitudes than if seen near an equatorial view, and 3) rotational mapping is insensitive to rotationally symmetric heterogeneities such as bands.
\item We find that the fraction of variables is similar for mid-L, late-L, and mid-T spectral types. In our smaller sub-sample we do not find any confidently variable early T dwarf, the spectral type where most of the highly variables are currently known, but because of our short observations we also cannot exclude high-amplitude variability if the sources have long periods. 
\item In some cases the variability is limited to the flux peak in the J or H band (but not necessarily both) or to absorption regions, suggesting that broad-band photometric surveys may miss a fraction of variable brown dwarfs.
\item We find 4 objects with significant broad band variability that may be well suited for ground-based follow-up studies: 2M0559-14 in H band, 2M1750-00 in J band, and 2M0825+21 and 2M0624-45 in both J and H bands. These sources are likely to have peak-to-peak amplitudes $A \gtrsim 1\%$ if periods are longer than $3-4$ hours and the light curve shapes are sinusoidal.
\end{itemize}

Variable brown dwarfs have already provided a unique window into the atmospheric structure and the process of cloud dispersal at the L/T transition for a handful of objects. Our survey shows that brown dwarfs with low-level variability of $\sim 1\%$ at some wavelengths are common, but precision-limited broand-band photometric surveys are likely missing some of these objects. It is not yet clear if these brown dwarfs with low-level variability represent a different population in terms of cloud structure than the known broad-band highly variables. Finally, our study demonstrates that patchy photospheres are a frequent characteristic for many brown dwarfs and should be accounted for in atmospheric models.

\acknowledgements
We thank the staff at Space Telescope Science Institute (STScI), in particular Tricia Royle, for the efficient coordination and scheduling of the observations. We thank the anonymous referee for useful comments and suggestions. Based on observations made with the NASA/ESA Hubble Space Telescope, obtained at the Space Telescope Science Institute, which is operated by the Association of Universities for Research in Astronomy, Inc., under NASA contract NAS 5-26555. These observations are associated with program \#12550. EB was partially supported by the Swiss National Science Foundation (SNSF). This research has benefitted from the M, L, T, and Y dwarf compendium housed at DwarfArchives.org. This research has made use of the SIMBAD database, operated at CDS, Strasbourg, France. This research has made use of NASA's Astrophysics Data System Bibliographic Services.

\appendix 

\clearpage

\begin{figure*}
\epsscale{1.0}
\plotone{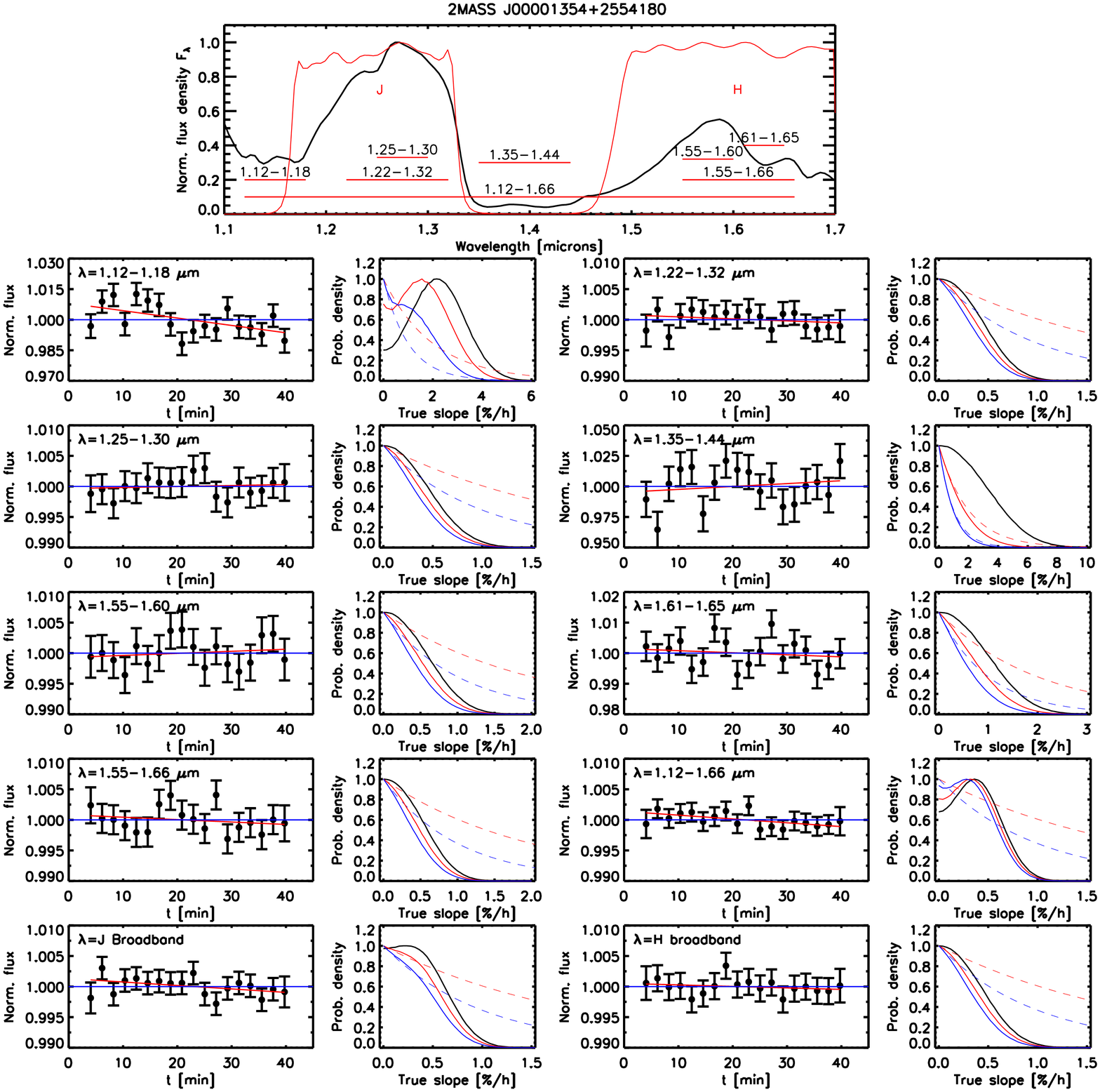}
\caption{Observations (left) and best fit slope (red) for all selected wavelength regions for 2M0000+25. The right panel gives the probability for the value of the true slope calculated from Bayes' theorem. The black line is the likelihood function for the slope, the dotted red and blue line two prior distributions, and the red and blue solid lines the
corresponding posterior distribution functions.  
\label{fig:v01}}
\end{figure*}

\begin{figure*}
\epsscale{1.0}
\plotone{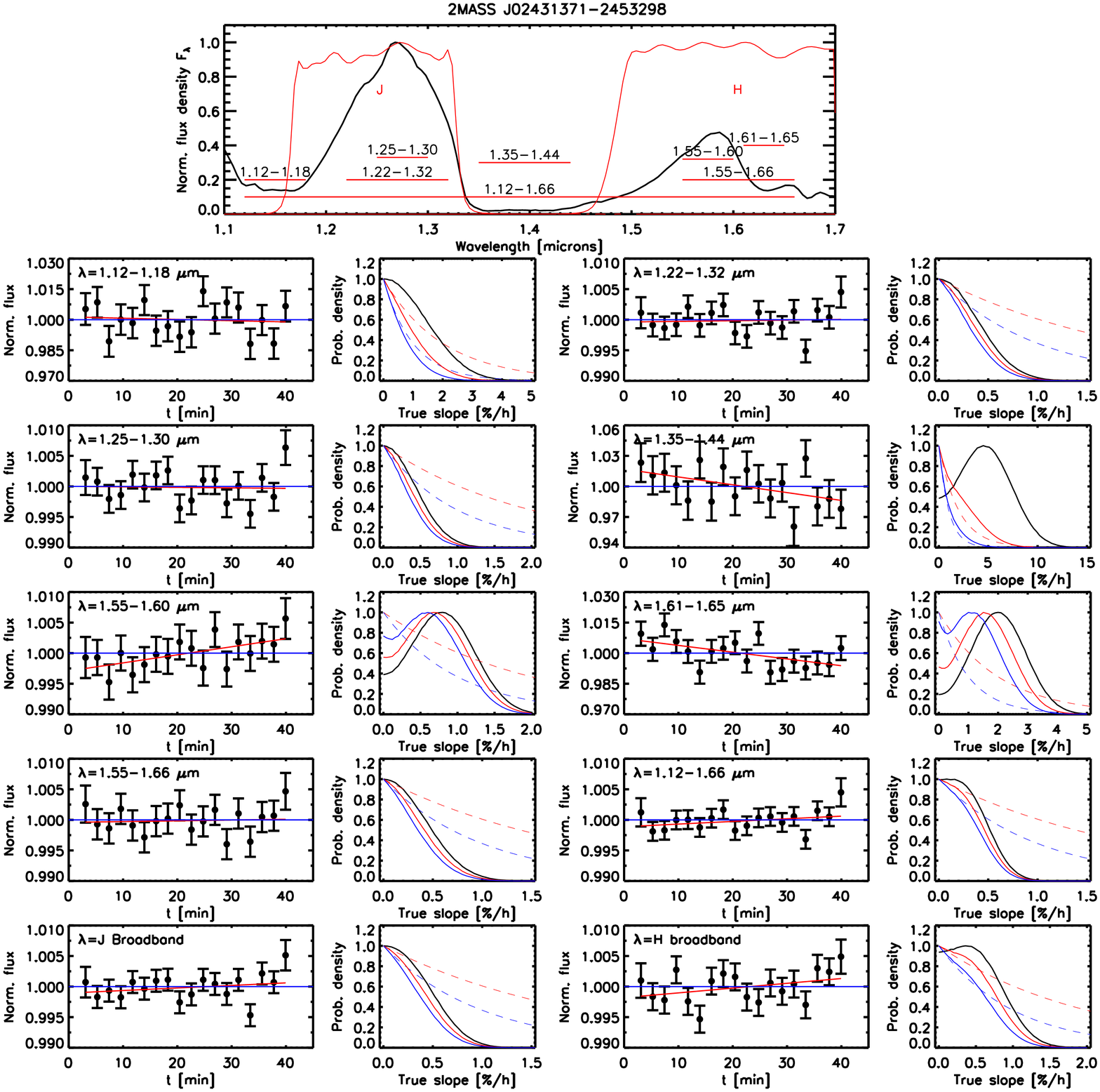}
\caption{Like Fig. \ref{fig:v01} but for 2M0243-24.  
\label{fig:v10}}
\end{figure*}

\begin{figure*}
\epsscale{1.0}
\plotone{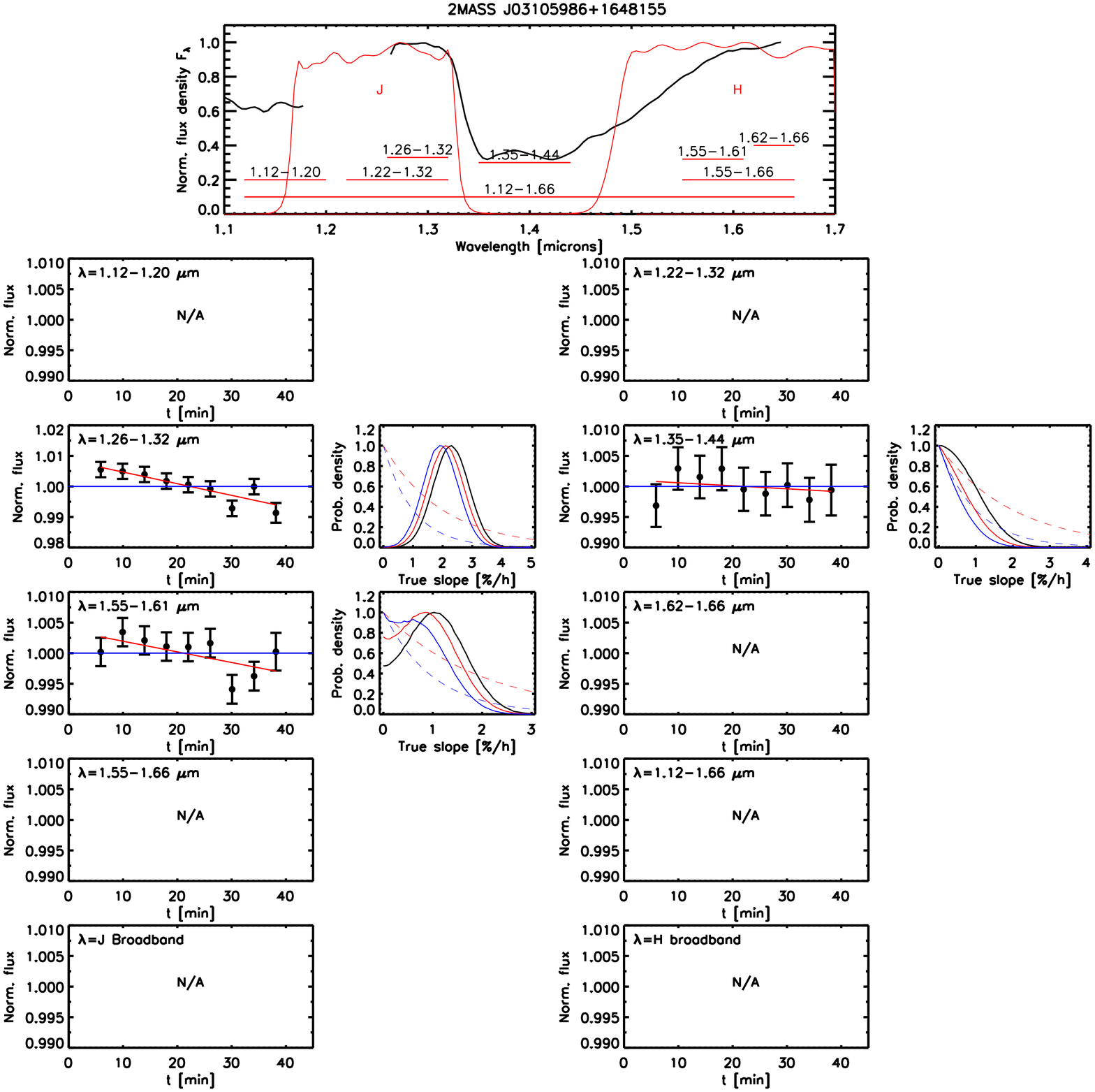}
\caption{Like Fig. \ref{fig:v01} but for 2M0310+16. Missing panels are wavelength regions where this source had a large number of bad pixels.
\label{fig:v12}}
\end{figure*}

\begin{figure*}
\epsscale{1.0}
\plotone{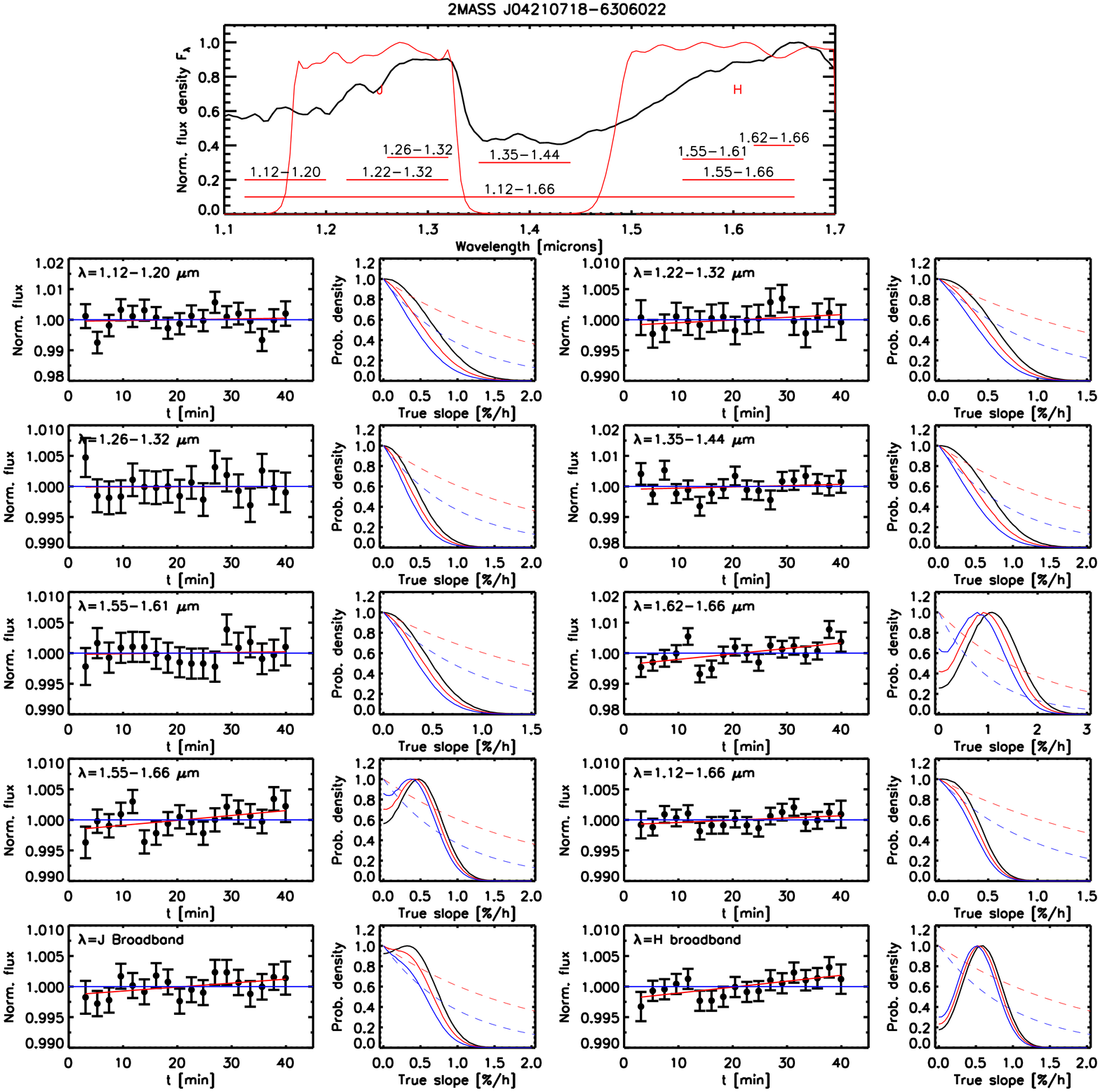}
\caption{Like Fig. \ref{fig:v01} but for 2M0421-63.  
\label{fig:v16}}
\end{figure*}

\begin{figure*}
\epsscale{1.0}
\plotone{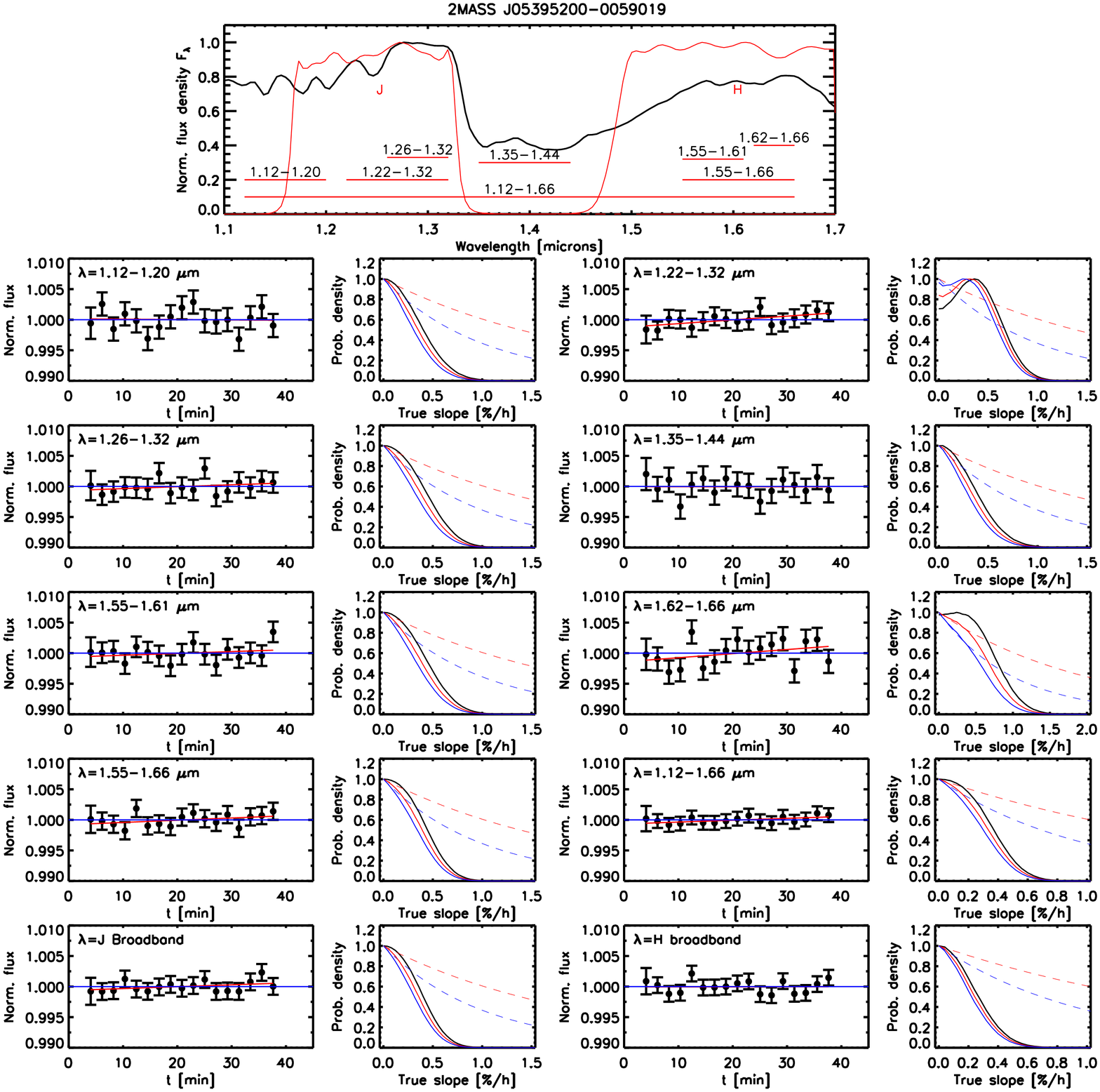}
\caption{Like Fig. \ref{fig:v01} but for 2M0539-00.  
\label{fig:v17}}
\end{figure*}

\begin{figure*}
\epsscale{1.0}
\plotone{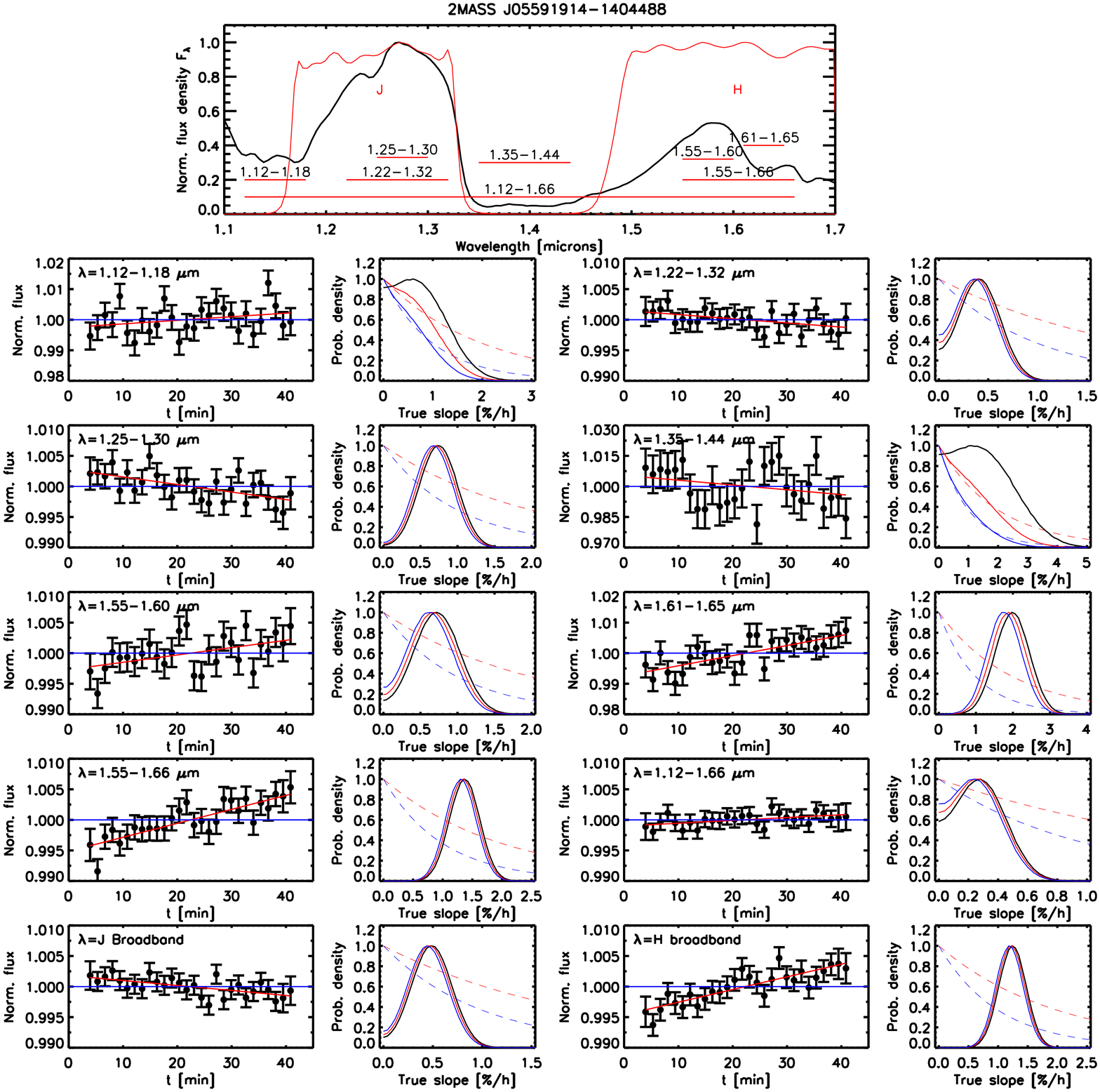}
\caption{Like Fig. \ref{fig:v01} but for 2M0559-14.  
\label{fig:v18}}
\end{figure*}

\begin{figure*}
\epsscale{1.0}
\plotone{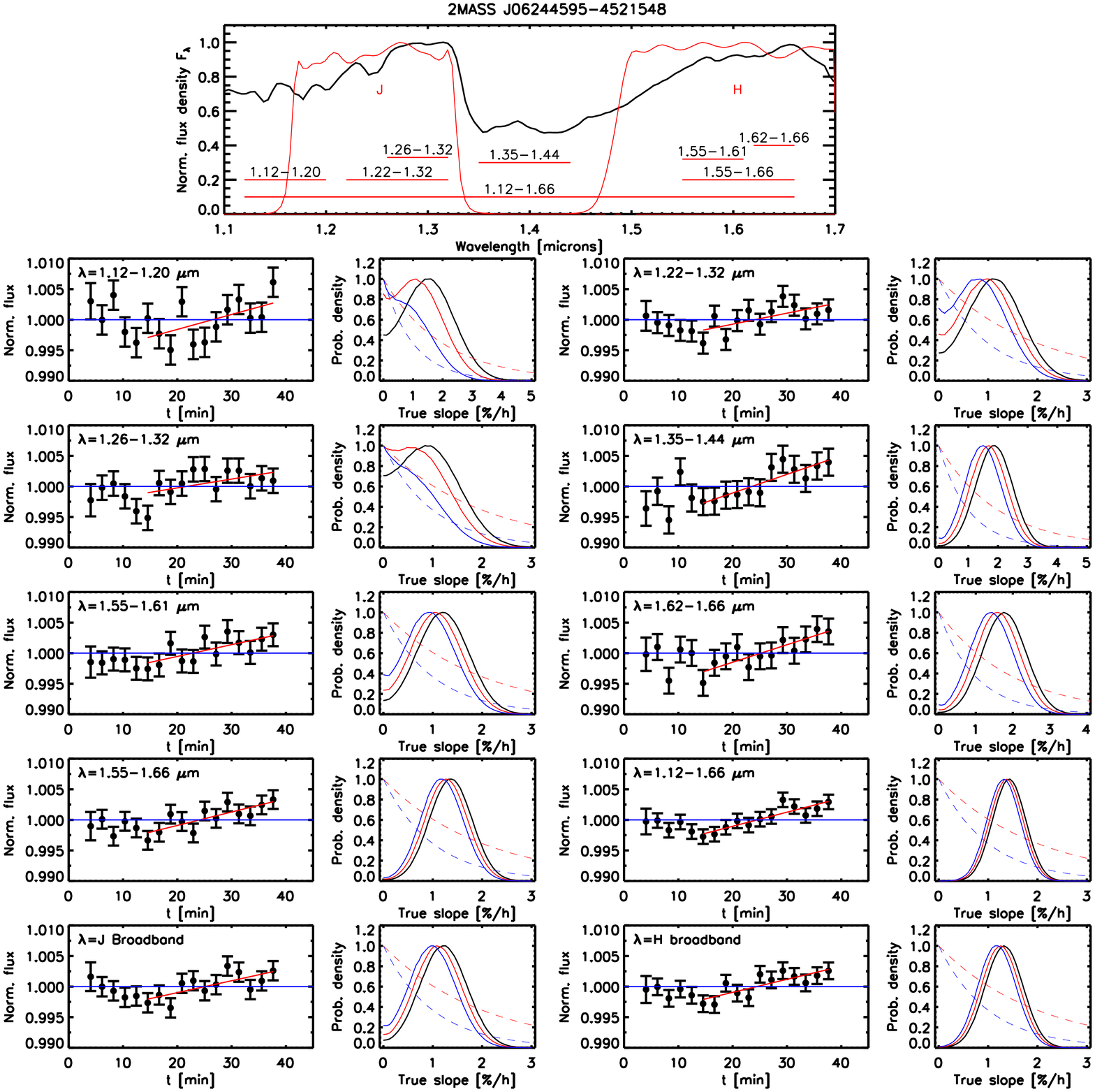}
\caption{Like Fig. \ref{fig:v01} but for 2M0624-45.  
\label{fig:v19}}
\end{figure*}

\begin{figure*}
\epsscale{1.0}
\plotone{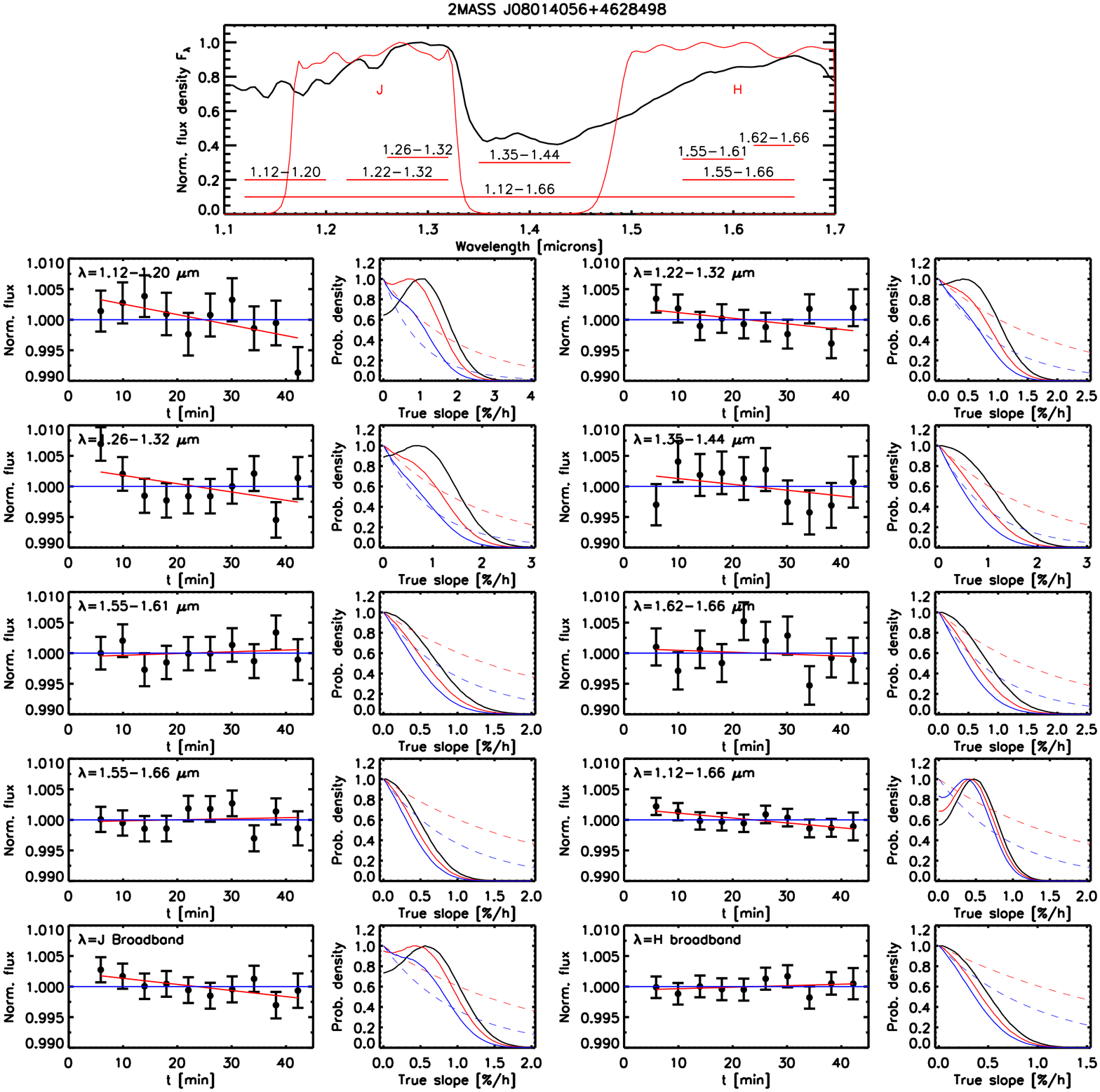}
\caption{Like Fig. \ref{fig:v01} but for 2M0801+46.  
\label{fig:v23}}
\end{figure*}

\begin{figure*}
\epsscale{1.0}
\plotone{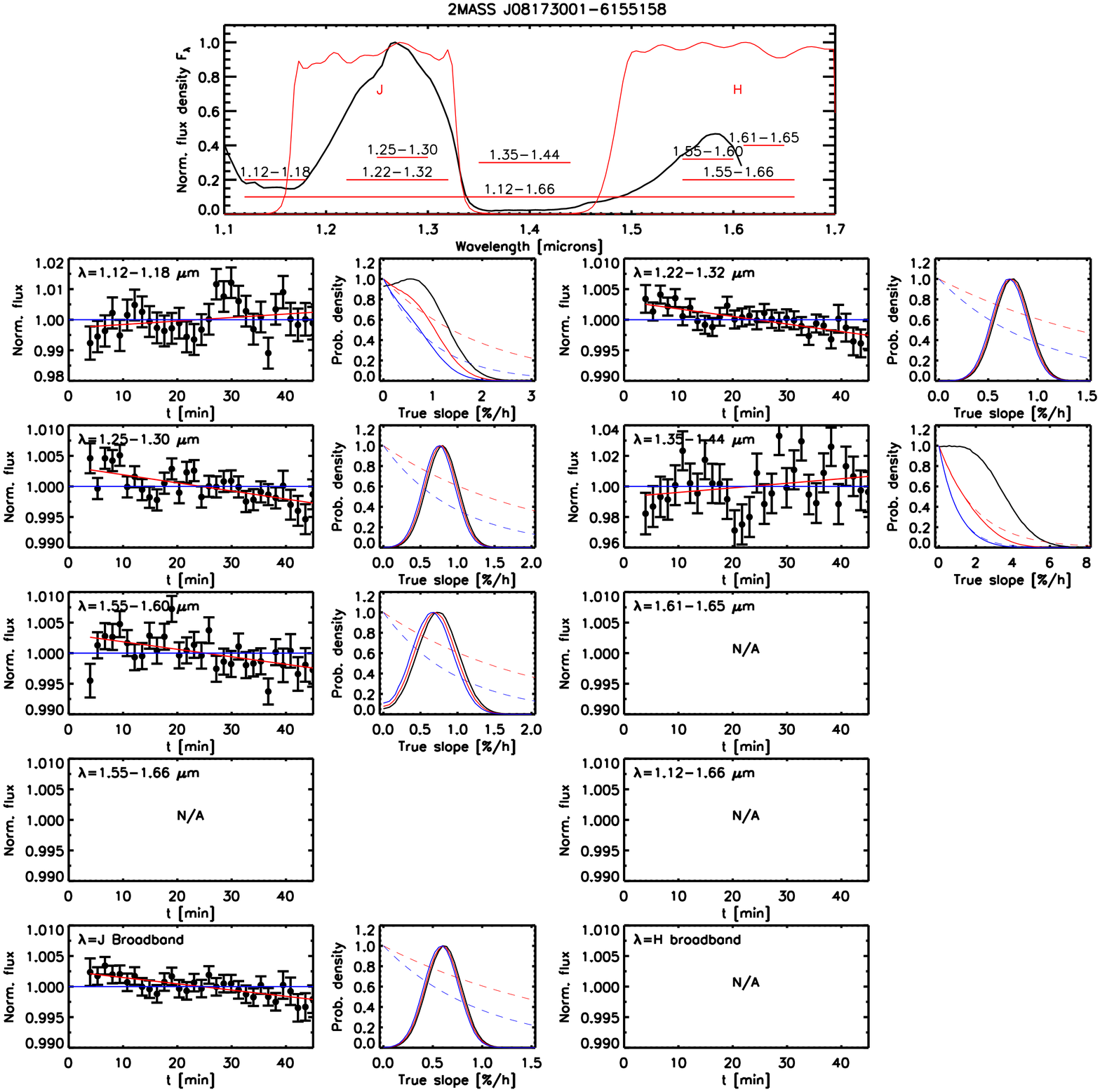}
\caption{Like Fig. \ref{fig:v01} but for 2M0817-61. Missing panels are for wavelengths that were missing in this spectrum due to cut off at the edge of the detector.  
\label{fig:v24}}
\end{figure*}

\clearpage

\begin{figure*}
\epsscale{1.0}
\plotone{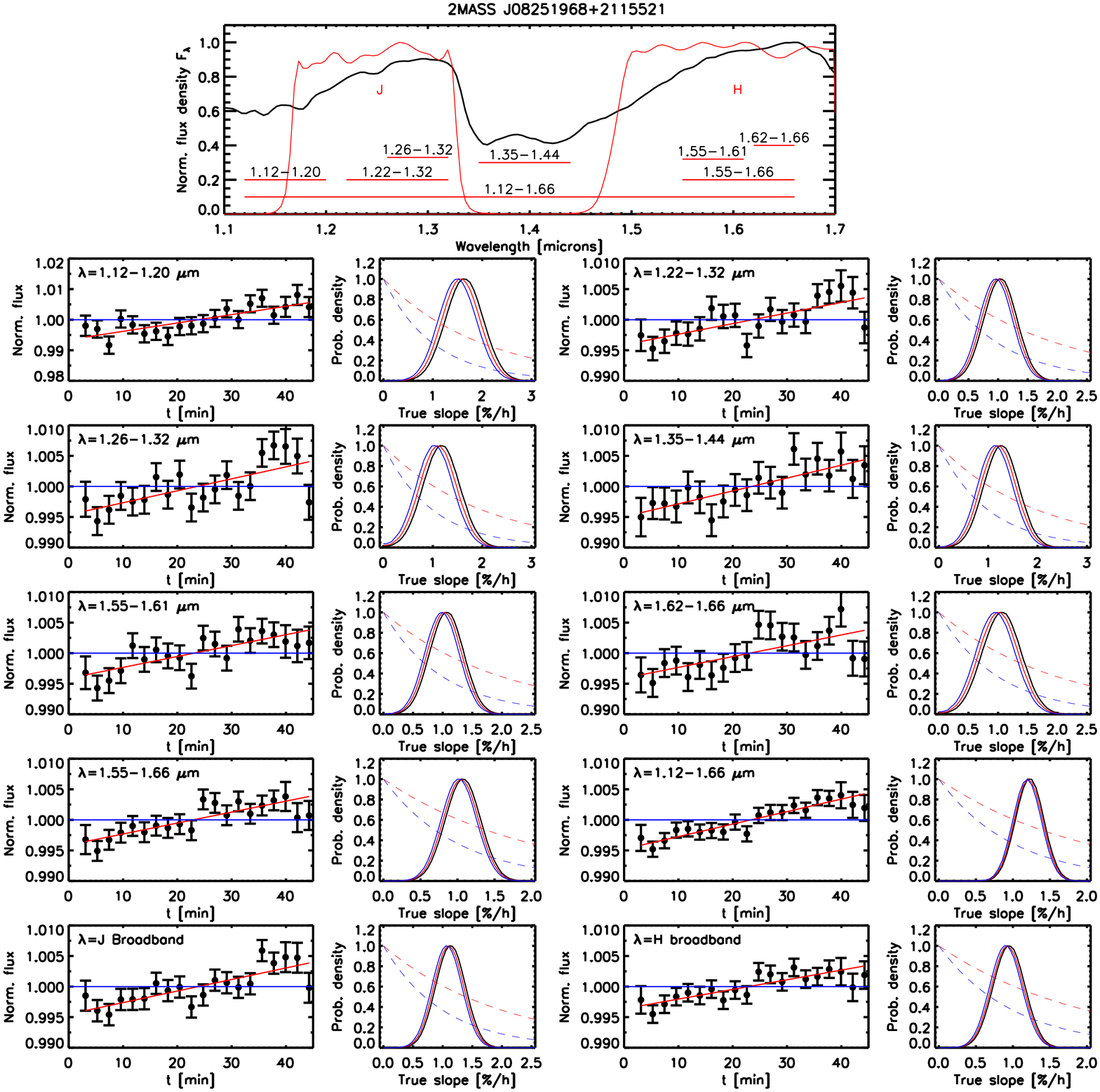}
\caption{Like Fig. \ref{fig:v01} but for 2M0825+21.   
\label{fig:v26}}
\end{figure*}

\begin{figure*}
\epsscale{1.0}
\plotone{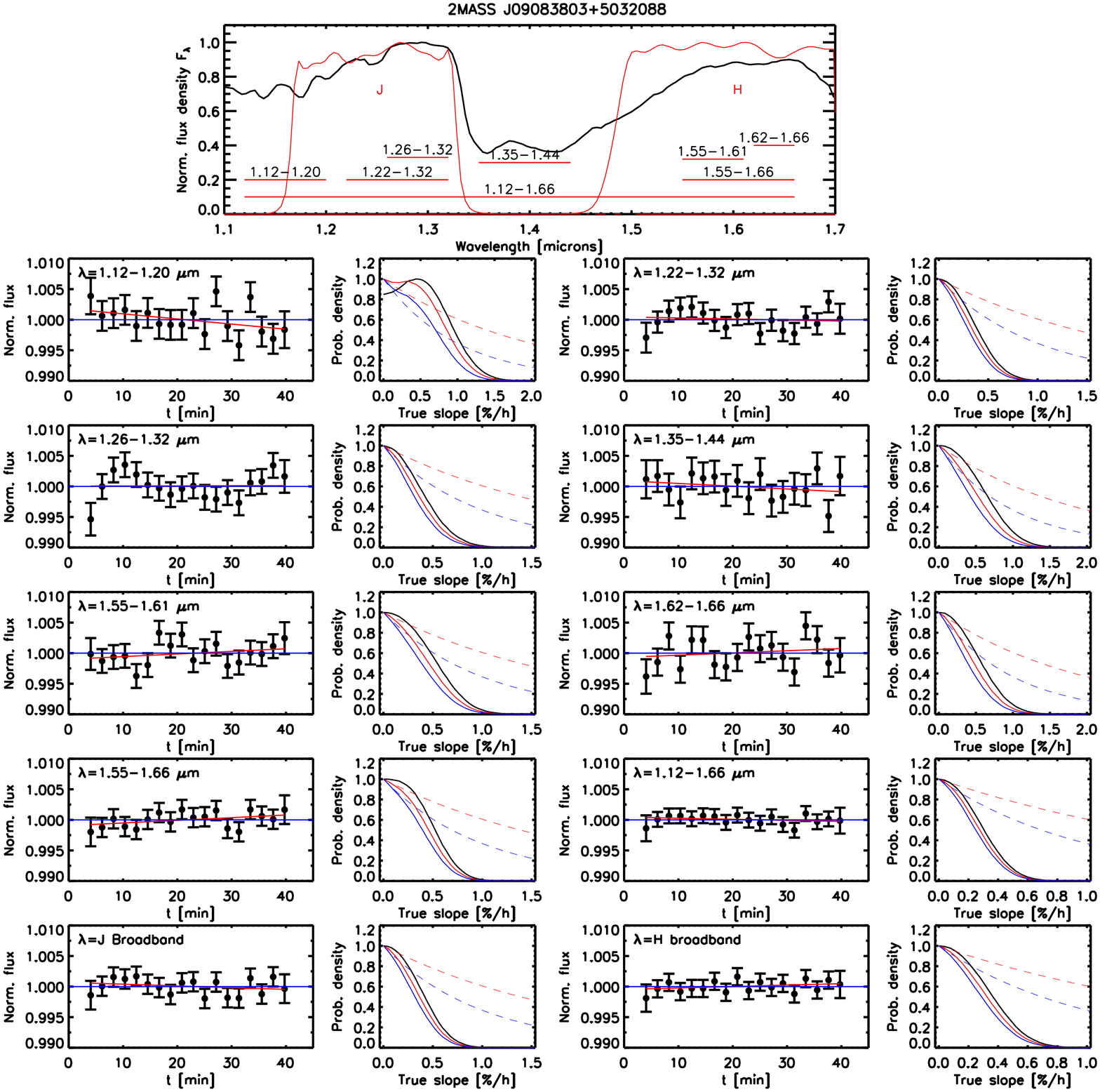}
\caption{Like Fig. \ref{fig:v01} but for 2M0908+50.
\label{fig:v31}}
\end{figure*}

\begin{figure*}
\epsscale{1.0}
\plotone{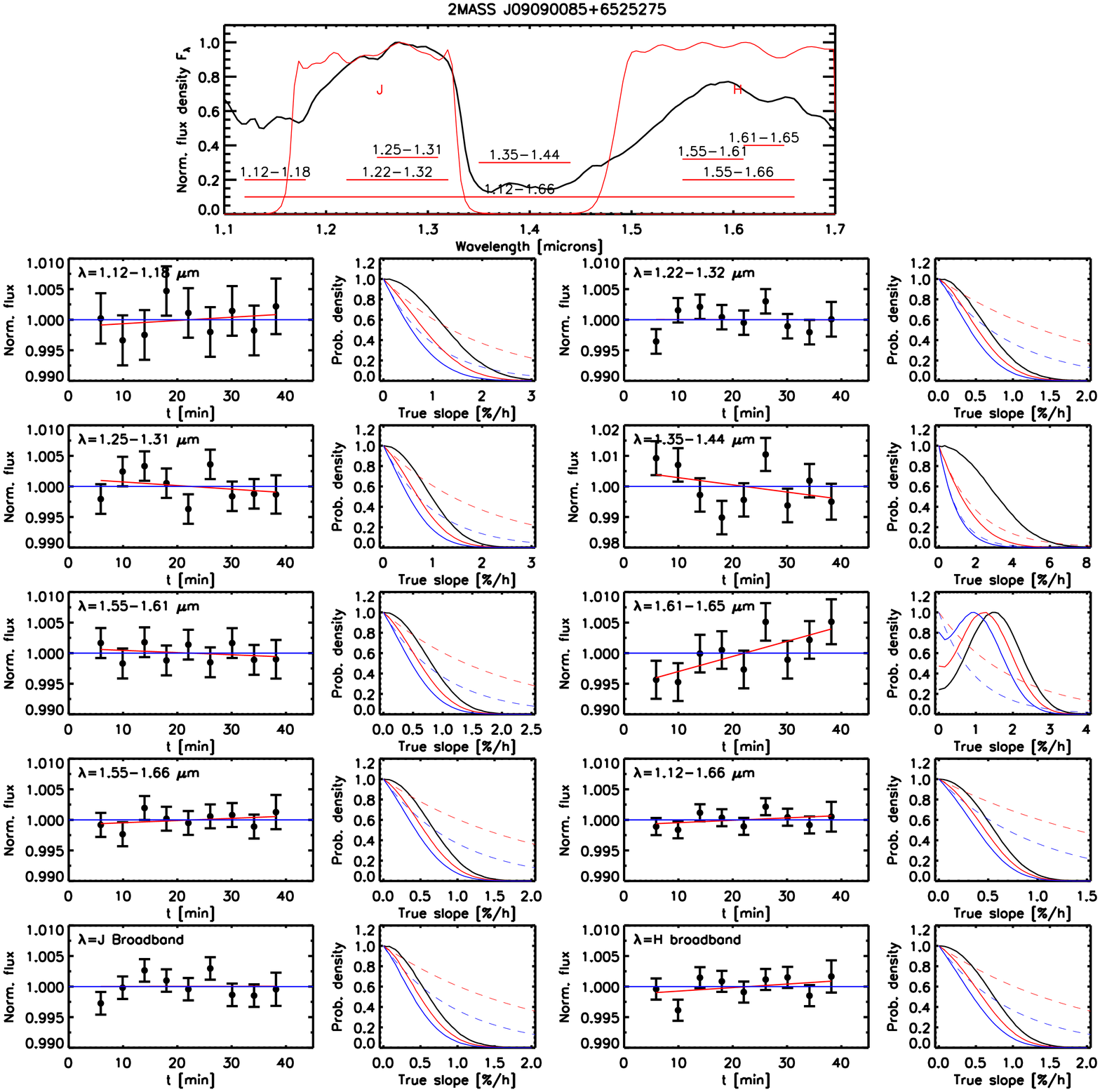}
\caption{Like Fig. \ref{fig:v01} but for 2M0909+65.
\label{fig:v32}}
\end{figure*}

\begin{figure*}
\epsscale{1.0}
\plotone{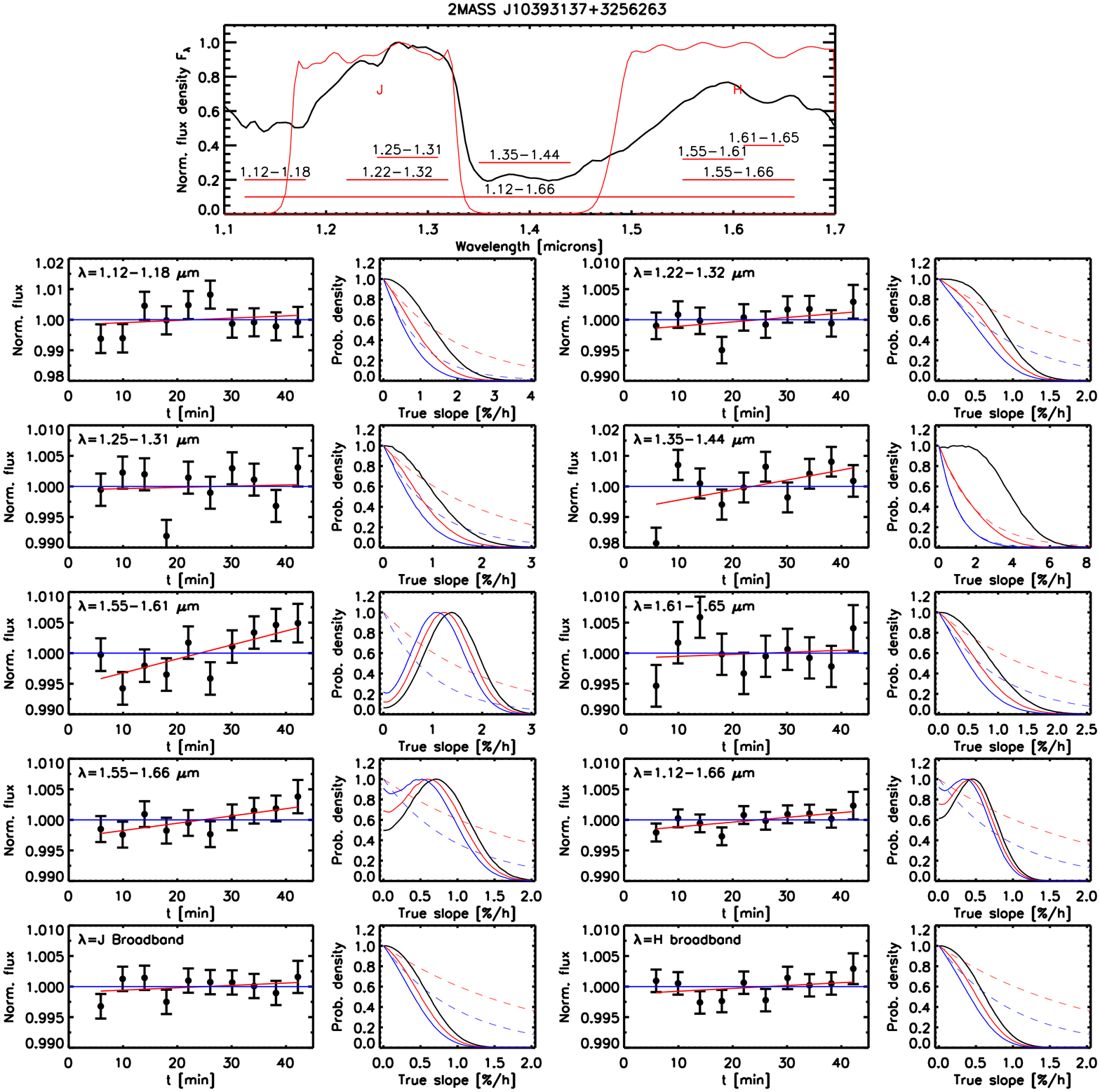}
\caption{Like Fig. \ref{fig:v01} but for 2M1039+32. 
\label{fig:v35}}
\end{figure*}

\begin{figure*}
\epsscale{1.0}
\plotone{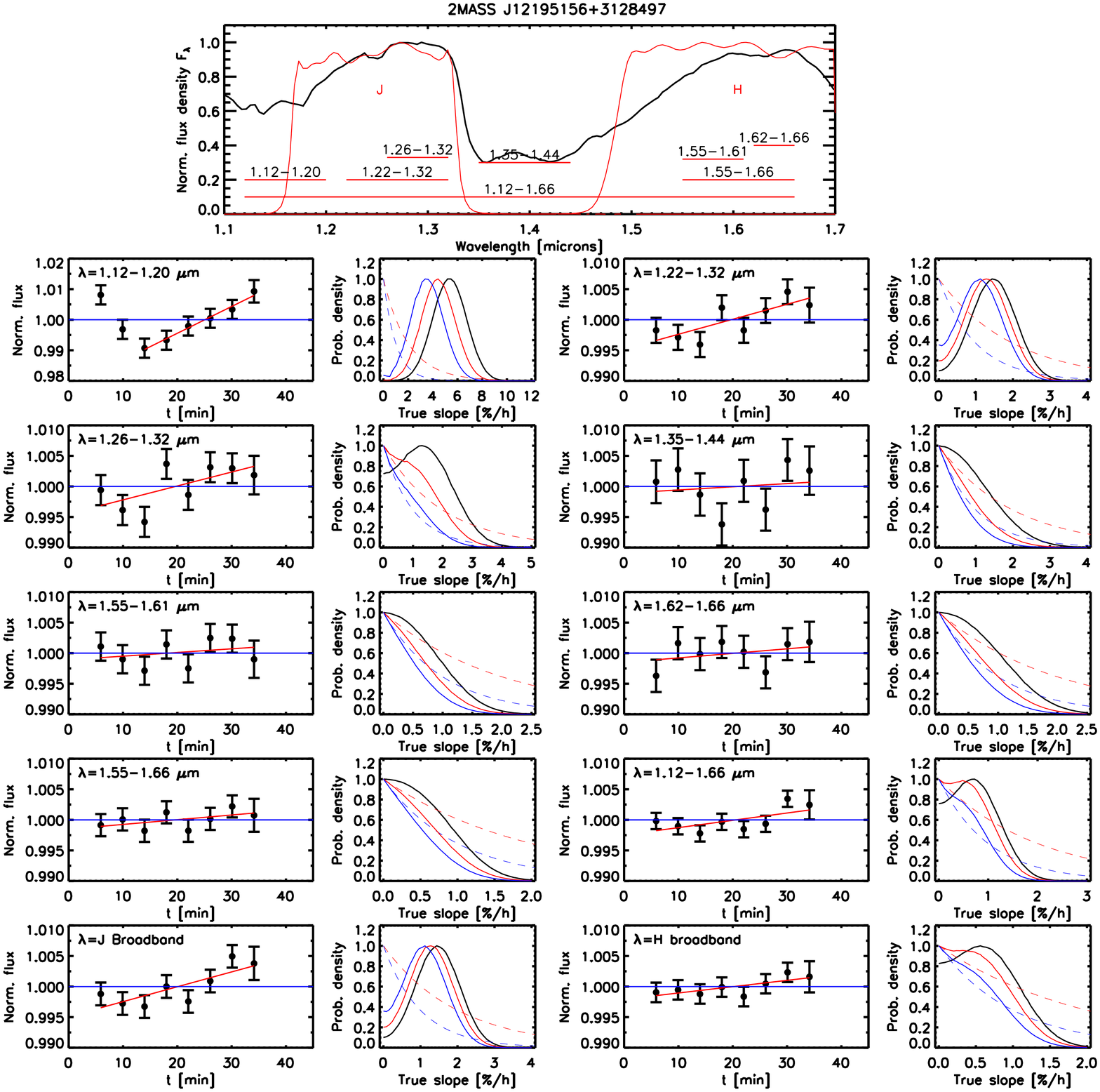}
\caption{Like Fig. \ref{fig:v01} but for 2M1219+31.
\label{fig:v44}}
\end{figure*}

\begin{figure*}
\epsscale{1.0}
\plotone{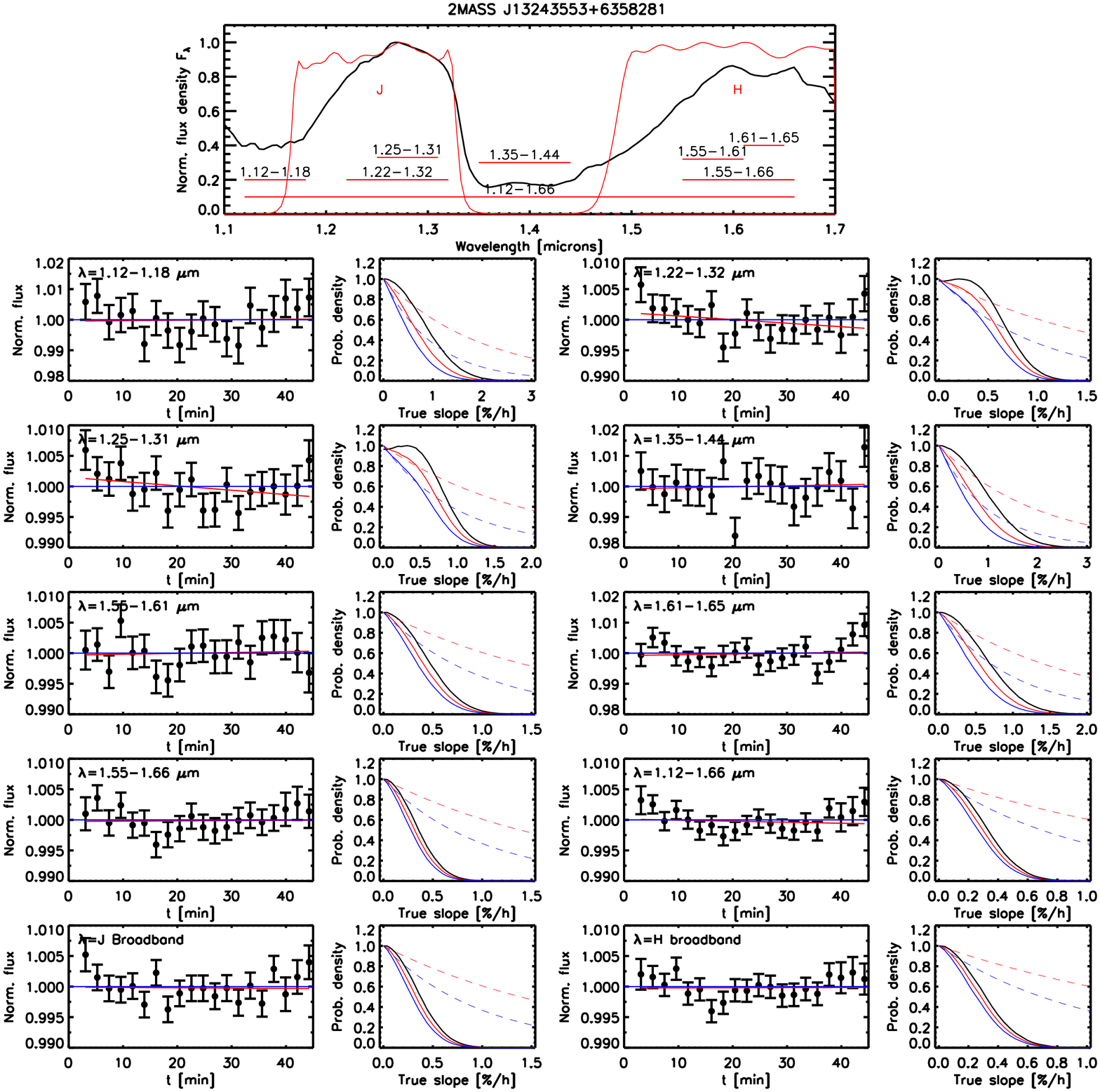}
\caption{Like Fig. \ref{fig:v01} but for 2M1324+63.
\label{fig:v46}}
\end{figure*}

\begin{figure*}
\epsscale{1.0}
\plotone{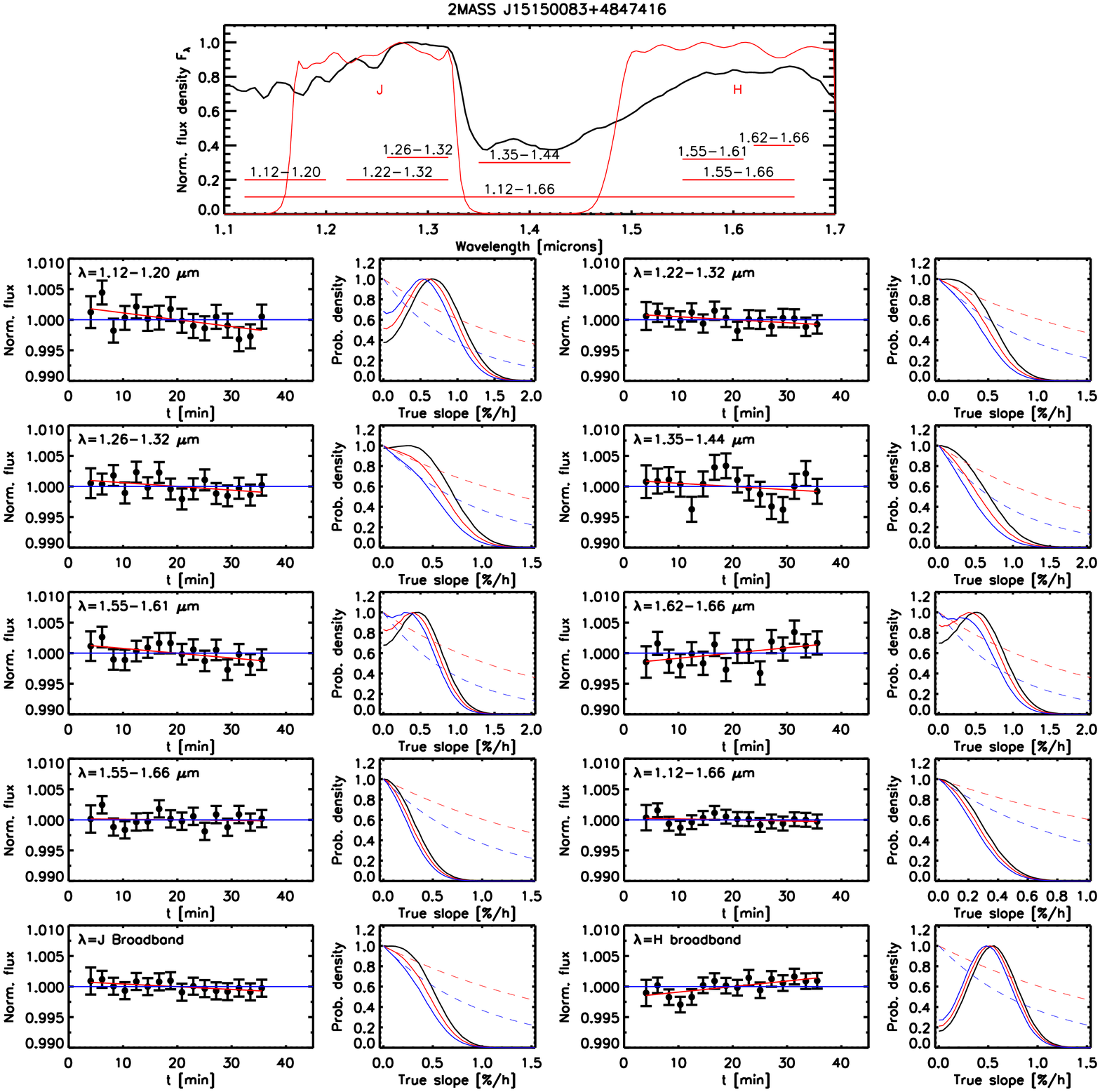}
\caption{Like Fig. \ref{fig:v01} but for 2M1515+48. 
\label{fig:v52}}
\end{figure*}

\begin{figure*}
\epsscale{1.0}
\plotone{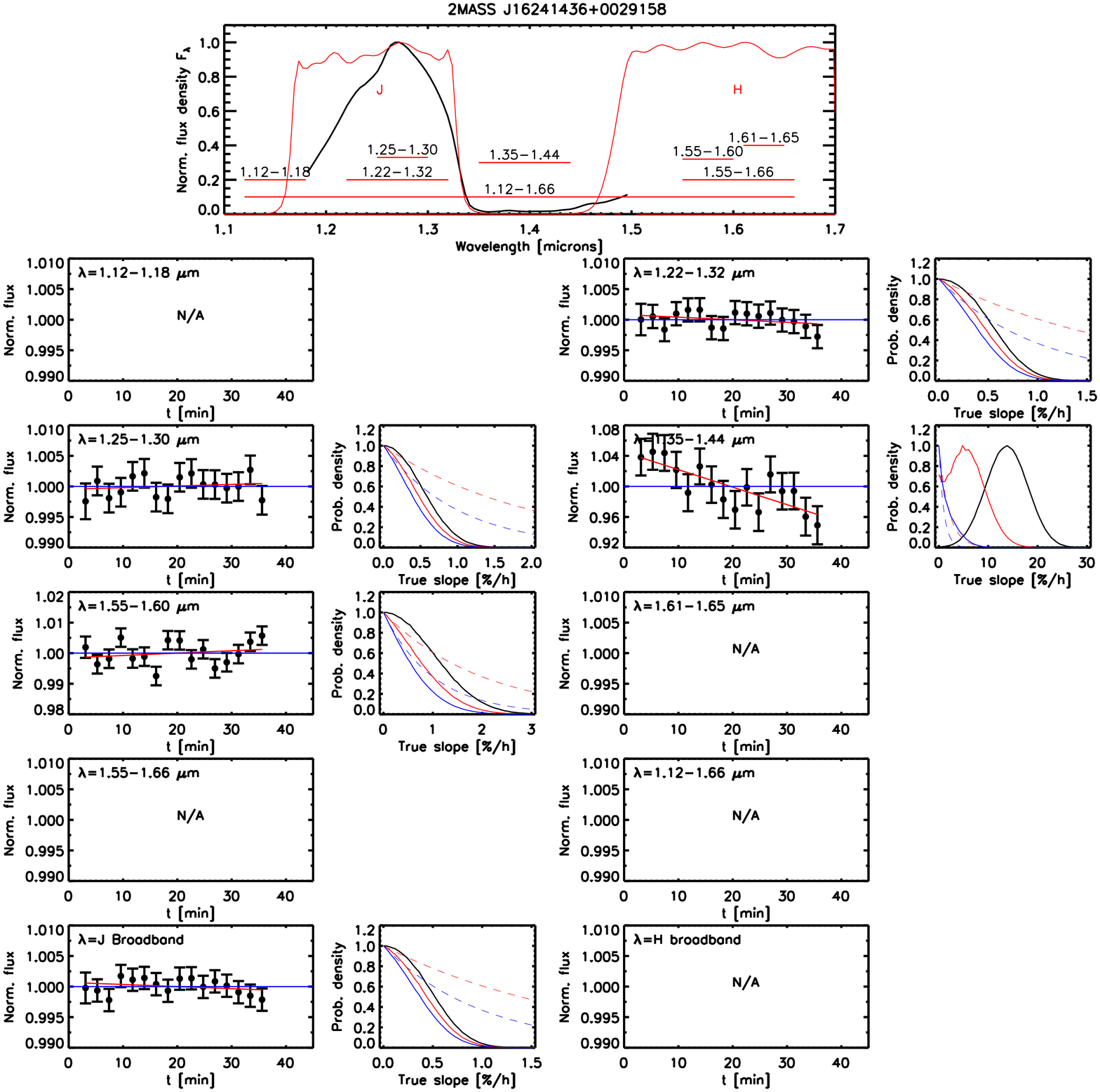}
\caption{Like Fig. \ref{fig:v01} but for 2M1624+00. Missing panels are wavelengths where the brown dwarf spectrum was contaminated by a background star spectrum.  
\label{fig:v57}}
\end{figure*}

\begin{figure*}
\epsscale{1.0}
\plotone{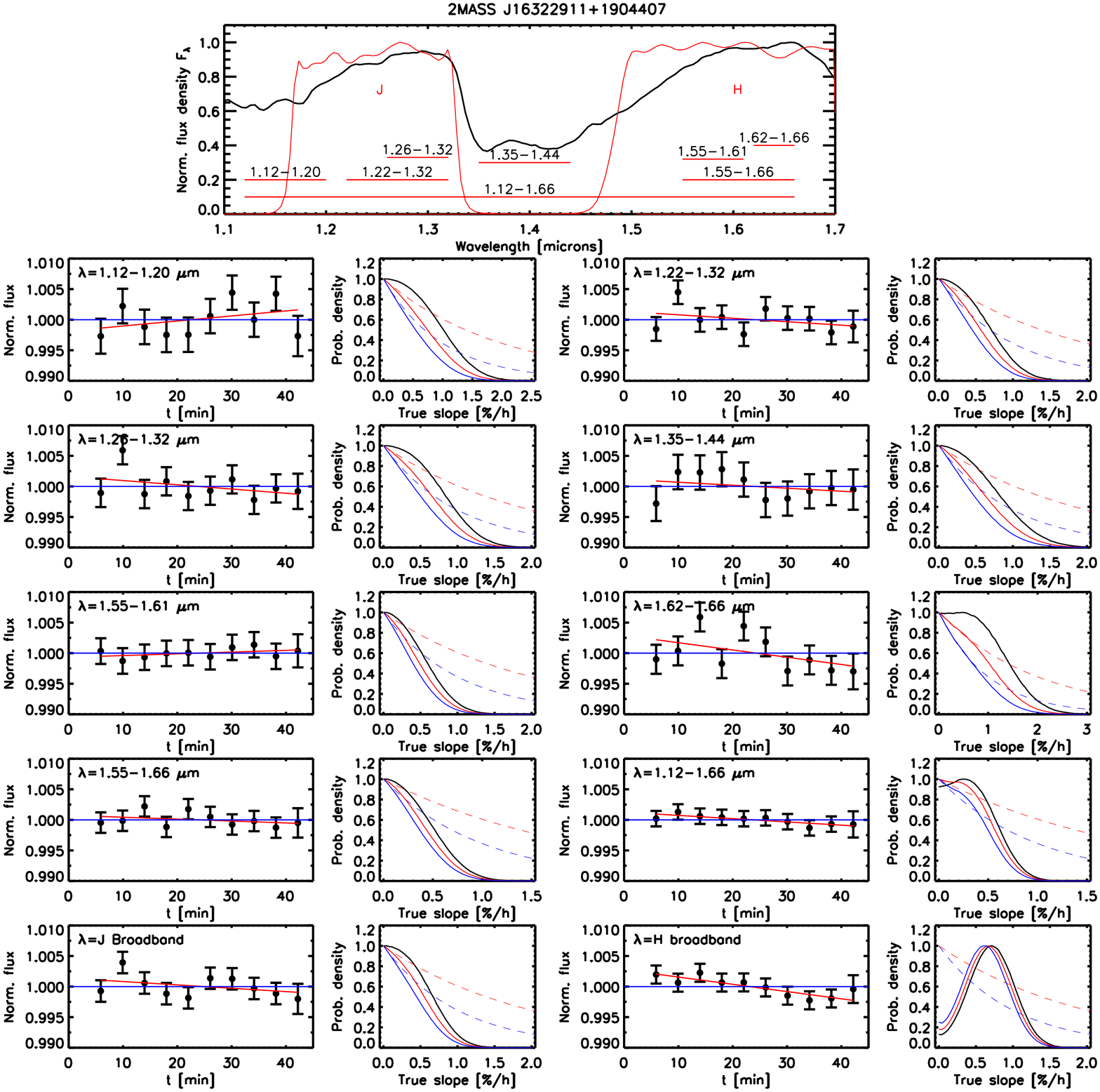}
\caption{Like Fig. \ref{fig:v01} but for 2M1632+19. 
\label{fig:v58}}
\end{figure*}

\begin{figure*}
\epsscale{1.0}
\plotone{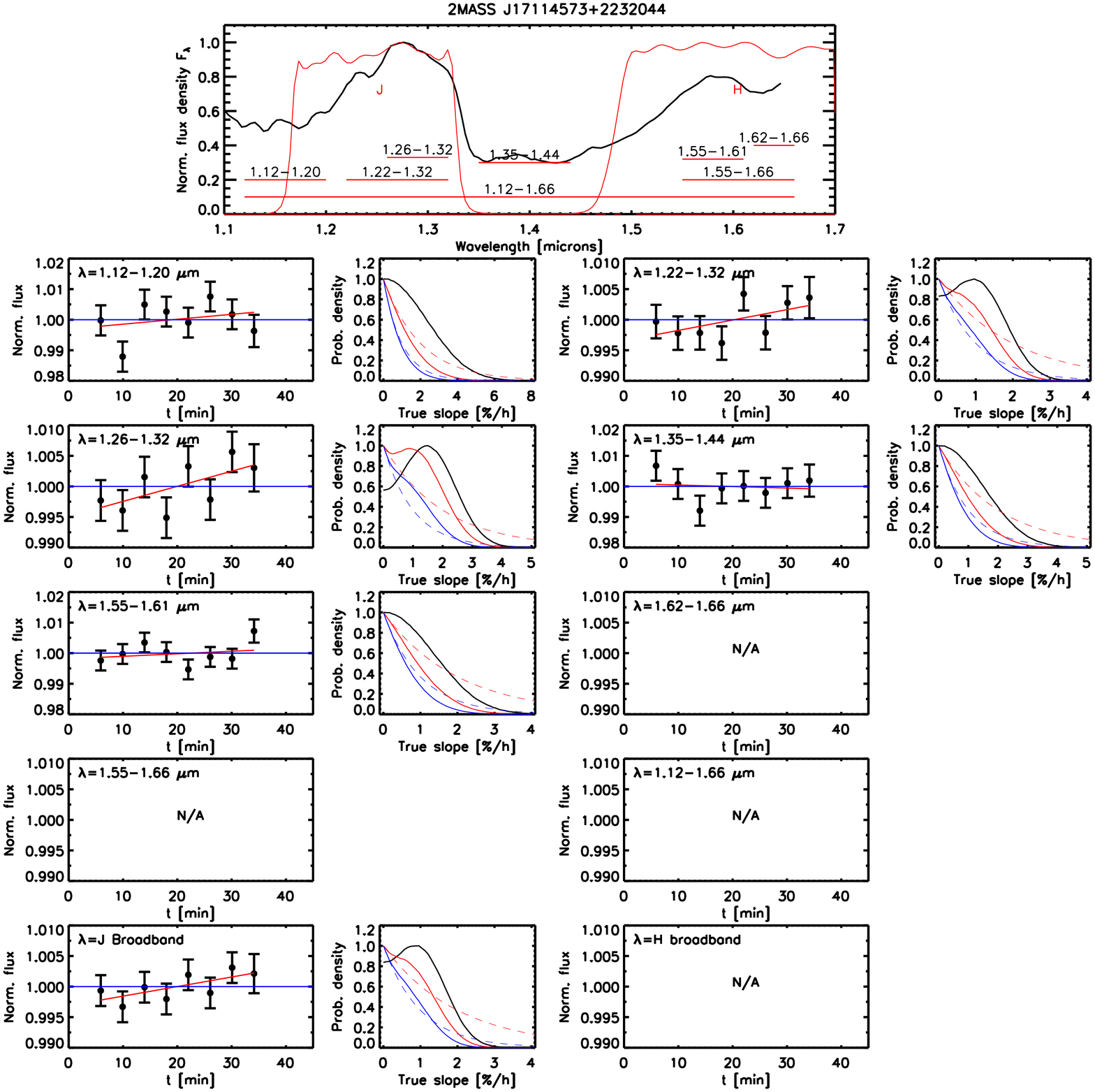}
\caption{Like Fig. \ref{fig:v01} but for 2M1711+22. 
\label{fig:v59}}
\end{figure*}

\clearpage

\begin{figure*}
\epsscale{1.0}
\plotone{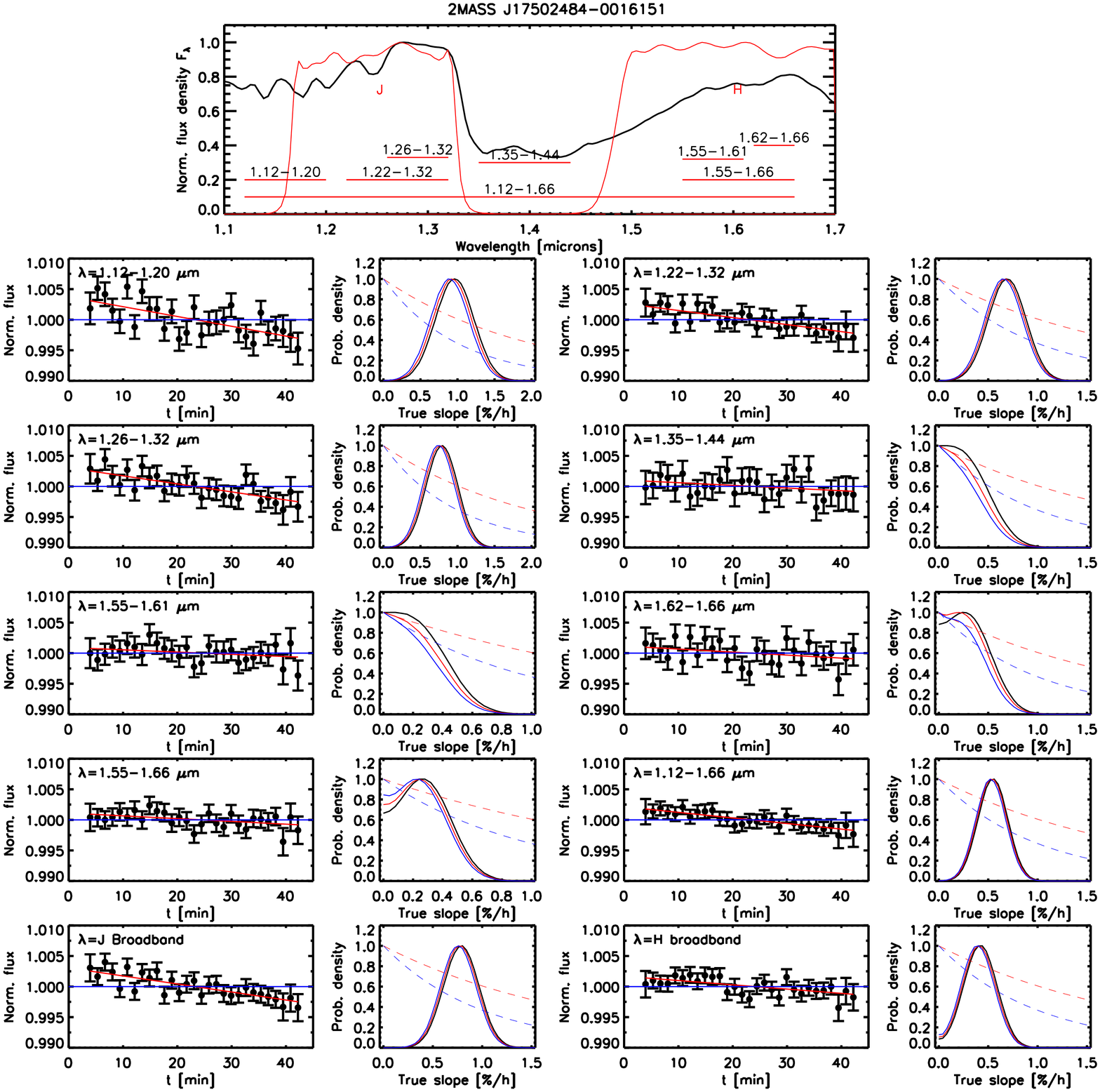}
\caption{Like Fig. \ref{fig:v01} but for 2M1750-00. 
\label{fig:v61}}
\end{figure*}

\begin{figure*}
\epsscale{1.0}
\plotone{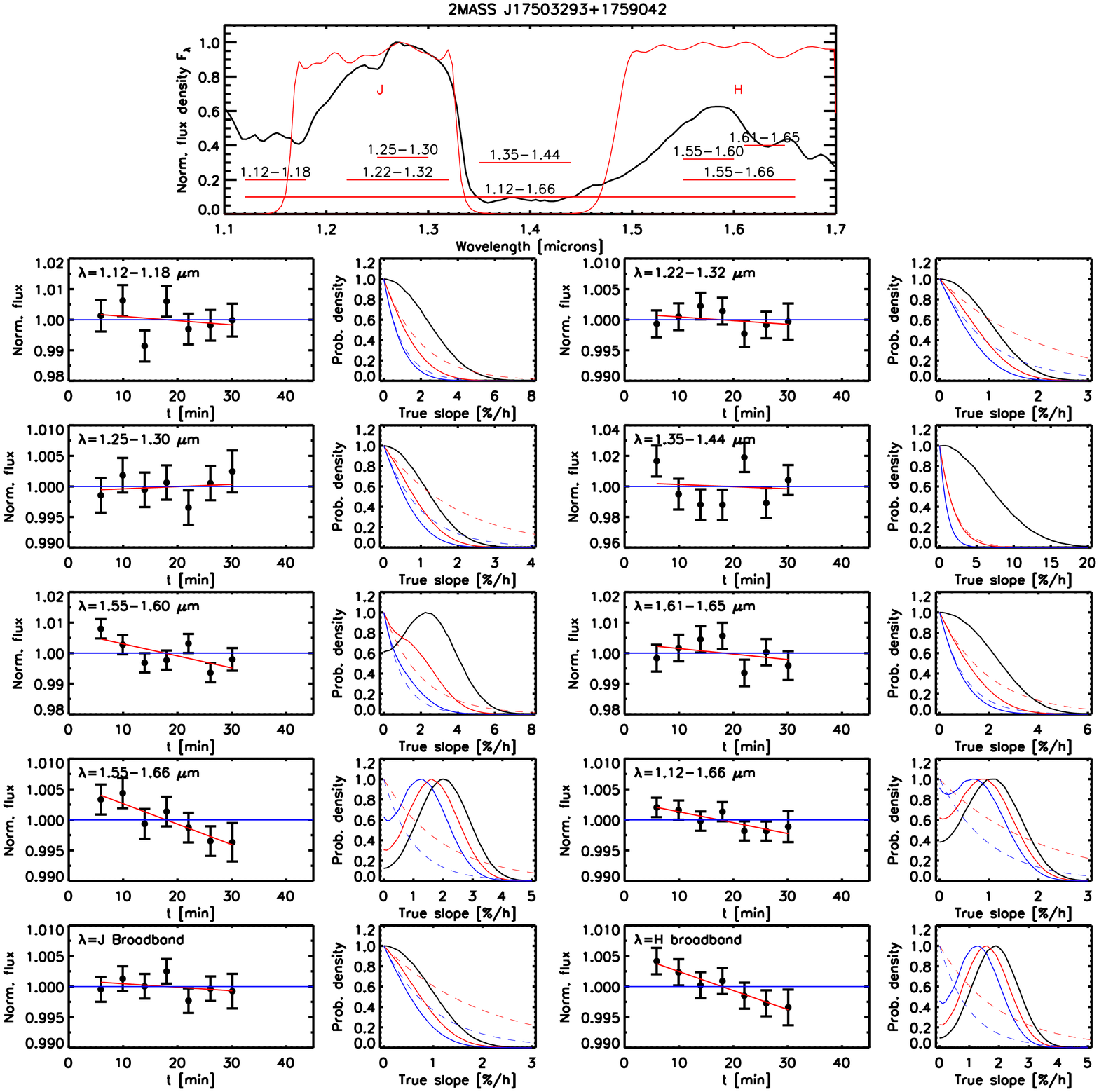}
\caption{Like Fig. \ref{fig:v01} but for 2M1750+17.
\label{fig:v62}}
\end{figure*}

\begin{figure*}
\epsscale{1.0}
\plotone{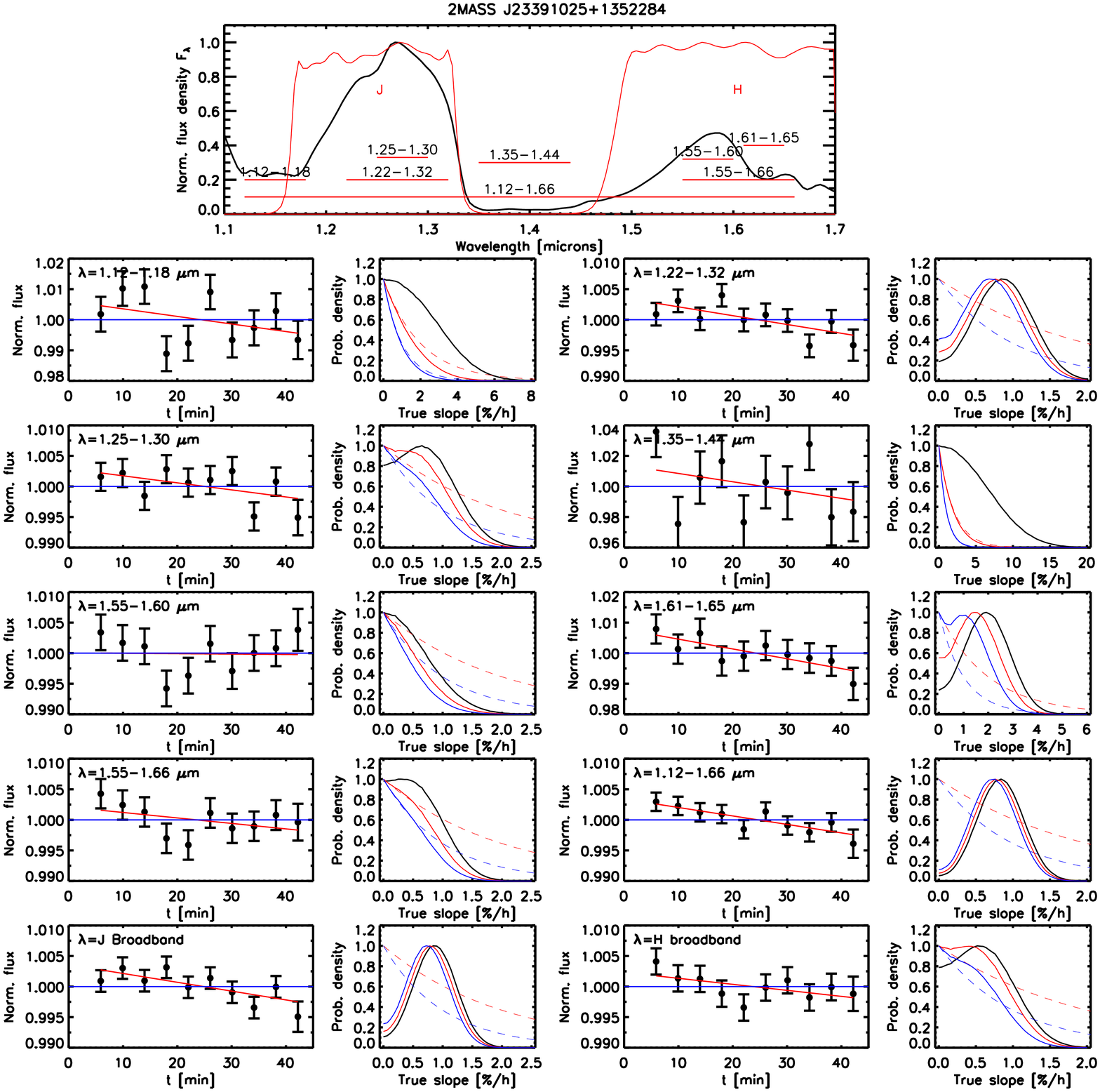}
\caption{Like Fig. \ref{fig:v01} but for 2M2339+13.
\label{fig:v67}}
\end{figure*}

\begin{figure*}
\epsscale{1.0}
\plotone{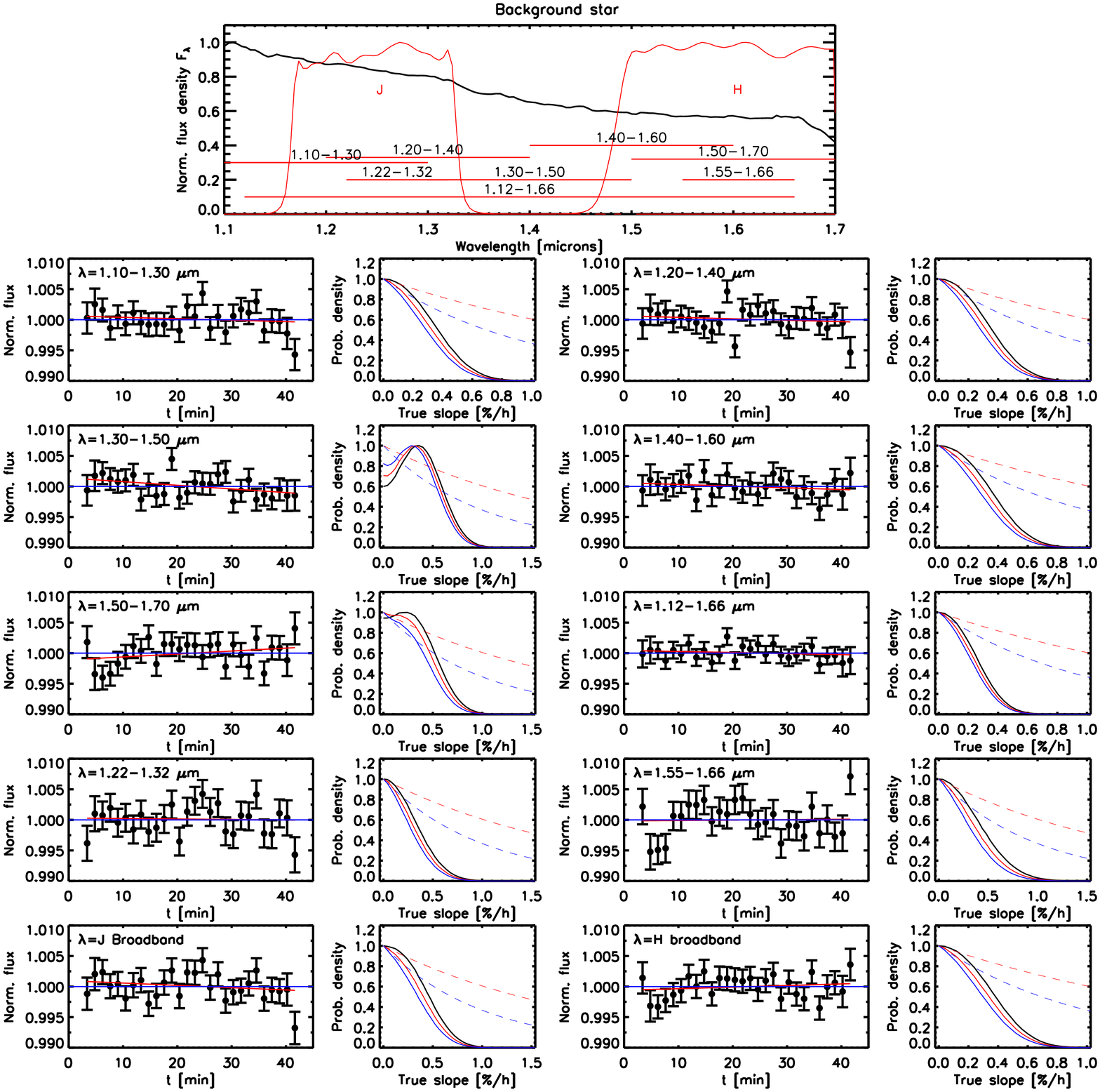}
\caption{Like Fig. \ref{fig:v01} but for a non-variable background star used to test the data reduction and calibration. 
\label{fig:v98}}
\end{figure*}

\clearpage

\LongTables

\begin{deluxetable*}{lcccccccccccccc}
\tablecaption{Maximum likelihood and confidence intervals for slopes for all sources and all wavelength bands \label{tab:all}} 
\tablehead{
\colhead{Target name} &  \colhead{Filter} &\colhead{ML} & \colhead{L95}  & \colhead{L68}  &  \colhead{Max}  & \colhead{U68} & \colhead{U95} & \colhead{L95}  & \colhead{L68}  &  \colhead{Max}  & \colhead{U68} & \colhead{U95} & \colhead{p} & \colhead{Red.}\\
\colhead{           } &  [$\mu$m] &  &    \colhead{P1}		&    \colhead{P1}	 &    \colhead{P1}   &  \colhead{P1}  &  \colhead{P1}  &  \colhead{P2}  &  \colhead{P2}  &  \colhead{P2}  &  \colhead{P2}  &  \colhead{P2} & \colhead{null} & \colhead{$\chi^2$}
}
\startdata
2MASS J00001354+2554180 & 1.12-1.18 & 2.16 & 0.00 & 0.40 & 1.56 & 2.46 & 3.40 & 0.00 & 0.00 & 0.00 & 1.58 & 2.84  & 0.01  &  1.42  \\
 & 1.22-1.32 & 0.00 & 0.00 & 0.00 & 0.00 & 0.36 & 0.70 & 0.00 & 0.00 & 0.00 & 0.33 & 0.66 & 0.94  &  0.50 \\
 & 1.25-1.30 & 0.00 & 0.00 & 0.00 & 0.00 & 0.38 & 0.76 & 0.00 & 0.00 & 0.00 & 0.35 & 0.71 & 0.98  &  0.41 \\
 & 1.35-1.44 & 0.00 & 0.00 & 0.00 & 0.00 & 1.61 & 3.67 & 0.00 & 0.00 & 0.00 & 1.01 & 2.48  & 0.20  &  1.24 \\
 & 1.55-1.60 & 0.00 & 0.00 & 0.00 & 0.00 & 0.48 & 0.94 & 0.00 & 0.00 & 0.00 & 0.42 & 0.87 & 0.85  &  0.63  \\
 & 1.61-1.65 & 0.00 & 0.00 & 0.00 & 0.00 & 0.74 & 1.51 & 0.00 & 0.00 & 0.00 & 0.61 & 1.32 & 0.29  &  1.14 \\
 & 1.55-1.66 & 0.00 & 0.00 & 0.00 & 0.00 & 0.43 & 0.83 & 0.00 & 0.00 & 0.00 & 0.39 & 0.78  & 0.69  &  0.77 \\
 & 1.12-1.66 & 0.36 & 0.00 & 0.04 & 0.32 & 0.51 & 0.76 & 0.00 & 0.00 & 0.28 & 0.43 & 0.73 & 0.89  &  0.42 \\
 & J   & 0.24 & 0.00 & 0.00 & 0.00 & 0.45 & 0.79 & 0.00 & 0.00 & 0.00 & 0.41 & 0.75 & 0.82  &  0.57   \\
 & H   & 0.00 & 0.00 & 0.00 & 0.00 & 0.36 & 0.71 & 0.00 & 0.00 & 0.00 & 0.33 & 0.67   & 0.99  &  0.34   \\
2MASS J02431371-2453298 & 1.12-1.18 & 0.00 & 0.00 & 0.00 & 0.00 & 1.01 & 2.16 & 0.00 & 0.00 & 0.00 & 0.76 & 1.75  & 0.24  &  1.21 \\
& 1.22-1.32 & 0.00 & 0.00 & 0.00 & 0.00 & 0.35 & 0.70 & 0.00 & 0.00 & 0.00 & 0.32 & 0.66   & 0.16  &  1.33 \\
& 1.25-1.30 & 0.00 & 0.00 & 0.00 & 0.00 & 0.38 & 0.77 & 0.00 & 0.00 & 0.00 & 0.35 & 0.72   & 0.24  &  1.22 \\
& 1.35-1.44 & 4.50 & 0.00 & 0.00 & 0.00 & 2.85 & 5.76 & 0.00 & 0.00 & 0.00 & 1.40 & 3.51  & 0.34  &  0.87 \\
& 1.55-1.60 & 0.80 & 0.00 & 0.25 & 0.68 & 1.09 & 1.41 & 0.00 & 0.13 & 0.60 & 0.94 & 1.32  & 0.69  &  0.52 \\
& 1.61-1.65 & 2.00 & 0.00 & 0.65 & 1.50 & 2.44 & 3.10 & 0.00 & 0.00 & 1.00 & 1.62 & 2.69    & 0.054  &  1.14 \\
& 1.55-1.66 & 0.00 & 0.00 & 0.00 & 0.00 & 0.37 & 0.74 & 0.00 & 0.00 & 0.00 & 0.34 & 0.69    & 0.73  &  0.76 \\
& 1.12-1.66 & 0.08 & 0.00 & 0.00 & 0.00 & 0.35 & 0.64 & 0.00 & 0.00 & 0.00 & 0.33 & 0.61  & 0.46  &  0.90 \\
& J   & 0.00 & 0.00 & 0.00 & 0.00 & 0.38 & 0.73 & 0.00 & 0.00 & 0.00 & 0.35 & 0.69    & 0.25  &  1.14 \\
& H   & 0.36 & 0.00 & 0.00 & 0.00 & 0.59 & 1.05 & 0.00 & 0.00 & 0.00 & 0.53 & 0.98    & 0.14  &  1.22 \\
2MASS J03105986+1648155 & 1.26-1.32 & 2.30 & 0.97 & 1.52 & 2.10 & 2.70 & 3.28 & 0.79 & 1.35 & 1.90 & 2.52 & 3.10 & 0.0004  &  0.96 \\
 & 1.35-1.44 & 0.08 & 0.00 & 0.00 & 0.00 & 0.75 & 1.53 & 0.00 & 0.00 & 0.00 & 0.62 & 1.34 & 0.89  &  0.39 \\
 & 1.55-1.61 & 1.02 & 0.00 & 0.18 & 0.84 & 1.32 & 1.88 & 0.00 & 0.00 & 0.00 & 0.99 & 1.70 & 0.077  &  1.15  \\
2MASS J04210718-6306022 & 1.12-1.20 & 0.00 & 0.00 & 0.00 & 0.00 & 0.51 & 1.03 & 0.00 & 0.00 & 0.00 & 0.45 & 0.94   & 0.57  &  0.89 \\
 & 1.22-1.32 & 0.00 & 0.00 & 0.00 & 0.00 & 0.42 & 0.80 & 0.00 & 0.00 & 0.00 & 0.38 & 0.76   & 0.96  &  0.42 \\
 & 1.26-1.32 & 0.00 & 0.00 & 0.00 & 0.00 & 0.38 & 0.76 & 0.00 & 0.00 & 0.00 & 0.34 & 0.71   & 0.92  &  0.55 \\
 & 1.35-1.44 & 0.00 & 0.00 & 0.00 & 0.00 & 0.50 & 0.99 & 0.00 & 0.00 & 0.00 & 0.45 & 0.91   & 0.42  &  1.01 \\
 & 1.55-1.61 & 0.00 & 0.00 & 0.00 & 0.00 & 0.37 & 0.73 & 0.00 & 0.00 & 0.00 & 0.33 & 0.69   & 0.97  &  0.44 \\
 & 1.62-1.66 & 1.08 & 0.00 & 0.40 & 0.90 & 1.44 & 1.81 & 0.00 & 0.26 & 0.78 & 1.26 & 1.67   & 0.017  &  1.34 \\
 & 1.55-1.66 & 0.48 & 0.00 & 0.11 & 0.44 & 0.67 & 0.93 & 0.00 & 0.02 & 0.36 & 0.55 & 0.89   & 0.37  &  0.86 \\
 & 1.12-1.66 & 0.00 & 0.00 & 0.00 & 0.00 & 0.31 & 0.56 & 0.00 & 0.00 & 0.00 & 0.29 & 0.54   & 0.85  &  0.57 \\
 & J   & 0.32 & 0.00 & 0.00 & 0.00 & 0.50 & 0.87 & 0.00 & 0.00 & 0.00 & 0.46 & 0.83   & 0.83 &  0.55 \\
 & H   & 0.60 & 0.02 & 0.28 & 0.52 & 0.81 & 1.00 & 0.00 & 0.25 & 0.52 & 0.77 & 0.95   & 0.49  &  0.56 \\
2MASS J05395200-0059019 & 1.12-1.20 & 0.00 & 0.00 & 0.00 & 0.00 & 0.31 & 0.62 & 0.00 & 0.00 & 0.00 & 0.29 & 0.58   & 0.55  &  0.91 \\
 & 1.22-1.32 & 0.36 & 0.00 & 0.03 & 0.32 & 0.49 & 0.75 & 0.00 & 0.00 & 0.24 & 0.43 & 0.72   & 0.94  &  0.31 \\
 & 1.26-1.32 & 0.00 & 0.00 & 0.00 & 0.00 & 0.34 & 0.64 & 0.00 & 0.00 & 0.00 & 0.31 & 0.61   & 0.92  &  0.50 \\
 & 1.35-1.44 & 0.00 & 0.00 & 0.00 & 0.00 & 0.31 & 0.62 & 0.00 & 0.00 & 0.00 & 0.29 & 0.59   & 0.93  &  0.52 \\
 & 1.55-1.61 & 0.00 & 0.00 & 0.00 & 0.00 & 0.33 & 0.64 & 0.00 & 0.00 & 0.00 & 0.31 & 0.61   & 0.81  &  0.64 \\
 & 1.62-1.66 & 0.24 & 0.00 & 0.00 & 0.00 & 0.52 & 0.94 & 0.00 & 0.00 & 0.00 & 0.47 & 0.88   & 0.19  &  1.17 \\
 & 1.55-1.66 & 0.00 & 0.00 & 0.00 & 0.00 & 0.32 & 0.60 & 0.00 & 0.00 & 0.00 & 0.30 & 0.57   & 0.91  &  0.49 \\
 & 1.12-1.66 & 0.00 & 0.00 & 0.00 & 0.00 & 0.27 & 0.49 & 0.00 & 0.00 & 0.00 & 0.25 & 0.47   & 0.99  &  0.17 \\
 & J   & 0.00 & 0.00 & 0.00 & 0.00 & 0.30 & 0.56 & 0.00 & 0.00 & 0.00 & 0.28 & 0.54   & 0.96  &  0.41 \\
 & H   & 0.00 & 0.00 & 0.00 & 0.00 & 0.21 & 0.42 & 0.00 & 0.00 & 0.00 & 0.20 & 0.41   & 0.78  &  0.71 \\
2MASS J05591914-1404488 & 1.12-1.18 & 0.60 & 0.00 & 0.00 & 0.00 & 0.84 & 1.51 & 0.00 & 0.00 & 0.00 & 0.71 & 1.37  & 0.11  &  1.25 \\
 & 1.22-1.32 & 0.40 & 0.00 & 0.17 & 0.40 & 0.59 & 0.74 & 0.00 & 0.15 & 0.36 & 0.57 & 0.71  & 0.70  &  0.65 \\
 & 1.25-1.30 & 0.72 & 0.22 & 0.46 & 0.72 & 0.97 & 1.21 & 0.18 & 0.42 & 0.68 & 0.93 & 1.17  & 0.17  &  0.82 \\
 & 1.35-1.44 & 1.10 & 0.00 & 0.00 & 0.00 & 1.42 & 2.73 & 0.00 & 0.00 & 0.00 & 1.01 & 2.20  & 0.28  &  1.06 \\
 & 1.55-1.60 & 0.72 & 0.05 & 0.36 & 0.68 & 0.98 & 1.23 & 0.00 & 0.31 & 0.64 & 0.93 & 1.14  & 0.10  &  1.06 \\
 & 1.61-1.65 & 1.97 & 0.99 & 1.42 & 1.88 & 2.34 & 2.78 & 0.89 & 1.31 & 1.70 & 2.23 & 2.68  & 0.0095  &  0.79 \\
 & 1.55-1.66 & 1.36 & 0.85 & 1.08 & 1.36 & 1.59 & 1.84 & 0.81 & 1.05 & 1.30 & 1.56 & 1.80  & 0.0003  &  0.70 \\
 & 1.12-1.66 & 0.28 & 0.00 & 0.07 & 0.24 & 0.41 & 0.55 & 0.00 & 0.05 & 0.24 & 0.38 & 0.54  & 0.95  &  0.46 \\
 & J   & 0.48 & 0.07 & 0.27 & 0.48 & 0.67 & 0.85 & 0.05 & 0.25 & 0.44 & 0.65 & 0.82  & 0.64  &  0.58 \\
 & H   & 1.24 & 0.77 & 0.99 & 1.24 & 1.45 & 1.67 & 0.74 & 0.96 & 1.18 & 1.42 & 1.64  & 0.0004  &  0.67 \\
2MASS J06244595-4521548 & 1.12-1.20 & 1.60 & 0.00 & 0.12 & 1.10 & 1.68 & 2.59 & 0.00 & 0.00 & 0.00 & 1.24 & 2.23  & 0.021  &  1.44 \\
 & 1.22-1.32 & 1.08 & 0.00 & 0.40 & 1.02 & 1.50 & 1.90 & 0.00 & 0.23 & 0.84 & 1.28 & 1.74  & 0.14  &  1.08 \\
 & 1.26-1.32 & 0.84 & 0.00 & 0.00 & 0.00 & 1.03 & 1.77 & 0.00 & 0.00 & 0.00 & 0.86 & 1.58\ & 0.15  &  0.97 \\
 & 1.35-1.44 & 1.90 & 0.49 & 1.07 & 1.70 & 2.29 & 2.88 & 0.32 & 0.88 & 1.50 & 2.10 & 2.68  & 0.031  &  0.36 \\
 & 1.55-1.61 & 1.20 & 0.03 & 0.55 & 1.02 & 1.58 & 1.96 & 0.00 & 0.41 & 0.96 & 1.43 & 1.79  & 0.24  &  0.71 \\
 & 1.62-1.66 & 1.76 & 0.46 & 1.01 & 1.60 & 2.18 & 2.74 & 0.30 & 0.84 & 1.44 & 1.99 & 2.55  & 0.14  &  0.41 \\
 & 1.55-1.66 & 1.38 & 0.45 & 0.85 & 1.26 & 1.70 & 2.10 & 0.37 & 0.76 & 1.20 & 1.61 & 2.01  & 0.046  &  0.73 \\
 & 1.12-1.66 & 1.44 & 0.73 & 1.04 & 1.38 & 1.70 & 2.02 & 0.68 & 0.99 & 1.32 & 1.64 & 1.96  & 0.002  &  0.61 \\
 & J   & 1.26 & 0.18 & 0.63 & 1.08 & 1.57 & 2.00 & 0.05 & 0.52 & 1.02 & 1.46 & 1.82  & 0.11  &  0.93 \\
 & H   & 1.32 & 0.52 & 0.87 & 1.26 & 1.63 & 1.99 & 0.45 & 0.80 & 1.14 & 1.56 & 1.92  & 0.022  &  0.85 \\
2MASS J08014056+4628498 & 1.12-1.20 & 1.12 & 0.00 & 0.00 & 0.64 & 1.14 & 1.94 & 0.00 & 0.00 & 0.00 & 0.95 & 1.73   & 0.44  &  0.61 \\
 & 1.22-1.32 & 0.40 & 0.00 & 0.00 & 0.00 & 0.68 & 1.22 & 0.00 & 0.00 & 0.00 & 0.60 & 1.12   & 0.43  &  0.77 \\
 & 1.26-1.32 & 0.72 & 0.00 & 0.00 & 0.00 & 0.94 & 1.70 & 0.00 & 0.00 & 0.00 & 0.78 & 1.51   & 0.12  &  1.25 \\
 & 1.35-1.44 & 0.06 & 0.00 & 0.00 & 0.00 & 0.77 & 1.51 & 0.00 & 0.00 & 0.00 & 0.64 & 1.34   & 0.61  &  0.68 \\
 & 1.55-1.61 & 0.04 & 0.00 & 0.00 & 0.00 & 0.53 & 1.05 & 0.00 & 0.00 & 0.00 & 0.46 & 0.96   & 0.86  &  0.47 \\
 & 1.62-1.66 & 0.06 & 0.00 & 0.00 & 0.00 & 0.59 & 1.20 & 0.00 & 0.00 & 0.00 & 0.51 & 1.08   & 0.39  &  1.04 \\
 & 1.55-1.66 & 0.00 & 0.00 & 0.00 & 0.00 & 0.41 & 0.82 & 0.00 & 0.00 & 0.00 & 0.37 & 0.77   & 0.56  &  0.84 \\
 & 1.12-1.66 & 0.48 & 0.00 & 0.11 & 0.40 & 0.67 & 0.92 & 0.00 & 0.04 & 0.36 & 0.57 & 0.88   & 0.65  &  0.30 \\
 & J   & 0.56 & 0.00 & 0.00 & 0.40 & 0.72 & 1.21 & 0.00 & 0.00 & 0.00 & 0.64 & 1.13   & 0.70  &  0.35 \\
 & H   & 0.00 & 0.00 & 0.00 & 0.00 & 0.38 & 0.75 & 0.00 & 0.00 & 0.00 & 0.35 & 0.71   & 0.93  &  0.35 \\
2MASS J08173001-6155158 & 1.12-1.18 & 0.54 & 0.00 & 0.00 & 0.00 & 0.82 & 1.48 & 0.00 & 0.00 & 0.00 & 0.70 & 1.34 & 0.22  &  1.11 \\
 & 1.22-1.32 & 0.76 & 0.39 & 0.55 & 0.72 & 0.90 & 1.07 & 0.37 & 0.54 & 0.72 & 0.89 & 1.06 & 0.070  &  0.58  \\
 & 1.25-1.30 & 0.80 & 0.36 & 0.56 & 0.76 & 0.99 & 1.19 & 0.33 & 0.54 & 0.76 & 0.96 & 1.17 & 0.005  &  1.08  \\
 & 1.35-1.44 & 0.56 & 0.00 & 0.00 & 0.00 & 1.62 & 3.39 & 0.00 & 0.00 & 0.00 & 1.04 & 2.48 & 0.053  &  1.37 \\
 & 1.55-1.60 & 0.72 & 0.16 & 0.42 & 0.68 & 0.97 & 1.23 & 0.12 & 0.38 & 0.68 & 0.93 & 1.19 & 0.017  &  1.20 \\
 & J   & 0.60 & 0.27 & 0.43 & 0.60 & 0.77 & 0.94 & 0.25 & 0.41 & 0.60 & 0.76 & 0.92 & 0.37  &  0.47 \\
 2MASS J08251968+2115521 & 1.12-1.20 & 1.60 & 0.85 & 1.19 & 1.53 & 1.95 & 2.31 & 0.78 & 1.13 & 1.53 & 1.87 & 2.23  & 0.002  &  0.85 \\
  & 1.22-1.32 & 1.02 & 0.47 & 0.73 & 1.02 & 1.28 & 1.54 & 0.43 & 0.69 & 0.97 & 1.24 & 1.50  & 0.003  &  1.04  \\
  & 1.26-1.32 & 1.15 & 0.39 & 0.74 & 1.09 & 1.47 & 1.83 & 0.32 & 0.67 & 1.04 & 1.41 & 1.76  & 0.001  &  1.25 \\
  & 1.35-1.44 & 1.24 & 0.52 & 0.85 & 1.18 & 1.55 & 1.88 & 0.47 & 0.79 & 1.12 & 1.49 & 1.82  & 0.068  &  0.56 \\
  & 1.55-1.61 & 1.04 & 0.48 & 0.75 & 1.04 & 1.31 & 1.58 & 0.45 & 0.71 & 0.99 & 1.27 & 1.54  & 0.012  &  0.82 \\
  & 1.62-1.66 & 1.04 & 0.38 & 0.68 & 0.99 & 1.33 & 1.64 & 0.33 & 0.63 & 0.94 & 1.28 & 1.59  & 0.010  &  1.06 \\
  & 1.55-1.66 & 1.05 & 0.59 & 0.81 & 1.05 & 1.28 & 1.50 & 0.57 & 0.78 & 1.00 & 1.25 & 1.48  & 0.001  &  0.72 \\
  & 1.12-1.66 & 1.20 & 0.84 & 1.02 & 1.20 & 1.40 & 1.58 & 0.83 & 1.00 & 1.20 & 1.38 & 1.56  & $<10^{-4}$  &  0.53 \\
  & J   & 1.11 & 0.62 & 0.85 & 1.11 & 1.35 & 1.59 & 0.59 & 0.82 & 1.05 & 1.32 & 1.56  & 0.0004  &  0.96 \\
  & H   & 0.93 & 0.52 & 0.71 & 0.93 & 1.14 & 1.34 & 0.50 & 0.69 & 0.93 & 1.11 & 1.32  & 0.003  &  0.65 \\
2MASS J09083803+5032088 & 1.12-1.20 & 0.44 & 0.00 & 0.00 & 0.00 & 0.62 & 1.07 & 0.00 & 0.00 & 0.00 & 0.56 & 1.00  & 0.46  &  0.85 \\
 & 1.22-1.32 & 0.00 & 0.00 & 0.00 & 0.00 & 0.30 & 0.58 & 0.00 & 0.00 & 0.00 & 0.28 & 0.55  & 0.61  &  0.86 \\
 & 1.26-1.32 & 0.00 & 0.00 & 0.00 & 0.00 & 0.33 & 0.65 & 0.00 & 0.00 & 0.00 & 0.30 & 0.61  & 0.35  &  1.10 \\
 & 1.35-1.44 & 0.00 & 0.00 & 0.00 & 0.00 & 0.47 & 0.90 & 0.00 & 0.00 & 0.00 & 0.42 & 0.84  & 0.83  &  0.62 \\
 & 1.55-1.61 & 0.00 & 0.00 & 0.00 & 0.00 & 0.39 & 0.74 & 0.00 & 0.00 & 0.00 & 0.35 & 0.70  & 0.51  &  0.89 \\
 & 1.62-1.66 & 0.00 & 0.00 & 0.00 & 0.00 & 0.42 & 0.83 & 0.00 & 0.00 & 0.00 & 0.38 & 0.77  & 0.29  &  1.13 \\
 & 1.55-1.66 & 0.04 & 0.00 & 0.00 & 0.00 & 0.36 & 0.66 & 0.00 & 0.00 & 0.00 & 0.33 & 0.63  & 0.88  &  0.53 \\
 & 1.12-1.66 & 0.00 & 0.00 & 0.00 & 0.00 & 0.23 & 0.45 & 0.00 & 0.00 & 0.00 & 0.22 & 0.43  & 0.99  &  0.33 \\
 & J   & 0.00 & 0.00 & 0.00 & 0.00 & 0.31 & 0.59 & 0.00 & 0.00 & 0.00 & 0.29 & 0.56  & 0.81  &  0.65 \\
 & H   & 0.00 & 0.00 & 0.00 & 0.00 & 0.27 & 0.52 & 0.00 & 0.00 & 0.00 & 0.26 & 0.50  & 0.98  &  0.38 \\
2MASS J09090085+6525275 & 1.12-1.18 & 0.00 & 0.00 & 0.00 & 0.00 & 0.83 & 1.71 & 0.00 & 0.00 & 0.00 & 0.67 & 1.46  & 0.86  &  0.445\\
 & 1.22-1.32 & 0.00 & 0.00 & 0.00 & 0.00 & 0.49 & 1.00 & 0.00 & 0.00 & 0.00 & 0.43 & 0.91  & 0.29  &  1.22 \\
 & 1.25-1.31 & 0.00 & 0.00 & 0.00 & 0.00 & 0.66 & 1.35 & 0.00 & 0.00 & 0.00 & 0.56 & 1.20  & 0.24  &  1.24 \\
 & 1.35-1.44 & 0.32 & 0.00 & 0.00 & 0.00 & 1.50 & 3.34 & 0.00 & 0.00 & 0.00 & 0.98 & 2.36  & 0.04  &  1.85 \\
 & 1.55-1.61 & 0.00 & 0.00 & 0.00 & 0.00 & 0.57 & 1.14 & 0.00 & 0.00 & 0.00 & 0.49 & 1.03  & 0.87  &  0.43 \\
 & 1.61-1.65 & 1.52 & 0.00 & 0.50 & 1.28 & 1.91 & 2.41 & 0.00 & 0.20 & 0.96 & 1.52 & 2.16  & 0.17  &  0.67 \\
 & 1.55-1.66 & 0.00 & 0.00 & 0.00 & 0.00 & 0.49 & 0.97 & 0.00 & 0.00 & 0.00 & 0.43 & 0.89 & 0.84  &  0.45 \\
 & 1.12-1.66 & 0.00 & 0.00 & 0.00 & 0.00 & 0.40 & 0.77 & 0.00 & 0.00 & 0.00 & 0.37 & 0.73  & 0.51  &  0.80 \\
 & J   & 0.00 & 0.00 & 0.00 & 0.00 & 0.46 & 0.93 & 0.00 & 0.00 & 0.00 & 0.41 & 0.85  & 0.30  &  1.20 \\
 & H   & 0.00 & 0.00 & 0.00 & 0.00 & 0.52 & 1.01 & 0.00 & 0.00 & 0.00 & 0.46 & 0.93  & 0.28  &  1.10 \\
 2MASS J10393137+3256263 & 1.12-1.18 & 0.00 & 0.00 & 0.00 & 0.00 & 0.87 & 1.80 & 0.00 & 0.00 & 0.00 & 0.69 & 1.52  & 0.34  &  1.09 \\
  & 1.22-1.32 & 0.04 & 0.00 & 0.00 & 0.00 & 0.57 & 1.06 & 0.00 & 0.00 & 0.00 & 0.50 & 0.98  & 0.41  &  0.86 \\
  & 1.25-1.31 & 0.00 & 0.00 & 0.00 & 0.00 & 0.74 & 1.53 & 0.00 & 0.00 & 0.00 & 0.60 & 1.32  & 0.055  &  1.90 \\
  & 1.35-1.44 & 1.28 & 0.00 & 0.00 & 0.00 & 1.70 & 3.58 & 0.00 & 0.00 & 0.00 & 1.07 & 2.57  & 0.004 &  2.14 \\
  & 1.55-1.61 & 1.38 & 0.22 & 0.71 & 1.20 & 1.75 & 2.24 & 0.08 & 0.57 & 1.08 & 1.61 & 2.03  & 0.029  &  0.98 \\
  & 1.61-1.65 & 0.00 & 0.00 & 0.00 & 0.00 & 0.63 & 1.28 & 0.00 & 0.00 & 0.00 & 0.54 & 1.14  & 0.39  &  1.04 \\
  & 1.55-1.66 & 0.72 & 0.00 & 0.17 & 0.60 & 0.97 & 1.32 & 0.00 & 0.00 & 0.52 & 0.74 & 1.24  & 0.51  &  0.42 \\
  & 1.12-1.66 & 0.44 & 0.00 & 0.08 & 0.40 & 0.64 & 0.92 & 0.00 & 0.00 & 0.32 & 0.52 & 0.87  & 0.45  &  0.57 \\
  & J   & 0.04 & 0.00 & 0.00 & 0.00 & 0.44 & 0.86 & 0.00 & 0.00 & 0.00 & 0.40 & 0.81  & 0.62  &  0.72 \\
  & H   & 0.00 & 0.00 & 0.00 & 0.00 & 0.44 & 0.85 & 0.00 & 0.00 & 0.00 & 0.40 & 0.79  & 0.48  &  0.85 \\
 2MASS J12195156+3128497 & 1.12-1.20 & 5.46 & 1.83 & 3.08 & 4.42 & 5.77 & 7.10 & 1.03 & 2.22 & 3.38 & 4.84 & 6.16  & 0.0001  &  0.09 \\
  & 1.22-1.32 & 1.44 & 0.10 & 0.69 & 1.28 & 1.89 & 2.38 & 0.00 & 0.52 & 1.12 & 1.70 & 2.11  & 0.024  &  1.03 \\
  & 1.26-1.32 & 1.30 & 0.00 & 0.00 & 0.00 & 1.41 & 2.53 & 0.00 & 0.00 & 0.00 & 1.07 & 2.12  & 0.028  &  1.50 \\
  & 1.35-1.44 & 0.00 & 0.00 & 0.00 & 0.00 & 0.91 & 1.92 & 0.00 & 0.00 & 0.00 & 0.71 & 1.60  & 0.29  &  1.20 \\
  & 1.55-1.61 & 0.00 & 0.00 & 0.00 & 0.00 & 0.67 & 1.34 & 0.00 & 0.00 & 0.00 & 0.57 & 1.20  & 0.44  &  0.91 \\
  & 1.62-1.66 & 0.00 & 0.00 & 0.00 & 0.00 & 0.75 & 1.51 & 0.00 & 0.00 & 0.00 & 0.62 & 1.32  & 0.56  &  0.74 \\
  & 1.55-1.66 & 0.00 & 0.00 & 0.00 & 0.00 & 0.63 & 1.21 & 0.00 & 0.00 & 0.00 & 0.54 & 1.10  & 0.65  &  0.53 \\
  & 1.12-1.66 & 0.72 & 0.00 & 0.00 & 0.00 & 0.84 & 1.45 & 0.00 & 0.00 & 0.00 & 0.73 & 1.32  & 0.050  &  1.38 \\
  & J   & 1.44 & 0.09 & 0.68 & 1.28 & 1.88 & 2.35 & 0.00 & 0.51 & 1.12 & 1.68 & 2.09  & 0.011  &  1.15 \\
  & H   & 0.56 & 0.00 & 0.00 & 0.00 & 0.74 & 1.29 & 0.00 & 0.00 & 0.00 & 0.65 & 1.19  & 0.58  &  0.39 \\
2MASS J13243553+6358281 & 1.12-1.18 & 0.00 & 0.00 & 0.00 & 0.00 & 0.64 & 1.30 & 0.00 & 0.00 & 0.00 & 0.54 & 1.16  & 0.62  &  0.87 \\
 & 1.22-1.32 & 0.20 & 0.00 & 0.00 & 0.00 & 0.46 & 0.82 & 0.00 & 0.00 & 0.00 & 0.42 & 0.78  & 0.42  &  0.93 \\
 & 1.25-1.31 & 0.32 & 0.00 & 0.00 & 0.00 & 0.54 & 0.97 & 0.00 & 0.00 & 0.00 & 0.49 & 0.91  & 0.48  &  0.88 \\
 & 1.35-1.44 & 0.06 & 0.00 & 0.00 & 0.00 & 0.70 & 1.42 & 0.00 & 0.00 & 0.00 & 0.58 & 1.25  & 0.42  &  1.02 \\
 & 1.55-1.61 & 0.00 & 0.00 & 0.00 & 0.00 & 0.36 & 0.71 & 0.00 & 0.00 & 0.00 & 0.33 & 0.67  & 0.63  &  0.86 \\
 & 1.61-1.65 & 0.00 & 0.00 & 0.00 & 0.00 & 0.50 & 1.01 & 0.00 & 0.00 & 0.00 & 0.44 & 0.93  & 0.15  &  1.33 \\
 & 1.55-1.66 & 0.00 & 0.00 & 0.00 & 0.00 & 0.28 & 0.55 & 0.00 & 0.00 & 0.00 & 0.26 & 0.53  & 0.75  &  0.76 \\
 & 1.12-1.66 & 0.00 & 0.00 & 0.00 & 0.00 & 0.25 & 0.50 & 0.00 & 0.00 & 0.00 & 0.24 & 0.48  & 0.34  &  1.09 \\
 & J   & 0.00 & 0.00 & 0.00 & 0.00 & 0.29 & 0.58 & 0.00 & 0.00 & 0.00 & 0.27 & 0.55  & 0.40  &  1.05 \\
 & H   & 0.00 & 0.00 & 0.00 & 0.00 & 0.25 & 0.49 & 0.00 & 0.00 & 0.00 & 0.23 & 0.47  & 0.62  &  0.87 \\
2MASS J15150083+4847416 & 1.12-1.20 & 0.64 & 0.00 & 0.23 & 0.60 & 0.95 & 1.21 & 0.00 & 0.15 & 0.52 & 0.85 & 1.15  & 0.45 &  0.68 \\
 & 1.22-1.32 & 0.00 & 0.00 & 0.00 & 0.00 & 0.40 & 0.72 & 0.00 & 0.00 & 0.00 & 0.36 & 0.69  & 0.98  &  0.29 \\
 & 1.26-1.32 & 0.24 & 0.00 & 0.00 & 0.00 & 0.47 & 0.85 & 0.00 & 0.00 & 0.00 & 0.43 & 0.80  & 0.82  &  0.53 \\
 & 1.35-1.44 & 0.00 & 0.00 & 0.00 & 0.00 & 0.48 & 0.94 & 0.00 & 0.00 & 0.00 & 0.43 & 0.87  & 0.25  &  1.15 \\
 & 1.55-1.61 & 0.48 & 0.00 & 0.04 & 0.40 & 0.61 & 0.94 & 0.00 & 0.00 & 0.28 & 0.53 & 0.90  & 0.70  &  0.56 \\
 & 1.62-1.66 & 0.52 & 0.00 & 0.00 & 0.40 & 0.63 & 1.04 & 0.00 & 0.00 & 0.00 & 0.57 & 0.98  & 0.41  &  0.84 \\
 & 1.55-1.66 & 0.00 & 0.00 & 0.00 & 0.00 & 0.27 & 0.53 & 0.00 & 0.00 & 0.00 & 0.25 & 0.51  & 0.73  &  0.74 \\
 & 1.12-1.66 & 0.00 & 0.00 & 0.00 & 0.00 & 0.24 & 0.47 & 0.00 & 0.00 & 0.00 & 0.23 & 0.45  & 0.97  &  0.40 \\
 & J   & 0.00 & 0.00 & 0.00 & 0.00 & 0.37 & 0.67 & 0.00 & 0.00 & 0.00 & 0.34 & 0.64  & 0.99  &  0.18 \\
 & H   & 0.56 & 0.03 & 0.27 & 0.52 & 0.76 & 0.94 & 0.00 & 0.24 & 0.48 & 0.73 & 0.89  & 0.37  &  0.60 \\
2MASS J16241436+0029158 & 1.22-1.32 & 0.00 & 0.00 & 0.00 & 0.00 & 0.41 & 0.78 & 0.00 & 0.00 & 0.00 & 0.37 & 0.74  & 0.93  &  0.45 \\
 & 1.25-1.30 & 0.00 & 0.00 & 0.00 & 0.00 & 0.43 & 0.85 & 0.00 & 0.00 & 0.00 & 0.39 & 0.79 & 0.90  &  0.55  \\
 & 1.35-1.44 & 13.80 & 0.00 & 1.44 & 4.80 & 8.67 & 12.12 & 0.00 & 0.00 & 0.00 & 2.61 & 6.06 & 0.058  &  0.62  \\
 & 1.55-1.60 & 0.06 & 0.00 & 0.00 & 0.00 & 0.77 & 1.60 & 0.00 & 0.00 & 0.00 & 0.63 & 1.38  & 0.037  &  1.70 \\
 & J   & 0.00 & 0.00 & 0.00 & 0.00 & 0.38 & 0.73 & 0.00 & 0.00 & 0.00 & 0.35 & 0.69 & 0.93  &  0.48 \\
 2MASS J16322911+1904407 & 1.12-1.20 & 0.00 & 0.00 & 0.00 & 0.00 & 0.66 & 1.28 & 0.00 & 0.00 & 0.00 & 0.56 & 1.16  & 0.37  &  0.96 \\
  & 1.22-1.32 & 0.04 & 0.00 & 0.00 & 0.00 & 0.49 & 0.95 & 0.00 & 0.00 & 0.00 & 0.44 & 0.88  & 0.28  &  1.10 \\
  & 1.26-1.32 & 0.04 & 0.00 & 0.00 & 0.00 & 0.57 & 1.08 & 0.00 & 0.00 & 0.00 & 0.50 & 0.99  & 0.34  &  0.99 \\
  & 1.35-1.44 & 0.04 & 0.00 & 0.00 & 0.00 & 0.58 & 1.14 & 0.00 & 0.00 & 0.00 & 0.50 & 1.04  & 0.78  &  0.55 \\
  & 1.55-1.61 & 0.00 & 0.00 & 0.00 & 0.00 & 0.43 & 0.85 & 0.00 & 0.00 & 0.00 & 0.39 & 0.80  & 0.99  &  0.14 \\
  & 1.62-1.66 & 0.48 & 0.00 & 0.00 & 0.00 & 0.86 & 1.61 & 0.00 & 0.00 & 0.00 & 0.71 & 1.42  & 0.053  &  1.53 \\
  & 1.55-1.66 & 0.00 & 0.00 & 0.00 & 0.00 & 0.38 & 0.73 & 0.00 & 0.00 & 0.00 & 0.35 & 0.69  & 0.83  &  0.49 \\
  & 1.12-1.66 & 0.24 & 0.00 & 0.00 & 0.00 & 0.41 & 0.72 & 0.00 & 0.00 & 0.00 & 0.39 & 0.69  & 0.93  &  0.13 \\
  & J   & 0.00 & 0.00 & 0.00 & 0.00 & 0.46 & 0.86 & 0.00 & 0.00 & 0.00 & 0.41 & 0.81  & 0.34  &  0.99 \\
  & H   & 0.72 & 0.07 & 0.36 & 0.68 & 0.97 & 1.22 & 0.01 & 0.32 & 0.60 & 0.92 & 1.13  & 0.27  &  0.31 \\
2MASS J17114573+2232044 & 1.12-1.20 & 0.00 & 0.00 & 0.00 & 0.00 & 1.38 & 3.06 & 0.00 & 0.00 & 0.00 & 0.93 & 2.23  & 0.11  &  1.61 \\
 & 1.22-1.32 & 0.96 & 0.00 & 0.00 & 0.00 & 1.12 & 2.02 & 0.00 & 0.00 & 0.00 & 0.90 & 1.76 & 0.21  &  1.01  \\
 & 1.26-1.32 & 1.50 & 0.00 & 0.00 & 0.00 & 1.52 & 2.60 & 0.00 & 0.00 & 0.00 & 1.17 & 2.21 & 0.15  &  0.99  \\
 & 1.35-1.44 & 0.00 & 0.00 & 0.00 & 0.00 & 1.04 & 2.21 & 0.00 & 0.00 & 0.00 & 0.79 & 1.78 & 0.55  &  0.81 \\
 & 1.55-1.61 & 0.00 & 0.00 & 0.00 & 0.00 & 0.97 & 2.03 & 0.00 & 0.00 & 0.00 & 0.74 & 1.68  & 0.20  &  1.35  \\
 & J   & 0.96 & 0.00 & 0.00 & 0.00 & 1.06 & 1.88 & 0.00 & 0.00 & 0.00 & 0.86 & 1.66  & 0.52  &  0.47  \\
2MASS J17502484-0016151 & 1.12-1.20 & 0.96 & 0.43 & 0.67 & 0.92 & 1.19 & 1.43 & 0.39 & 0.64 & 0.87 & 1.15 & 1.40  & 0.003  &  1.12 \\
 & 1.22-1.32 & 0.68 & 0.31 & 0.49 & 0.68 & 0.86 & 1.04 & 0.29 & 0.47 & 0.64 & 0.84 & 1.03  & 0.27  &  0.47 \\
 & 1.26-1.32 & 0.80 & 0.36 & 0.56 & 0.76 & 0.98 & 1.18 & 0.34 & 0.54 & 0.76 & 0.96 & 1.16  & 0.13  &  0.61 \\
 & 1.35-1.44 & 0.08 & 0.00 & 0.00 & 0.00 & 0.36 & 0.66 & 0.00 & 0.00 & 0.00 & 0.34 & 0.63  & 0.95  &  0.55 \\
 & 1.55-1.61 & 0.08 & 0.00 & 0.00 & 0.00 & 0.31 & 0.55 & 0.00 & 0.00 & 0.00 & 0.29 & 0.53  & 0.96  &  0.53 \\
 & 1.62-1.66 & 0.24 & 0.00 & 0.00 & 0.20 & 0.38 & 0.66 & 0.00 & 0.00 & 0.00 & 0.36 & 0.63  & 0.73  &  0.74 \\
 & 1.55-1.66 & 0.24 & 0.00 & 0.05 & 0.24 & 0.39 & 0.56 & 0.00 & 0.01 & 0.24 & 0.34 & 0.54  & 0.94  &  0.50 \\
 & 1.12-1.66 & 0.56 & 0.25 & 0.39 & 0.52 & 0.69 & 0.83 & 0.24 & 0.38 & 0.52 & 0.67 & 0.82  & 0.57  &  0.25 \\
 & J   & 0.80 & 0.43 & 0.60 & 0.76 & 0.95 & 1.12 & 0.41 & 0.58 & 0.76 & 0.93 & 1.10 & 0.018  &  0.65 \\
 & H   & 0.44 & 0.08 & 0.24 & 0.40 & 0.57 & 0.73 & 0.07 & 0.22 & 0.40 & 0.56 & 0.71  & 0.62  &  0.58 \\
 2MASS J17503293+1759042 & 1.12-1.18 & 0.00 & 0.00 & 0.00 & 0.00 & 1.38 & 3.05 & 0.00 & 0.00 & 0.00 & 0.93 & 2.23  & 0.27 &  1.20 \\
  & 1.22-1.32 & 0.00 & 0.00 & 0.00 & 0.00 & 0.75 & 1.51 & 0.00 & 0.00 & 0.00 & 0.62 & 1.32  & 0.72  &  0.52 \\
  & 1.25-1.30 & 0.00 & 0.00 & 0.00 & 0.00 & 0.83 & 1.72 & 0.00 & 0.00 & 0.00 & 0.67 & 1.47  & 0.73  &  0.55 \\
  & 1.35-1.44 & 0.80 & 0.00 & 0.00 & 0.00 & 2.02 & 5.04 & 0.00 & 0.00 & 0.00 & 1.10 & 2.84  & 0.048  &  2.22 \\
  & 1.55-1.60 & 2.24 & 0.00 & 0.00 & 0.00 & 2.06 & 3.80 & 0.00 & 0.00 & 0.00 & 1.33 & 2.87  & 0.014  &  1.56 \\
  & 1.61-1.65 & 0.00 & 0.00 & 0.00 & 0.00 & 1.30 & 2.78 & 0.00 & 0.00 & 0.00 & 0.91 & 2.12  & 0.30  &  1.06 \\
  & 1.55-1.66 & 2.00 & 0.00 & 0.78 & 1.60 & 2.47 & 3.06 & 0.00 & 0.43 & 1.30 & 2.04 & 2.70  & 0.11  &  0.33 \\
  & 1.12-1.66 & 1.08 & 0.00 & 0.29 & 0.84 & 1.40 & 1.86 & 0.00 & 0.00 & 0.66 & 1.02 & 1.70  & 0.28  &  0.35 \\
  & J   & 0.00 & 0.00 & 0.00 & 0.00 & 0.71 & 1.43 & 0.00 & 0.00 & 0.00 & 0.59 & 1.27  & 0.64  &  0.62 \\
  & H   & 1.90 & 0.07 & 0.81 & 1.60 & 2.32 & 2.90 & 0.00 & 0.54 & 1.30 & 2.01 & 2.55 & 0.13  &  0.10 \\
2MASS J23391025+1352284 & 1.12-1.18 & 0.00 & 0.00 & 0.00 & 0.00 & 1.50 & 3.30 & 0.00 & 0.00 & 0.00 & 0.98 & 2.36  & 0.020  &  1.96 \\
 & 1.22-1.32 & 0.84 & 0.00 & 0.39 & 0.76 & 1.18 & 1.44 & 0.00 & 0.31 & 0.68 & 1.09 & 1.36  & 0.045  &  1.08 \\
 & 1.25-1.30 & 0.65 & 0.00 & 0.00 & 0.00 & 0.82 & 1.42 & 0.00 & 0.00 & 0.00 & 0.71 & 1.30  & 0.13  &  1.18 \\
 & 1.35-1.44 & 0.00 & 0.00 & 0.00 & 0.00 & 2.02 & 4.98 & 0.00 & 0.00 & 0.00 & 1.10 & 2.84  & 0.081  &  1.60 \\
 & 1.55-1.60 & 0.00 & 0.00 & 0.00 & 0.00 & 0.58 & 1.20 & 0.00 & 0.00 & 0.00 & 0.50 & 1.08  & 0.27  &  1.24 \\
 & 1.61-1.65 & 1.92 & 0.00 & 0.55 & 1.44 & 2.36 & 3.04 & 0.00 & 0.00 & 0.00 & 1.54 & 2.63  & 0.32  &  0.44 \\
 & 1.55-1.66 & 0.25 & 0.00 & 0.00 & 0.00 & 0.67 & 1.23 & 0.00 & 0.00 & 0.00 & 0.58 & 1.13  & 0.28  &  1.01 \\
 & 1.12-1.66 & 0.84 & 0.18 & 0.47 & 0.76 & 1.09 & 1.38 & 0.14 & 0.42 & 0.76 & 1.04 & 1.33  & 0.082  &  0.55 \\
 & J   & 0.84 & 0.09 & 0.44 & 0.78 & 1.16 & 1.47 & 0.02 & 0.38 & 0.72 & 1.09 & 1.35  & 0.058  &  0.91 \\
 & H   & 0.52 & 0.00 & 0.00 & 0.00 & 0.71 & 1.22 & 0.00 & 0.00 & 0.00 & 0.63 & 1.13  & 0.41  &  0.72 
\enddata
\tablecomments{Columns are: Maximum likelihood for the slope, lower limit of 95\% and 68\% confidence interval, maximum of probability distribution, upper limit of  68\% and 95\% confidence interval (all confidence intervals are given for prior 1 and prior 2), p-value of null hypothesis from $\chi^2$ test and reduced $\chi^2$ value of the best fit. All slopes are given in \%/h. For 2M0624 (all wavelengths) and 2M1219 (only 1.12-1.20) the best-fit model was derived from the partial light curve due to non-linearity.}
\end{deluxetable*}

\end{document}